\documentclass[%
reprint,
superscriptaddress,
amssymb,
aps,
prl,
floatfix,
]{revtex4-2}

\usepackage{svg}
\usepackage{adjustbox}
\usepackage{tikz}
\usepackage{braket}
\usepackage{mathtools}
\usetikzlibrary{positioning, arrows.meta, calc, fit, backgrounds}
\definecolor{cTeal}{HTML}{1A6B5A}
\definecolor{cMaroon}{HTML}{5C1A1A}
\definecolor{cPurple}{HTML}{4E2A78}
\definecolor{cOlive}{HTML}{3D4A1F}
\definecolor{cSimBg}{HTML}{E8E6E0}
\definecolor{cCream}{HTML}{FFF8EC}

\usepackage{lineno}
\AtBeginEnvironment{figure}{\nolinenumbers}
\AtBeginEnvironment{figure*}{\nolinenumbers}
\AtBeginEnvironment{table}{\nolinenumbers}
\AtBeginEnvironment{table*}{\nolinenumbers}

\usepackage[svgnames]{xcolor}
\usepackage{graphicx}
\usepackage{caption}
\usepackage{subcaption}
\usepackage{dcolumn}
\usepackage{bm}
\usepackage{enumitem}
\setlist[itemize]{noitemsep, topsep=2pt, leftmargin=*}
\setlist[enumerate]{noitemsep, topsep=2pt, leftmargin=*}
\usepackage[colorlinks,linkcolor=blue,citecolor=blue,urlcolor=blue]{hyperref}
\usepackage{xr}
\newcommand{\rana}[1]{\textcolor{DarkOrange}
{#1}}

\newcommand{\YC}[1]{\textcolor{Crimson}{#1}}
\newcommand{\zach}[1]{\textcolor{purple}{Zach : #1}}

\newcommand{\su}[1]{\textcolor{Green}{#1}}

\newcommand{\SM}[1]{\textcolor{SteelBlue}{#1}}

\providecommand{\rana}[1]{}\renewcommand{\rana}[1]{}
\providecommand{\YC}[1]{}\renewcommand{\YC}[1]{}
\providecommand{\zach}[1]{}\renewcommand{\zach}[1]{}
\providecommand{\su}[1]{}\renewcommand{\su}[1]{}
\providecommand{\SM}[1]{}\renewcommand{\SM}[1]{}
\begin{document}


\title{Optimized Quantum States for Sensing in the Presence of Loss and Phase Noise}
\author{Shruti Maliakal}
\affiliation{Institute for Quantum Information and Matter, California Institute of Technology, Pasadena, CA, USA}
\author{Zachary Mann}
\affiliation{Institute for Quantum Information and Matter, California Institute of Technology, Pasadena, CA, USA}
\author{Christopher Wipf}
\affiliation{Institute for Quantum Information and Matter, California Institute of Technology, Pasadena, CA, USA}
\affiliation{LIGO Laboratory, California Institute of Technology, Pasadena, CA, USA}
\author{Rana X Adhikari}
\email{rana@caltech.edu}
\affiliation{Institute for Quantum Information and Matter, California Institute of Technology, Pasadena, CA, USA}
\affiliation{LIGO Laboratory, California Institute of Technology, Pasadena, CA, USA}
\author{Su Direkci}
\affiliation{Institute for Quantum Information and Matter, California Institute of Technology, Pasadena, CA, USA}
\author{Yanbei Chen}
\affiliation{Institute for Quantum Information and Matter, California Institute of Technology, Pasadena, CA, USA}

\date{\today}

\begin{abstract}
Squeezed vacuum lets gravitational-wave detectors and other quantum sensors surpass the standard quantum limit, and is optimal in the loss-limited regime; phase noise breaks this optimality. Numerically optimizing the quantum Fisher information across the loss--phase noise landscape, we identify non-Gaussian states that outperform any Gaussian state. These fall into three classes: Fock-like states, cubic-phase-like states, and states with discrete rotational symmetry. Limiting the average number of photons in the input state to $\bar n = 5$, with $1-\eta = 5\%$ photon loss, and 200 mrad phase noise, the non-Gaussian advantage reaches up to 2.2 dB. Furthermore, we observe that the non-Gaussian advantage can persist even when the measurement strategy is homodyne detection.
\end{abstract}

\maketitle




\textbf{\textit{Introduction.}}---Measuring a weak signal $\epsilon$ is a canonical task in quantum metrology. 
The estimation sensitivity can be enhanced by using quantum resources, which ultimately can achieve a error that scales as $1/N$, referred to as the Heisenberg limit \cite{Giovannetti2006}.
Quantum-noise-limited sensors (e.g. gravitational wave detectors) have used squeezed light to enhance their quantum-limited sensitivity~\cite{Aasi2013, Tse2019, Acernese2019,Caves81, Escher2011, Demkowicz2012, Giovannetti2006,  lang_optimal_2013, Demkowicz2009}. 

As losses are reduced, another decoherence mechanism becomes significant: phase noise or dephasing~\cite{Genoni2011, McCuller2021} of the quantum state.
Phase noise couples in \emph{more} noise than vacuum through the projection of the orthogonal (anti-squeezed) quadrature. 

The precision limits in the phase-insensitive case were analyzed in the context of force sensing~\cite{Grochowski25}, which showed a metrological advantage of non-Gaussian states, as was also seen in Refs.~\cite{Giovannetti2011, Demkowicz2012}.
The intermediate and experimentally relevant regime of \emph{finite} loss and \emph{finite} phase noise has, however, remained unexplored. 
Loss and dephasing favor opposing state structures, and the optimal state lives on their Pareto frontier. 
For example, photon loss removes excitations from the state and disrupts fixed-photon-number states such as Fock states, whereas dephasing applies a random phase-space rotation that suppresses the squeezing of directional states but leaves rotationally symmetric, number-definite states invariant. These competing tendencies make it difficult to anticipate the optimal state, motivating a direct search.

We construct states for this regime by numerically optimizing the quantum Fisher information (QFI), which describes the achievable precision given an input state.
We construct non-Gaussian states that provide a Signal-to-Noise Ratio (SNR) improvement compared to the best Gaussian states. 
These states  resemble Fock states, cubic phase states, or states with discrete rotational symmetry depending on the strength of the decoherence mechanisms.
This advantage applies to two-mode interferometric phase sensing in the limit where the common mode is strongly pumped, e.g. dark-port injection in high-power interferometers.

The sensitivity implied by the QFI is only achieved for the state-dependent, optimal measurement strategy, governed by the symmetric logarithmic derivative (SLD). 
The optimal readout may be very difficult to implement experimentally and so we examine here the performance of the states when they are paired with experimentally relevant readouts such as photon number resolving (PNR) readout, parity, and homodyne measurement. 
We observe that these states outperform the sensitivity achieved with Gaussian states and homodyne readout, even with sub-optimal measurements.

\begin{figure}[!t]
    \centering
    \includegraphics[width=\columnwidth]{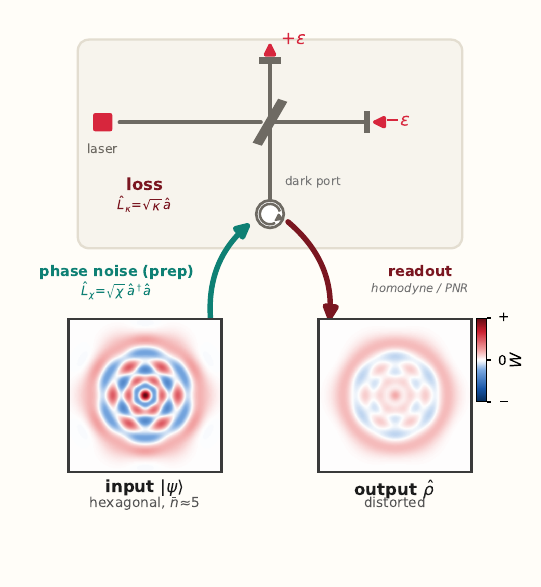}
    \caption{Optimized non-Gaussian probe states for displacement sensing in an
    interferometer limited by loss and phase noise. A state with discrete ($C_6$)
    rotational symmetry (with an average number of $\bar n \approx 5$ photons) is prepared and injected at the dark
    port through a circulator, and the returning field is read out at the same port.
    Phase noise acts during state preparation (dephasing,
    $\hat L_\chi = \sqrt{\chi}\,\hat a^\dagger \hat a$) and photon loss acts during
    sensing ($\hat L_\kappa = \sqrt{\kappa}\,\hat a$); the small displacement
    $\epsilon$ to be estimated enters at the end mirrors. Wigner functions show the
    input probe (left) and its distorted output (right), obtained by evolving the
    state through both channels with the Lindblad master equation.}
    \label{fig:channel_and_ifo}
\end{figure}



\textbf{\textit{Preliminaries.}}---%
Consider sensing a displacement $\epsilon = gt \ll 1$ imposed on a probe, via a Hamiltonian 
$\hat{\mathcal{H}} = g \left( \hat{a} + \hat{a} ^\dagger \right)$. 
In gravitational-wave interferometers, the probe is a light mode $\hat a$ at the dark port, which gains a coherent amplitude upon exiting the interferometer due to the beat between the carrier field and the differential mirror displacement. 
The 
probe is also 
subject to loss and phase noise (de-phasing), represented by their respective Lindblad operators:
%
$\hat{\mathcal{L}}_\kappa = \sqrt{\kappa} \hat a$ for amplitude-damping photon loss, and 
$\hat{\mathcal{L}}_\chi = \sqrt{\chi} \hat a ^\dagger \hat a$ for 
a Gaussian distributed random phase rotation with standard deviation $\sigma_\phi = \sqrt{\chi t}$, where $t$ is the evolution time.  
We set $t=1$, so that $\sigma_\phi^2 = \chi$ fixes the dephasing strength and $\epsilon=g$ is the displacement.
We specifically analyze the case where phase noise occurs \emph{before} the sensing process, and loss occurs \emph{during} sensing---a configuration particularly relevant to gravitational wave detectors, where dephasing originates in the probe state preparation while loss is distributed along the optical path (See Fig.~\ref{fig:channel_and_ifo}).

The QFI, calculated numerically using the spectral decomposition of the density matrix~\cite{Braunstein1994, gardner25_swe}, 
is maximized by a global optimizer over input states of a defined form \footnote{See the Supplementary Material}. 
We work in the infinitesimally small displacement limit, and calculate the QFI for $\epsilon = 0$.
In particular, we use three different forms (or ansatze) for the input states:
\begin{itemize}
    \item Squeezed vacuum superpositions: $\sum_k c_k |r_k \rangle$ where $|r_k \rangle = \hat S (r_k) |0 \rangle$ is a squeezed vacuum state with complex squeeze parameter $r_k$ and the squeeze operator $\hat S (r) = \exp( \frac{1}{2} (r^* \hat a^2 - r \hat a^{\dagger 2}))$.
    \item Displaced squeezed vacuum superpositions: $\sum_k c_k |\alpha_k, r_k \rangle$ where $|\alpha, r \rangle = \hat D (\alpha) \hat S (r) |0 \rangle$, with the displacement operator $\hat D (\alpha) = \exp ( \alpha \hat a^\dagger - \alpha^* \hat a)$.
    \item Fock basis superpositions: $\sum_k c_k |k \rangle$ where $\{|k \rangle \}$ are the number states or Fock states.
\end{itemize}
All of these states are non-Gaussian, beyond the trivial single-component Gaussian case. The (displaced) squeezed-vacuum superpositions are motivated by available preparation techniques~\cite{Ourjoumtsev2006, NeergaardNielsen2006, Takahashi2008},
whereas the Fock-superposition ansatz is motivated by Fock optimality in the phase-insensitive limit \cite{Grochowski25}\footnotemark[1]. We explicitly optimize over non-displaced squeezed vacuum superpositions due to the numerical advantage obtained from excluding the optimization over displacements: furthermore, these states are optimal for a single squeezed vacuum component in the absence of any loss and phase noise. For each ansatz, the complex coefficients $\{c_k\}$ are optimized together with the state parameters: $\{r_k\}$ for squeezed vacuum, $\{r_k, \alpha_k\}$ for displaced squeezed vacuum, and integer indices $\{k\}$ for Fock superpositions \footnotemark[1].
We also define the \textit{best Gaussian state}, which is the Gaussian state with the maximal QFI for a given set of noise parameters. We optimize over the rotation, squeezing, and displacement parameters of this state.

For a given efficiency $\eta = 1-L$ (where $L$ is the loss in power) and phase noise $\sigma_\phi$, the optimizer was run for each ansatz under a constraint on the input mean photon number, $\langle \hat{a}^\dagger \hat{a} \rangle = \bar{n} \leq N_{\rm target}$, enforced via a penalty in the cost function \footnotemark[1]. We swept the number of superpositions $n_{\rm sup} = 1, 2, 3, \ldots$, limiting ourselves to $n_{\rm sup} \leq 5$ due to the computational cost.

\textbf{\textit{Results and discussion.}}---%
We first perform a coarse exploration of the loss and phase noise $(\eta-\sigma_\phi)$ parameter space for optimized states, followed by a finer exploration of the low loss ($\eta>0.9$) and moderate-to-low phase noise ($\sigma_\phi<200$ mrad) region. We compare the optimized states with the best Gaussian states which are, in general, \emph{displaced} squeezed vacuum states \footnotemark[1] (with higher phase noise, a displacement in the input state is advantageous). We identify trends and summarize these results in the following sections.

\begin{figure*}[t]
    \centering
    \includegraphics[width=0.95\textwidth]{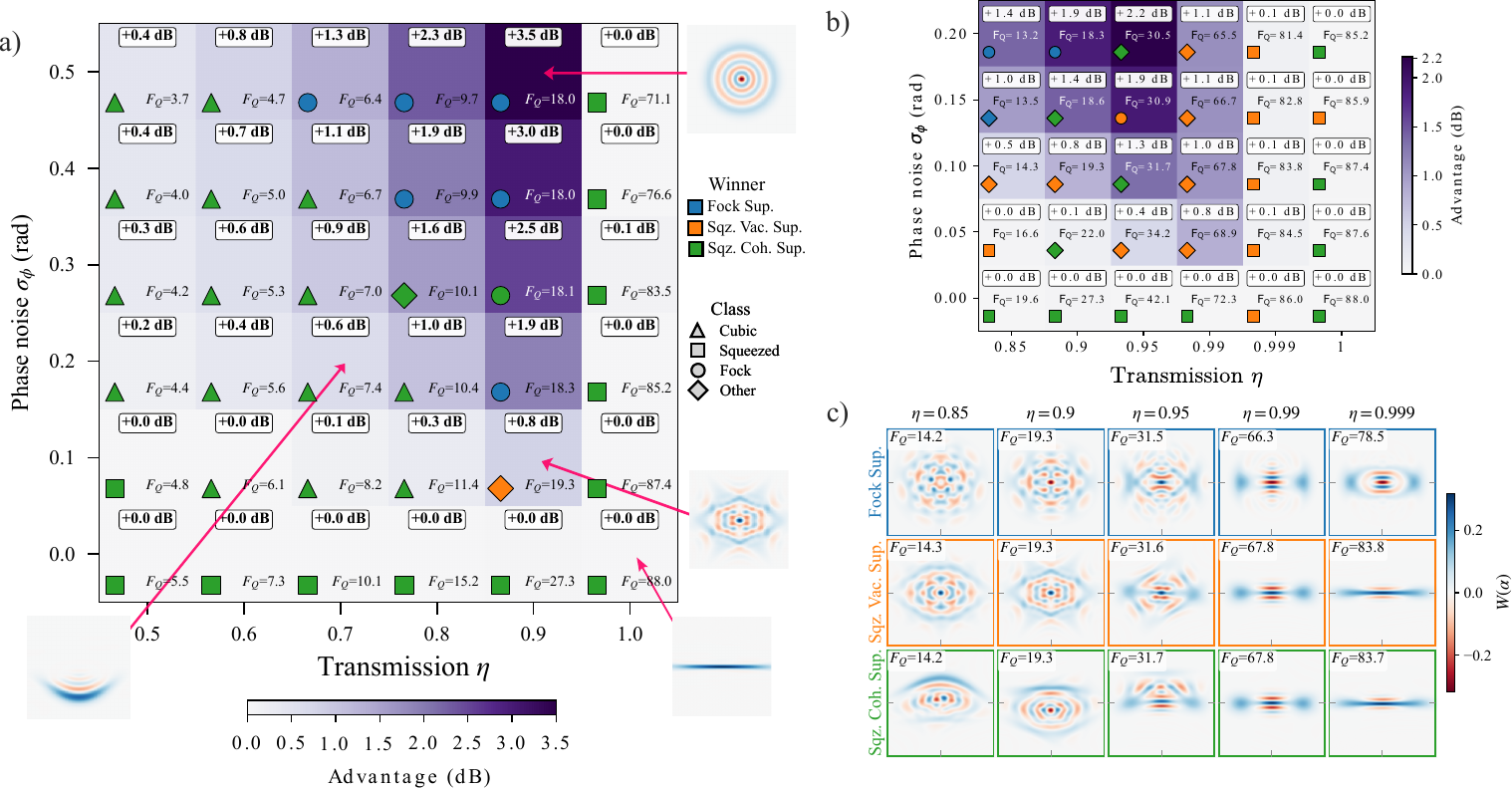}

    \begin{subfigure}{0pt}\phantomsubcaption\label{fig:ng_adv_heatmap}\end{subfigure}%
    \begin{subfigure}{0pt}\phantomsubcaption\label{fig:lowloss_heatmap}\end{subfigure}%
    \begin{subfigure}{0pt}\phantomsubcaption\label{fig:lowloss_symmetry}\end{subfigure}%

    \caption{\textbf{Non-Gaussian advantage}, $N_{\rm target}=5$. (a) dB advantage of the optimized state over the best Gaussian state with equal photon number across the $(\eta, \sigma_\phi)$ grid; marker shape indicates the state class (square=squeezed, circle=Fock, triangle=cubic, diamond=other) and marker color the winning ansatz, with insets showing Wigner functions of representative optimized states. (b) A finer zoom into the experimentally relevant low-loss, low-phase-noise region, $\eta\in[0.85, 0.999]$, $\sigma_\phi\in[0, 0.2]$. Apparent QFI values in the lowest-loss cells carry a ${\sim}1\%$ uncertainty from Hilbert-space truncation, which can produce small apparent violations of the analytic bounds (see main text). (c) Wigner functions of the optimized states across $\eta$ at $\sigma_\phi=0.1$ for the three ansatze (rows: Fock, squeezed vacuum, and displaced squeezed vacuum superpositions); the discrete rotational symmetry steps down in order $6\to4\to2$ as loss decreases.}
    \label{fig:ng_adv}
\end{figure*}

\textit{Emergence of non-Gaussianity.}---%
For $\bar n=5$, with $0.5\le \eta \le 1$ and $0\le \sigma_\phi\le 0.5$, and step size 0.1, we examined a coarse grid of the loss and phase noise parameter space in Fig.~\ref{fig:ng_adv_heatmap}.
This Figure shows that there are regions where non-Gaussian states perform better than the best Gaussian state.
We quantify the non-Gaussian advantage as
\begin{align}
    \mathrm{Advantage} = 10 \log_{10} \left( \frac{\mathcal{F}_Q^{\text{opt}}}{\mathcal{F}_Q^{\text{gauss}}} \right),
\end{align}
where $\mathcal{F}_Q^{\text{opt}}$ is the QFI of the optimized state, and $\mathcal{F}_Q^{\text{gauss}}$ is the QFI of the best Gaussian state, for a given configuration of noise parameters. With the optimal measurement, this is the expected gain in SNR. The Wigner functions of the resulting states are shown in Fig.~\ref{fig:ng_adv_heatmap}.
To indicate the type of optimized state, 
we assign a colored symbol in each box, with color indicating the superposition ansatz that achieves this maximum, and shape that indicates the class. Each optimized state is assigned the class label whose best-fit member, maximized over that class's parameters, achieves the highest fidelity with respect to the state; states for which no class achieves sufficiently high fidelity are labeled as ``other."

As we examine the Wigner functions of the optimized states shown in Fig.~\ref{fig:ng_adv_heatmap}, four classes of states emerge, in four regions. First, in the zero-phase-noise case ($\sigma_\phi=0$), we find that squeezed vacuum (shown with the box marker) recovers the highest QFI, as expected. In the lossless case ($\eta=1$), the optimized states again resemble a squeezed vacuum state (region I). This can be understood by noting that squeezed vacuum carries only even photon-number components. This holds true even after the state is subjected to phase noise. On the other hand, displacements introduce odd photon-number components, changing the parity. These states are therefore sensitive to displacement, while being robust against loss. The classical Fisher information (CFI) results support this interpretation (see End Matter).
States that resemble Fock states (shown with the circle marker), being sparse Fock-superpositions, are recovered in the limit of high phase noise and low loss (region II). For a proof of Fock optimality in this limit, see \footnotemark[1]. In region III, where the two effects are both moderate and are in competition, states that resemble cubic phase states \cite{Yanagimoto2020} emerge (shown with the triangle marker). Due to their displacement from the origin and the arc-like phase-space support, the displacement sensed by these states is less sensitive to the random rotation imposed by dephasing than the sharply directional quadrature of a squeezed state is. Furthermore, these states are more resilient to photon loss compared to Fock state superpositions, as the latter encodes the displacement in number-definite Fock states that are directly corrupted by loss.
Finally, for comparable  levels of loss and phase noise, we found in region IV non-Gaussian states with less regular morphology (shown with the diamond marker). One such state shows up in Fig.~\ref{fig:ng_adv_heatmap} for $\eta=0.9$ and $\sigma_\phi=0.1$ (the box with the orange, diamond marker).  This is consistent with the expectation that states with rotational symmetry are favored when loss and phase noise are both significant: their rotational structure resists the random dephasing rotation, while their phase-space localization resists loss. At $\eta=1$, by contrast, the parity protection of squeezed vacuum suffices.


\textit{Low loss and moderate phase noise.}---%
We examine a finer grid for the experimentally more relevant, low-to-moderate loss of $\eta\ge 0.85$ and the low phase noise regime of $\sigma_\phi<0.2$  (region IV) with the same energy constraint of five photons. The results are shown in Fig.~\ref{fig:lowloss_heatmap}.  Here, we found further non-Gaussian, discrete-rotational states, with significant QFI enhancement compared to their best Gaussian state counterpart. For example, for $\sigma_\phi=0.2$, $\eta = 0.95$, the non-Gaussian advantage is 2.2(2) dB.
We also see that for the very low loss regime of $\eta >0.999$, squeezing becomes optimal again, as this is the loss-limited input state.

In Fig.~\ref{fig:lowloss_symmetry}, we work in the region of $\sigma_\phi=0.1$ and $0.85\le \eta\le 1$, and examine the Wigner functions achieved by all three types of superposition states that we used for optimization. This not only demonstrates the transition from squeezed vacuum at $\eta =1$ to non-Gaussian states at lower $\eta$ values, but also shows approximate agreement between the three ansatze. See \footnotemark[1] for the convergence of the QFI with number of superpositions. As the loss decreases, the discrete rotational symmetry order of the optimal state steps down accordingly: the Wigner function is hexagonal (six-fold) at $\eta=0.85$, becomes four-fold at intermediate loss, and two-fold and cat-like by $\eta=0.99$, converging to a squeezed vacuum in the very-low-loss limit. 

Fig. \ref{fig:ng_adv} shows that for values of loss and phase noise relevant to gravitational-wave detectors, e.g., $<10\%$ loss and $<200$ mrad of phase noise, there exist non-Gaussian states that perform better than the best Gaussian baseline.


\begin{figure*}[!t]
    \centering
    \begin{subfigure}{0.32\linewidth}
        \includegraphics[width=\linewidth]{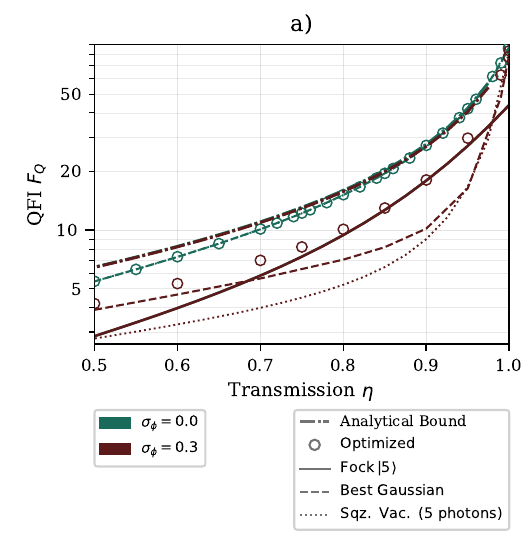}
        \phantomcaption\label{fig:trend_eta}
    \end{subfigure}\hfill
    \begin{subfigure}{0.32\linewidth}
        \includegraphics[width=\linewidth]{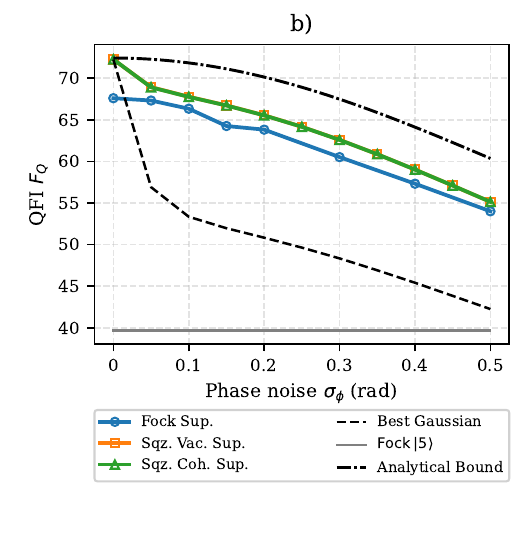}
        \phantomcaption
        \label{fig:trend_pn}
    \end{subfigure}\hfill
    \begin{subfigure}{0.32\linewidth}
        \includegraphics[width=\linewidth]{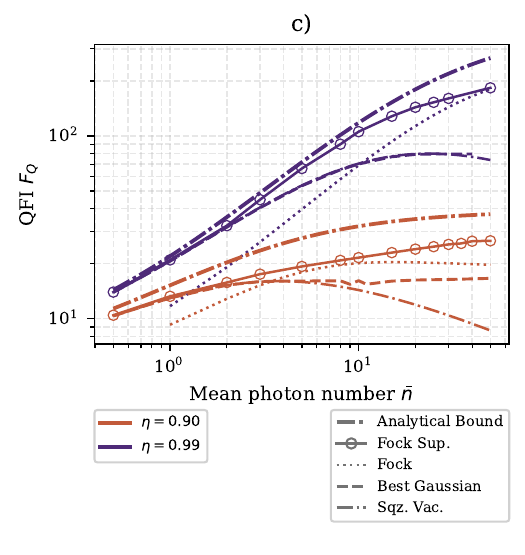}
        \phantomcaption
        \label{fig:trend_nt}
    \end{subfigure}
    \caption{\textbf{QFI trends.} Optimized QFI $\mathcal{F}_Q$ (markers) versus loss,
dephasing, and energy, compared with the Fock, squeezed-vacuum, and best-Gaussian
states and the analytical upper bound.
\textbf{(a)}~$\mathcal{F}_Q$ vs.\ $\eta$ at $\sigma_\phi\in\{0,0.3\}$,
$N_{\mathrm{target}}=5$.
\textbf{(b)}~$\mathcal{F}_Q$ vs.\ $\sigma_\phi$ at $\eta=0.99$,
$N_{\mathrm{target}}=5$.
\textbf{(c)}~$\mathcal{F}_Q$ vs.\ $N_{\mathrm{target}}$ at $\sigma_\phi=0.1$ for
$\eta\in\{0.9,0.99\}$.
Solid: Fock; dotted: squeezed vacuum; dashed: best Gaussian within the energy
limit; markers: best optimized state.}
    \label{fig:trends}
\end{figure*}

\textit{Trends in the loss and phase noise parameter space.}---%
Having mapped the regions where each state class dominates, we now examine how the QFI itself varies along one-dimensional cuts through the parameter space. The results can be found in Fig. \ref{fig:trends}.
We derive analytical upper bounds to the QFI in order to evaluate the performance of our numerical optimization, and plot the smallest QFI provided by these bounds for all $(\eta, \sigma_\phi, \bar n)$ as a reference. They consist of the convexity bound $\mathcal{F}_Q^\text{conv} = 4\frac{2\eta A(\bar n) +1}{1+4\eta(1-\eta)\bar{n}}$, the variance bound $\mathcal{F}_Q^\text{var} =4[1+2A(\eta \,\bar n) ]$, and the combined channel bound $\mathcal{F}_Q^\text{comb}=4\frac{1+2A(\bar n)}{1+2(1-\eta)A(\bar n)}$, where $A(x) = x+ e^{-2\sigma_\phi^2}\sqrt{x(x+1)}$.
For a detailed discussion about the derivations, see \footnotemark[1]. When examined more closely, the bottom right states of Figure~\ref{fig:lowloss_heatmap} display QFIs that seem to violate the analytical bounds. This can be explained by errors due to the truncation of the Hilbert space, which is done to make the problem computationally tractable. The violations of the bounds are consistent with relative errors in the QFI of order 1\%, which is the order of magnitude expected. The arguments for this can be found in \footnotemark[1].

First, in Fig. ~\ref{fig:trend_eta}, we fix $\sigma_\phi=0.3$ and vary $\eta$ from 0.5 to 1. 
The optimized states (round markers) in Fig.~\ref{fig:trend_eta} correspond to the third row from the top of Fig.~\ref{fig:ng_adv_heatmap}. Here, we examine the trend in loss with reference to the QFI for the Fock, squeezed vacuum, and best Gaussian states. As a reference, we also plot the case without phase noise ($\sigma_\phi=0$) in green to denote the ``loss limit'' where squeezed vacuum states are optimal. We observe that for each $\eta$, the class of the optimized state coincides with whichever single-class QFI is largest for that value.
For example, when the Fock state outperforms the ``Best Gaussian'' state, the optimized state is Fock-like. In particular, as the solid line is highest compared with the dashed, at around 0.9, that is when the ``class'' of the optimized state is the Fock superpositions. At both regimes of  very low loss ($\eta \sim 1$) and high loss ($\eta < 0.6$), the optimized state is either squeezed or cubic-phase shifted. We also compare the optimized QFI values to theoretical upper bounds, which we call convexity and variance bounds \footnotemark[1]. These bounds are generally loose, but get tighter as phase noise and loss are reduced.

 Next, in Fig.~\ref{fig:trend_pn}, we fix $\eta=0.99$ and vary $\sigma_\phi$ from 0 to 0.5. 
 This corresponds to the fourth column from left of Fig.~\ref{fig:lowloss_heatmap}.
 In this Figure, we again observe approximate agreement between the three numerical optimizations with the different superposition families (the squeezed vacuum, displaced squeezed vacuum, and Fock ansatze), which reassures the optimality of our search. It shows that Gaussian states on their own are quite insufficient in this case---at $\sigma_\phi=0.1$, the best optimized state attains a QFI $1.3\times$ that of the best Gaussian state, an advantage of $1.1$ dB---and that they will underperform with respect to Fock states if $\sigma_\phi$ further increases. However, the Fock state (with $\bar n=5$) does not become optimal for the $\sigma_\phi$'s evaluated here.

\textit{Trends with photon number.}---%
Results presented thus far correspond to an energy limit of five photons. We perform the optimization over Fock superpositions up to 50 mean photons (See Fig.~\ref{fig:trend_nt}). We only consider these superpositions due to the computational complexity of the displaced or undisplaced squeezed vacuum superpositions. From the Figure, we observe that the advantage in the QFI for the optimized Fock superpositions over the best Gaussian state persists at higher photon numbers. For $\eta = 0.9$ and $\sigma_\phi=0.1$, we see a saturation in QFI for Fock states, whereas Fock superpositions continue to improve and maintain an advantage over the best Gaussian state that widens with $\bar n$. When the loss is lowered to 1\% for the same phase noise, the optimal class itself depends on energy: the optimized states are Gaussian-like at low photon number and Fock-like at high photon number, with a transition region in between around $\bar n =10$ where the optimized states perform significantly better than both. This result further emphasizes the non-Gaussian advantage: as the loss strength weakens, Gaussian states outperform Fock states for larger mean photon numbers, however the non-Gaussian advantage gap gets larger once Fock states start outperforming the best Gaussian state for the respective noise parameters. For example, at $\eta = 0.99$ this advantage grows to approximately $4$ dB at $\bar n = 50$.



\textbf{\textit{Outlook.}}---%
In summary, these results show that non-Gaussian states (Fock-like, cubic-phase-like, and states with discrete rotational symmetry) outperform Gaussian states across the finite loss and finite phase noise landscape. Prior analyses only dealt with limiting cases: either for pure loss or complete dephasing with loss~\cite{Grochowski25}, where the optimal probe states are squeezed vacuum and Fock states, respectively. 
The strongest non-Gaussian advantage was achieved in the regime of low loss and moderate-to-high phase noise. For a mean photon number of $\bar n{=}5$, this corresponds to a transmission of $\eta = 0.95$, and a phase noise standard deviation of $\sigma_\phi = 0.2$.  For higher energy limits (i.e. larger $\bar n$), our results imply an increased advantage extended to lower loss.

In several regions of the $(\eta, \sigma_\phi)$ space, all three ansatze converged to nearly the same QFI (see \footnotemark[1] and Fig.~\ref{fig:lowloss_symmetry}), suggesting that we have located near-optimal states rather than merely the best within a restricted family. All four identified state classes are within current experimental reach: superpositions of (displaced) squeezed vacuum~\cite{Ourjoumtsev2006, NeergaardNielsen2006, Takahashi2008, Eaton2019} and GKP-class grid states~\cite{larsen_integrated_2025} using heralded preparation techniques, and recent realizations of cubic-phase states in microwave and optical platforms~\cite{Budinger2024, Eriksson2024}. 



For every optimized state, we also compare its QFI with the CFI of feasible measurements. Realizing the non-Gaussian advantage generally requires non-Gaussian measurements such as the PNR detection. Although, for cat and squeezed-cat states it is already accessible by a balanced homodyne measurement, with an advantage of up to $4$ dB over the best Gaussian state at $\eta=0.99$. Notably, even the best-Gaussian QFI is not reachable by homodyne alone (see End Matter).


Finally, the rotational symmetry and phase-space compactness of several optimized states suggests intrinsic back-action evasion, which for GW interferometers could obviate the filter cavities currently required for frequency-dependent squeezing.
A practical caveat remains for broadband sensors such as GW detectors: squeezed vacuum carries broadband squeezing automatically through time-frequency photon-pair correlations, but non-Gaussian states do not generically share this property and must be prepared at each frequency with the required resolution bandwidth.

\textit{Acknowledgments.}
We thank James W. Gardner, Senrui Chen, and Tuvia Gefen for helpful discussions.
The LIGO Laboratory is supported by the U.S. National Science Foundation award PHY‑2309200 and operated jointly by Caltech and MIT.
S.D.\ and Y.C.\ are also supported by the Simons Foundation (Award No. 568762)  and the National Science Foundation (via Grants No. PHY-2309211 and No. PHY-2309231).
R.X.A acknowledges support by the Woodnext Foundation and the John Templeton Foundation.
This work was made possible partially through the support of Grant 63405 from the John Templeton Foundation. 
The opinions expressed in this publication are those of the author(s) and do not necessarily reflect the views of the John Templeton Foundation.
\bibliography{apssamp}



\onecolumngrid
\bigskip
\begin{center}
{\large \textbf{End Matter}}
\end{center}
\twocolumngrid

\begin{figure*}
    \centering
    \includegraphics[width=\linewidth]{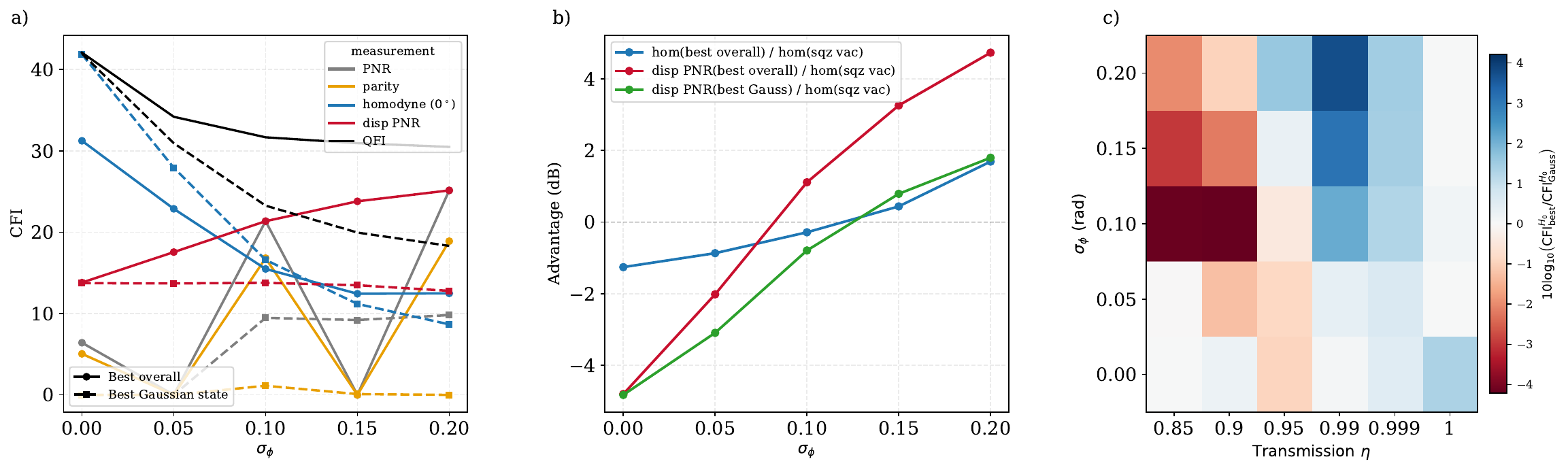}
    \begin{subfigure}{0pt}\phantomsubcaption\label{em_fig:cfi_lines_eta0p95}\end{subfigure}
    \begin{subfigure}{0pt}\phantomsubcaption\label{em_fig:cfi_advantage_eta0p95}\end{subfigure}
    \begin{subfigure}{0pt}\phantomsubcaption\label{em_fig:cfi_hom0_heatmap}\end{subfigure}
    \caption{(a) CFI of each measurement (PNR, parity, homodyne at optimized angle $\theta=0^\circ$, displaced PNR, displaced parity) and the QFI (dotted) versus phase noise $\sigma_\phi$ at $\eta=0.95$, $N_\text{target}=5$, for the best overall state and the best Gaussian state. (b) Metrological advantage (dB) of the displaced-PNR readout relative to homodyne of the squeezed vacuum, versus $\sigma_\phi$, for $\eta=0.95$. (c) Homodyne CFI of the state with the maximal QFI, relative to the best Gaussian state across the low-loss grid, $\Delta = 10\log_{10}\!\left(\mathrm{CFI}^{H_0}_{\rm best}/\mathrm{CFI}^{H_0}_{\rm Gauss}\right)$ in dB, as a function of the transmission coefficient $\eta$ and phase noise $\sigma_\phi$ for $N_\text{target}=5$ photons.}
    \label{fig:end_matter_fig}
\end{figure*}

\textit{Classical Fisher information.}--- While the symmetric logarithmic derivative (SLD) operator $\hat{L}_Q$, satisfying
\begin{equation}
\partial_\epsilon \hat\rho_\epsilon = \frac{1}{2} \left( \hat{L}_Q \hat\rho_\epsilon + \hat\rho_\epsilon \hat{L}_Q \right)
\end{equation}
corresponds to the optimal measurement that achieves the QFI for a given state and channel, it is not necessarily a practically feasible measurement. To gain insight into a measurement strategy that can be implemented in practice, we calculate the CFI for several practically feasible measurements.
We compare the CFI for each measurement with the QFI of the optimized state, to see how reachable the QFI is with current technology. Two reference baselines are used consistently throughout: balanced homodyne on the \emph{squeezed vacuum} --- the Gaussian probe currently injected at the dark port of operating gravitational-wave detectors --- which sets the \emph{deployed-practice} baseline against which realizable gains are quoted; and the homodyne CFI of the \emph{best Gaussian state}, used when asking how much of the non-Gaussian advantage survives under homodyne detection alone. We emphasize that the squeezed vacuum is, in the presence of phase noise, strictly weaker than the best Gaussian state (which additionally carries a displacement), so gains quoted against it are larger than those against the best Gaussian baseline. Both baselines are distinct from the QFI advantage of Fig.~\ref{fig:ng_adv}, which is referenced to the best Gaussian state.

The measurements considered are the following:
\begin{enumerate}
  \item \textbf{PNR}. Measures the populations
$p_n=\langle n|\hat\rho|n\rangle$, yielding a CFI of
$\sum_n |\partial_\epsilon p_n|^2/p_n$.
  \item \textbf{Parity}. The parity operator $\hat\Pi=(-1)^{\hat n}$ returns a binary
outcome with CFI $F=(\partial_\epsilon\langle\hat\Pi\rangle)^2/(1-\langle\hat\Pi\rangle^2)$,
and is a coarse-graining of PNR that retains only the photon-number parity.
  \item \textbf{Homodyne}. Balanced homodyne at homodyne quadrature phase $\theta$. The
        outcome density is the Radon transform of the Wigner function,
        $P(x|\epsilon, \theta) = \mathcal{R}_\theta[W_{\hat\rho}](x)$,
        and the CFI is evaluated directly in phase space as
        \begin{equation}\label{eq:cfi-hom-radon-rep}
          F_{\mathrm{hom}}(\theta) = \int \frac{\bigl(\mathcal{R}_\theta[
              W_{\partial_{\epsilon}\hat\rho}](x)\bigr)^2}
                                  {\mathcal{R}_\theta[W_{\hat\rho}](x)}\,dx,
        \end{equation}
        optimized over the quadrature direction $\theta\in[0,\pi/2]$.
  \item \textbf{Displaced PNR}. A displacement $\hat D(\beta)$ is applied before
photon counting, and the CFI is optimized over the displacement $\beta$.
  \item \textbf{Displaced parity}. The displaced parity observable
$\hat D(\beta)\hat\Pi\hat D^\dagger(\beta)$ is measured, again optimizing the CFI
over the displacement $\beta$.
\end{enumerate}

\textit{Feasible measurements}.---We find that homodyne measurement is optimal in the absence of phase noise \footnotemark[1], or when phase noise is negligible with high loss ($\eta=0.5$, $\sigma_\phi=0.1$), nearly or completely recovering the QFI. The optimized states remaining squeezed vacuum in the rest of region I.
Here, homodyne is exponentially worse with increasing phase noise, while \emph{displaced} PNR and parity measurements remain nearly optimal, recovering the QFI across the entire region. The advantage of displaced PNR over homodyne is 12 dB at $\sigma_\phi=0.5$ \footnotemark[1].
This can be understood in two ways. First, at $\epsilon=0$, the photon-counting probabilities vary only to second order in $\epsilon$ for the even-parity squeezed vacuum, whereas a prior displacement makes the variation first order, so displaced PNR is near-optimal while undisplaced PNR is not. Equivalently, the input state and the PNR measurement are both reflection-symmetric about the displacement axis and therefore cannot resolve its sign; the prior displacement breaks this symmetry, allowing the measurement to resolve both the direction and the amplitude of the displacement.

In region II, where the optimized states are Fock-like, none of the considered measurements recover the entire QFI, but displaced PNR performs best and beats squeezed vacuum with homodyne by a large margin. This suggests that the optimal measurement is more complex and requires further investigation.

In region III, where the optimized states are cubic-phase-like, we find that the best measurement varies. Displaced PNR is nearly optimal at high loss and phase noise. When $\sigma_\phi > 0.1$, displaced PNR is still the best among considered measurements, but the gap requires further investigation. When $\sigma_\phi = 0.1$, the non-Gaussian advantage is not realized by any of the considered measurements. Lastly, region IV requires further investigation as none of the considered measurements realize the non-Gaussian advantage, and the optimal measurement is likely to be more complex.





\textit{Observations in the low loss regime.}---
In this regime of finite loss and phase noise, displaced PNR is the best among the considered measurements (see Fig. \ref{em_fig:cfi_lines_eta0p95} and \ref{em_fig:cfi_advantage_eta0p95}). Here, switching to displaced PNR with the best non-Gaussian state results in 4.7 dB of SNR gain for a transmission coefficient of $\eta=0.95$ when compared to homodyne with the squeezed vacuum, as displayed in  Fig.~\ref{em_fig:cfi_advantage_eta0p95}.
There also exists a region where non-Gaussian advantage can even be realized by homodyne measurement, which is the case for the cat states and squeezed cat states. This is shown in Fig.~\ref{em_fig:cfi_hom0_heatmap}, where the advantage with homodyne measurement compared to the best Gaussian state is up to 4 dB at $\eta=0.99$. These homodyne gains exceed the QFI advantage of Fig.~\ref{fig:ng_adv} at the same noise levels because they compare the CFI rather than the QFI: the homodyne CFI of the best Gaussian state falls below its own QFI, therefore restricting the Gaussian baseline to homodyne lowers it and widens the gap.




\end{document}


\title{Supplemental Material for:\\Optimized Quantum States for Sensing in the Presence of Loss and Phase Noise}
\author{Shruti Maliakal}
\affiliation{Institute for Quantum Information and Matter, California Institute of Technology, Pasadena, CA, USA}
\author{Zachary Mann}
\affiliation{Institute for Quantum Information and Matter, California Institute of Technology, Pasadena, CA, USA}
\author{Christopher Wipf}
\affiliation{Institute for Quantum Information and Matter, California Institute of Technology, Pasadena, CA, USA}
\affiliation{LIGO Laboratory, California Institute of Technology, Pasadena, CA, USA}
\author{Rana X Adhikari}
\email{rana@caltech.edu}
\affiliation{Institute for Quantum Information and Matter, California Institute of Technology, Pasadena, CA, USA}
\affiliation{LIGO Laboratory, California Institute of Technology, Pasadena, CA, USA}
\author{Su Direkci}
\affiliation{Institute for Quantum Information and Matter, California Institute of Technology, Pasadena, CA, USA}
\author{Yanbei Chen}
\affiliation{Institute for Quantum Information and Matter, California Institute of Technology, Pasadena, CA, USA}


\maketitle
\tableofcontents

\vspace{2cm}

This document contains details of our theoretical framework, numerical implementation, as well as results from numerical simulations. It is organized as follows.  
Sec.~\ref{sup:pn_model_details} describes the phase noise model used in our simulations.
Sec.~\ref{app:limits_of_sense} discusses theory of quantum parameter sensing, and provides various theoretical bounds relevant to our problem.
Sec.~\ref{sup:opt_details} describes the optimization procedure, including the state parametrizations, noisy-channel simulation, QFI evaluation, objective function, photon-number constraint, warm starts, and basis-size choices. Here, we present convergence with respect to the truncated Fock basis, verifying that the reported QFI advantages are not numerical artifacts of the finite basis. 
Sec.~\ref{sup:ng_adv_calc} presents the main non-Gaussian advantage calculation
at $\bar N=5$, comparing optimized non-Gaussian probe states against the best Gaussian displaced-squeezed baseline in the presence of photon loss and phase noise. 
Section~\ref{sup:convergence_examples} analyzes the optimized states themselves and
shows that, as the number of superposition components is increased, the
solutions often converge toward recognizable non-Gaussian structures,
including cubic-phase-like states, binomial-amplitude states, sparse Fock superpositions, and states with discrete rotational or hexagonal symmetry.
Finally, Sec.~\ref{sup:cfi_section} provides an experimentally motivated benchmark by asking
whether a familiar cat state, measured with homodyne detection, can
already outperform squeezed vacuum in favorable low-loss regimes. Furthermore, the obtainable advantage with other types of experimentally available measurements, such as a photon number resolving (PNR) measurement, is studied.

\begin{table}[h]
\centering
\begin{tabular}{ccc}
\hline
Figures & Section &Purpose \\
\hline
\ref{fig:quad_var_phi_offset}-\ref{fig:quad_var_vs_phasenoise}
 & \ref{sup:pn_model_details}  & Lindblad approach for phase noise modeling. \\
\hline
\ref{fig:MZ_MM_modes}
 & \ref{app:limits_of_sense} & Evolution of annihilation operators for different interferometers. \\
 \ref{fig:bounds_comparison}
 &  & Comparison of the analytical upper bounds for the QFI. \\
\hline
\ref{fig:optimization} & \ref{sup:opt_details} & Optimization schematic. \\
\ref{fig:basis_convergence_nga5}--\ref{fig:basis_convergence_nga5_ll}
&   & Convergence with truncated Fock-basis size. \\ \hline
\ref{fig:best_fock_N5}--\ref{fig:best_sqzcoh_N5} & \ref{sup:ng_adv_calc} & Optimized Wigner Function from each ansatz for noise levels  in Fig.~2 of the main text.
\\
\ref{fig:displaced_gaussian_grid_theory}--\ref{fig:wigner_plots_disp_gaussian_theory}
& & Illustration of the advantage of displaced Gaussian states. \\
\hline
\ref{sup_fig:opt-overlap-cubic} & 
\ref{sup:convergence_examples}
& Convergences toward  cubic-phase states
\\
\ref{fig:binomial_amplitude_overlap_theory}--\ref{fig:binomial_photon_counting_grid_theory} & 

& Convergence toward binomial-amplitude distributions.
\\
\ref{sup_fig:opt-overlap-fock}--\ref{fig:fock_nsups_evo} &

& Convergence toward Fock states. \\
\ref{fig:hex_nsups_evo}&

& Convergence toward states with discrete rotational symmetry.
\\
\hline
\ref{fig:cat_homodyne_grid_cfi}--\ref{fig:cat_vs_sqzvac_benchmarks}
& \ref{sup:cfi_section}
 &  Cat-State with homodyne vs squeezed-state with homodyne, CFI comparison. \\
 \ref{em_fig:cfi_histograms_coarse}--\ref{em_fig:cfi_finite_loss} & & Obtainable CFI with other known experimental techniques. \\ \hline
\end{tabular}
\caption{Inventory of figures in the Supplemental Material that present numerical results.}
\label{tab:figure_inventory}
\end{table}

\section{Phase noise model \label{sup:pn_model_details}}

\begin{figure}[ht]
    \centering
    \includegraphics[width=0.8\linewidth]{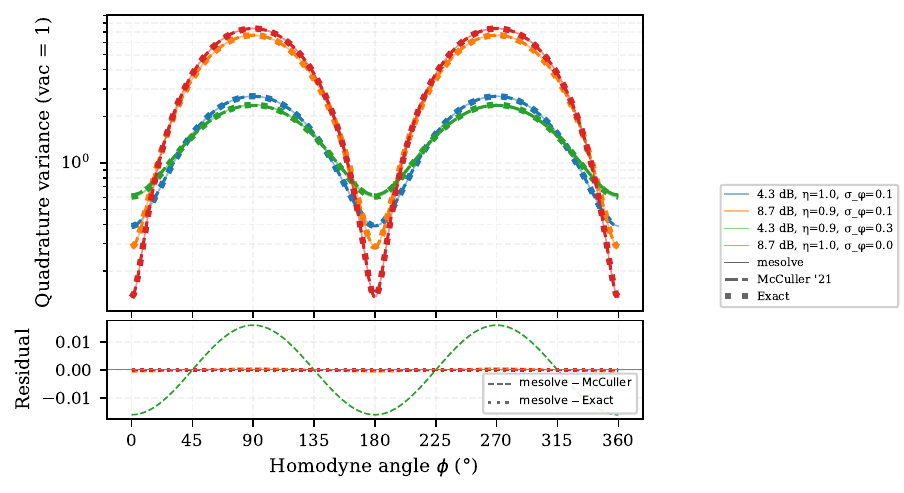}
    \caption{Quadrature variance as a function of homodyne angle $\phi$. The Lindblad-based \texttt{mesolve()} result matches the exact expression (Eq.~\eqref{eq:quad_var}), while it deviates from the Taylor expansion used in McCuller et al.~\cite{McCuller2021} at large phase noise.}
    \label{fig:quad_var_phi_offset}
\end{figure}

 We use the Lindblad operator $L_\chi = \sqrt{\chi} \hat a ^\dagger \hat a$ to model Gaussian-distributed random phase shifts with standard deviation $\chi$ for unit time~\cite{Genoni2011}. An alternative approach would be to construct a channel of random unitary phase shifts, in which case arbitrary distributions of phase noise can be modeled. However, the Lindblad approach was convenient for our optimization procedure. To validate our simulation of this model, we analyze it for squeezed vacuum states. For small phase noise, we can compare the results to~\cite{McCuller2021} (labelled McCuller in Fig.~\ref{fig:quad_var_phi_offset}).

\begin{figure}[ht]
    \centering
    \includegraphics[width=0.8\linewidth]{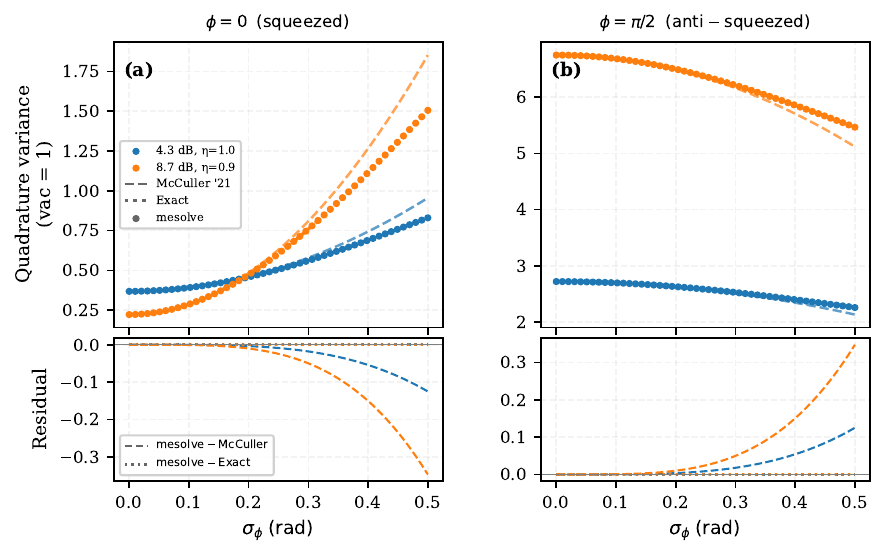}
    \caption{Comparison of the exact quadrature variance expression with the Lindblad-based \texttt{mesolve()} and Taylor expansion approximations. The Lindblad-based expression matches the exact expression, while it deviates from the Taylor expansion (McCuller et al.) at large phase noise.}
    \label{fig:quad_var_vs_phasenoise}
\end{figure}

We construct the squeezed vacuum state numerically in \texttt{QuantumToolbox.jl} and propagate it through a combined loss and phase-noise channel simulated by master equation evolution \texttt{mesolve()}; the quadrature variance is then computed from the resulting density matrix.
The exact quadrature variance at homodyne angle $\phi$ for a squeezed vacuum state with squeezing parameter $r$, transmission $\eta$, and RMS phase noise $\sigma_\phi$ is given by
\begin{align}
    V(\phi) = \eta\!\left(\cosh(2r) - e^{-2\sigma_\phi^2}\sinh(2r)\cos(2\phi)\right) + (1 - \eta).
    \label{eq:quad_var}
\end{align}
For small phase noise, this matches LIGO's quantum response to squeezed states~\cite{McCuller2021} which replaces the factor $e^{-2\sigma_\phi^2}$ with its second-order Taylor expansion $(1 - 2\sigma_\phi^2)$. A comparison of the exact (Eq. ~\eqref{eq:quad_var}), Lindblad-based \texttt{mesolve()}, and McCuller et al.~\cite{McCuller2021} is shown in Figures \ref{fig:quad_var_phi_offset} and \ref{fig:quad_var_vs_phasenoise}. Throughout this work, the squeezing level in decibels is related to the squeezing parameter by $S_\text{dB} = 20\,r\,\log_{10}(e) \approx 8.686\,r$. 

Motivated by the dominant source of phase noise in gravitational wave detectors, we only consider the phase noise prior to sensing, which is equivalent to a random phase shift on the input state before it enters the sensing channel.  For such detectors, this represents the relative phase noise of the state injected in the differential port with respect to the main laser. 


\section{Theory of Parameter Sensing \label{app:limits_of_sense}}

In this Section, we review the quantum limits of single-parameter estimation and establish the framework used throughout this work. The central question is: given an energy budget (i.e. maximal mean photon number $\bar n$), what is the best possible precision for estimating a small signal encoded in a quantum state, and how does decoherence, specifically photon loss and phase noise, modify it?

\subsection{Quantum Fisher Information and Classical Fisher Information}

We prepare an input state $\hat{\rho}$ that undergoes a completely positive and trace-preserving channel $\Lambda_\epsilon(\cdot)$, parametrized by $\epsilon$. The parameter is encoded in the output state through the evolution $\hat{\rho}_\epsilon = \Lambda_\epsilon(\hat{\rho})$. If one performs a positive operator-valued measure (POVM) $\{E_m\}$ to the output state to obtain outcomes $\{(m, p(m|\epsilon))\}$, where $p(m|\epsilon) = \text{Tr}(\hat{\rho}_\epsilon E_m)$, the classical Fisher information (CFI) can be written as
\begin{align}
    \mathcal{F}_C(\hat{\rho}_\epsilon) &\coloneqq E\left[\left(\frac{\partial}{\partial\epsilon} \ln{(p(m|\epsilon))} \right)^2\right] \nonumber \\
    &= \sum_m \frac{1}{p(m|\epsilon)} \left(\frac{\partial p(m|\epsilon)}{\partial\epsilon} \right)^2.
\end{align}
The precision in estimating $\epsilon$ is connected to the CFI via the Cramér-Rao bound, stated as
\begin{align}
    (\Delta\epsilon)^2 \geq \frac{1}{M \mathcal{F}_C(\hat{\rho}_\epsilon)}
\end{align}
where $M$ is the number of independent repetitions of the experiment, and $(\Delta\epsilon)^2$ is the variance in estimating $\epsilon$. The ultimate precision for a given input state is quantified by the quantum Fisher information (QFI) $\mathcal{F}_Q(\hat{\rho}_\epsilon)$. 
For a spectral decomposition $\hat{\rho}_\epsilon = \sum_k \lambda_i' \ket{\phi_i'} \bra{\phi_i'}$ of the state $\hat{\rho}$, the QFI about $\epsilon$ is given by
\begin{align}
    \mathcal{F}_Q(\hat{\rho}_\epsilon) \coloneqq 2\sum_{i, j} \frac{|\bra{\phi_i'} \partial_\epsilon \hat{\rho}_\epsilon  \ket{\phi_j'}|^2}{\lambda_i' + \lambda_j'}
\end{align}
where the sum excludes any pair $\lambda_i', \lambda_j'$ such that $\lambda_i' + \lambda_j' = 0$. The QFI states the precision obtainable when optimized over the POVM applied on the output state. Furthermore, the quantum Cramér-Rao bound states that
\begin{align}
    (\Delta\epsilon)^2 \geq \frac{1}{M \mathcal{F}_Q(\hat{\rho}_\epsilon)}.
\end{align}
The measurement strategy to attain the QFI is to project to the eigenbasis of the operator $\hat{L}$, defined as
\begin{align}
\label{eq:symm_log_derivative}
    \hat{L} \coloneqq 2\sum_{i, j} \frac{\bra{\phi_i'} \partial_\epsilon \hat{\rho}_\epsilon  \ket{\phi_j'}}{\lambda_i' + \lambda_j'} \ket{\phi_i'}\bra{\phi_j'}
\end{align}
also referred to as the symmetric logarithmic derivative. However, this POVM is often hard to implement experimentally. For a pure input state and unitary encoding channel $\Lambda_\epsilon(\cdot) = \hat{U} \cdot \hat{U}^\dagger$, with $\hat{U} = e^{-i \epsilon \hat{G}}$, the QFI expression simplifies to
\begin{align}
\label{eq:pure_state_qfi}
    \mathcal{F}_Q(\hat{\rho}_\epsilon) = 4(\langle \hat{G}^2\rangle - \langle \hat{G}\rangle^2) = 4 (\Delta \hat{G})^2
\end{align}
Here, $\hat{G}$ is the generator of the encoding. In contrast, for a mixed state $\hat{\rho}_\epsilon = \sum_k \lambda_i' \ket{\phi_i'} \bra{\phi_i'}$ evolving with the same unitary channel $U = e^{-i\epsilon \hat{G}}$, the QFI is found from
\begin{align}
    \label{eq:qfi_with_generator}\mathcal{F}_Q(\hat{\rho}_\epsilon) = 2 \sum_{k, l} \frac{(\lambda_k' - \lambda'_l)^2}{\lambda_k' + \lambda'_l} |\bra{\phi'_k} \hat{G} \ket{\phi'_l}|^2.
\end{align}
In general, when maximizing the QFI, we can make use of the convexity: for an output state $\hat{\rho}_\epsilon = \Lambda_\epsilon(\hat{\rho}) = \sum_i \lambda_i \Lambda_\epsilon(\ket{\phi_i} \bra{\phi_i})$, we have that
\begin{align}\label{eq:qfi_convexity_eq}
\mathcal{F}_Q\left(\sum_i \lambda_i \Lambda_\epsilon(\ket{\phi_i} \bra{\phi_i}) \right) \leq \sum_i \lambda_i \mathcal{F}_Q(\Lambda_\epsilon(\ket{\phi_i} \bra{\phi_i}))
\end{align}
Then, it is more advantageous to choose a pure input state. Because of this, all optimizations in this work are restricted to pure input states.

\subsection{Single mode displacement sensing as an approximation to two-mode phase sensing\label{app:two_mode}}

Most results presented in this paper were optimized for the single mode configuration, for the task of sensing a displacement, with signal encoded with the Hamiltonian $\hat{H} = g\,(\hat{a} + \hat{a}^\dagger)$.  We motivate this Hamiltonian in this section,  as the limiting case of sensing a phase shift in a Mach-Zehnder interferometer. Similar reductions have previously been explored in \cite{gardner25_swe,gorshenin_using_2024,gorshenin_using_2025,paris_displacement_1996}. 
Our  result also applies to Michelson interferometers such as the LIGO detectors.

\begin{figure}
    \centering
\includegraphics[width=0.4\linewidth]{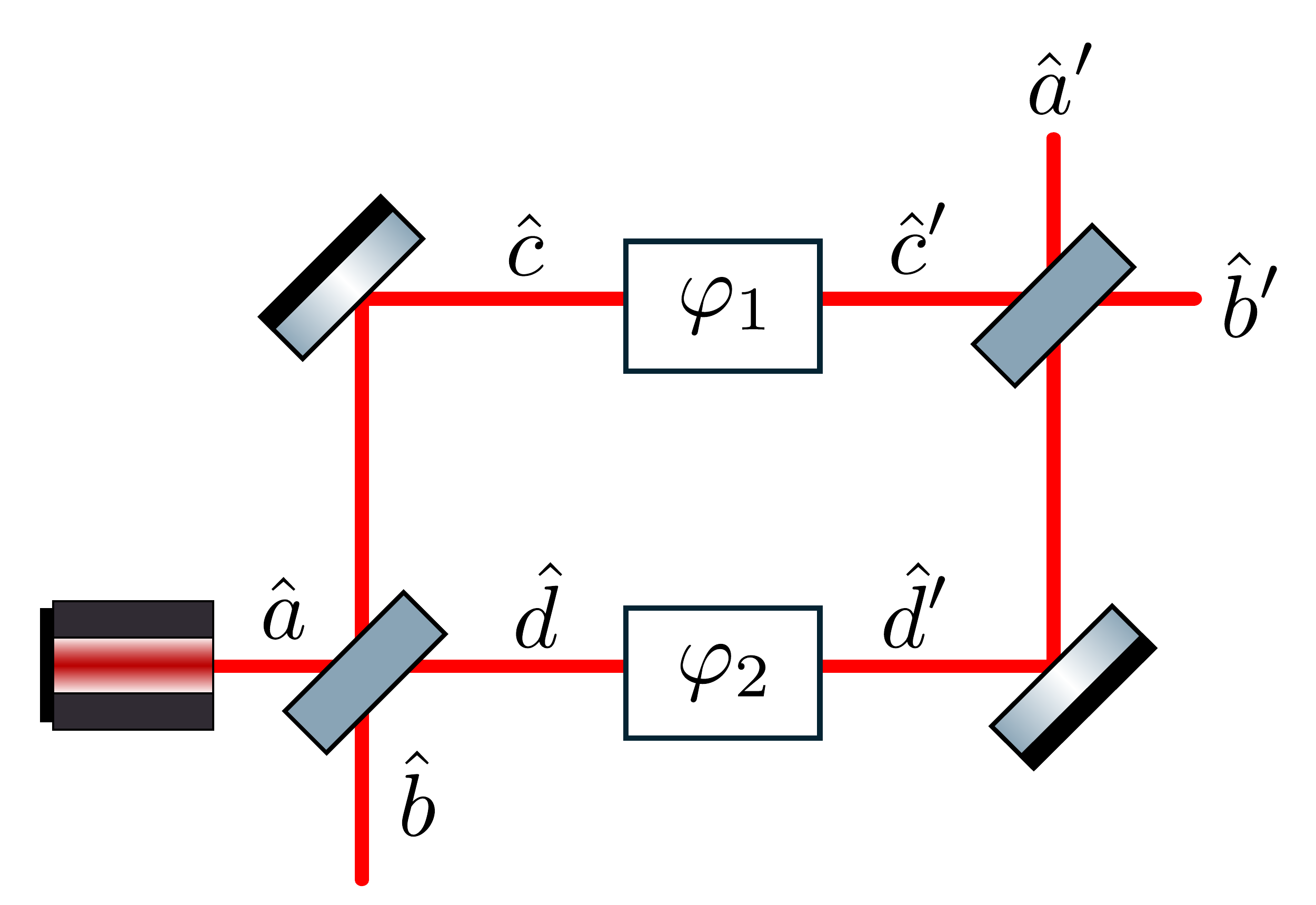}
        \label{fig:mach_zender_sup}
\includegraphics[width=0.4\linewidth]{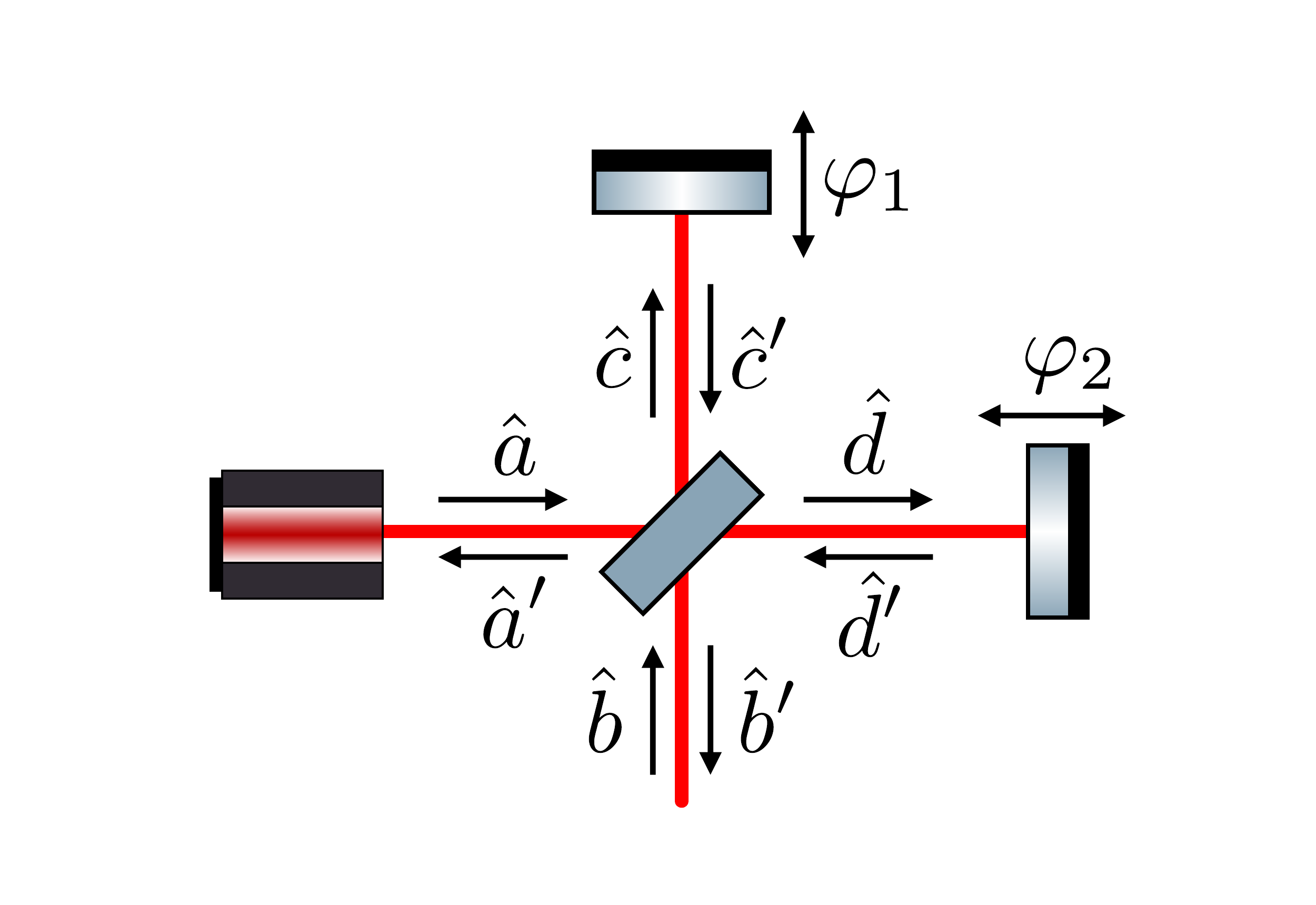}
    \caption{Evolution of annihilation operators in a Mach-Zehnder (left) and a Michelson (right) interferometer. }
    \label{fig:MZ_MM_modes}
\end{figure}

We consider the interferometer setup of Figure~\ref{fig:MZ_MM_modes}, where both beam splitters are 50/50. This setup is a special case of the interferometer explored in \cite{lang_optimal_2013}. From the input-output relations, we have that the modes in the arms after the beam splitter are given by 
\begin{equation*}
    \hat{d}=\frac{1}{\sqrt{2}}\left(\hat{a}+\hat{b}\right),\quad \hat{c}=\frac{1}{\sqrt{2}}\left(-\hat{a}+\hat{b}\right).
\end{equation*}
The light then undergoes phase shifts of $\varphi_1,\varphi_2$ respectively in the two arms
\begin{equation*}
    \hat{c}'=e^{-i\varphi_1}\hat{c}=e^{-i\varphi_1}\frac{1}{\sqrt{2}}\left(-\hat{a}+\hat{b}\right),
\quad    \hat{d}'=e^{-i\varphi_2}\hat{d}=e^{-i\varphi_2}\frac{1}{\sqrt{2}}\left(\hat{a}+\hat{b}\right).
\end{equation*}
Finally, the light is recombined at the second beam splitter
\begin{align*}
    \hat{b}'=\frac{1}{\sqrt{2}}\left(\hat{c}'+\hat{d}'\right)= 
    e^{-i\frac{\varphi_s}{2}}\left[-i\sin\left(\frac{\varphi_d}{2}\right)\hat{a}+\cos\left(\frac{\varphi_d}{2}\right)\hat{b}\right],
\end{align*}
where we have defined $\varphi_s=\varphi_1+\varphi_2$ and $\varphi_d=\varphi_1-\varphi_2$ for simplicity. We consider the scenario where the bright port (mode $\hat{a}$) is being pumped with a high-amplitude coherent state. In this limit, $\hat{a}\approx\alpha + \delta\hat{a}$ where $\alpha\in \mathbb{C}$, and $\delta\hat{a} \ll \alpha$ are the small fluctuations in $\hat{a}$. Plugging this into $\hat{b}'$, we obtain:
\begin{equation*}
    \hat{b}'\approx e^{-i\frac{\varphi_s}{2}}\left[-i\sin\left(\frac{\varphi_d}{2}\right)(\alpha + \delta\hat{a}) + \cos\left(\frac{\varphi_d}{2}\right)\hat{b}\right].
\end{equation*}
We also work in the limit of $\varphi_d\ll 1$. Taylor expanding $\hat{b}'$ with respect to $\varphi_d$, neglecting second order terms and higher,
\begin{equation*}
    \hat{b}'\approx e^{-i\frac{\varphi_s}{2}}\left(\hat{b}-i\frac{\varphi_d\alpha}{2}\right).
\end{equation*}
Under these assumptions, we see that the interferometer transforms mode $b$ by first displacing it by $-i\varphi_d\alpha/2$, then applying a phase shift of $-\varphi_s/2$ in the Heisenberg picture. 
The task of sensing the phase difference $\varphi_d$ is therefore equivalent to the task of sensing a single mode displacement in this regime.

Let $\epsilon=i\varphi_d\alpha/2$ be the displacement. The QFI $\mathcal{F}_{Q,\varphi_d}$ for sensing the phase difference $\varphi_d$ can be related to the QFI $\mathcal{F}_{Q,\epsilon}$ for sensing the displacement $\epsilon$ via the chain rule
\begin{align}
    \mathcal{F}_{Q,\varphi_d}(\rho)&=\left|\frac{\partial\epsilon}{\partial\varphi_d}\right|^2\mathcal{F}_{Q,\epsilon}(\rho) 
    =\frac{|\alpha|^2}{4}\mathcal{F}_{Q,\epsilon}(\rho).
\end{align}
The phase shift of $\varphi_s/2$ is a unitary transformation that does not depend on $\varphi_d$, and so does not change the QFI.

Given this correspondence, we now specialize to the sensing task studied throughout this work.
The Hamiltonian corresponding to single bosonic mode displacement is
\begin{align}
\label{eqn:ham_eps}
    \hat{H} = g\,(\hat{a} + \hat{a}^\dagger),
\end{align}
where $\hat{a}$ ($\hat{a}^\dagger$) is the annihilation (creation) operator of the mode, and $\epsilon \coloneqq g t$ parametrizes the displacement amplitude along the phase quadrature. The generator of the encoding is therefore $\hat{G} = \hat{a} + \hat{a}^\dagger$, and the QFI can be found from Eq. (\ref{eq:pure_state_qfi}).

\subsection{Limits of sensing}

In this paper, the combined quantum channel that incorporates both signal and noise is described by evolving the following master equation for a time interval of $t=1$: 
\begin{align}
    \frac{d\hat{\rho}}{dt} = -i [\hat{H}, \hat{\rho}] + \sum_i \left( \hat{L}_i \hat{\rho} \hat{L}_i^\dagger - \frac{1}{2} \{ \hat{L}_i^\dagger \hat{L}_i, \hat{\rho} \} \right)\,.
\end{align}
Here $\hat{H} = \epsilon(\hat{a}^\dagger + \hat{a})$ encodes the displacement/phase-sensing signal, while   $\{\hat{L}_i\}$ are the Lindblad operators, with loss operator given by $\hat{L} = \sqrt{\kappa} \hat{a}$ and  phase-noise operator given by $\hat{L} = \sqrt{\chi} \hat{a}^\dagger \hat{a}$, where $\kappa$ and $\chi$ are the loss and phase noise rates, respectively. For this channel,  the question that we address in this work then becomes: which single-mode state $\ket{\psi}$ with a given $\bra{\psi}\hat{a}^\dagger \hat{a}\ket{\psi}$ maximizes $\mathcal{F}_Q$ under realistic decoherence? 


The following Subsections answer this question in three asymptotic regimes that guide our numerical optimization procedure: noiseless phase-sensitive sensing (Sec. \ref{subsec:noiseless_phase_sensitive}), phase-sensitive sensing with loss (Sec. \ref{subsec:noisy_phase_sensitive}), and phase-insensitive sensing in the high-phase-noise limit (Sec. \ref{subsec:phase_insensitive}).

\subsubsection{Noiseless, phase-sensitive sensing (squeezed vacuum states are optimal)}
\label{subsec:noiseless_phase_sensitive}
In the absence of any loss, an input pure state is mapped to an output pure state with the encoding in Eq. \ref{eqn:ham_eps}.
The QFI for a given pure state $\ket{\psi}$ can be computed from Eq. (\ref{eq:pure_state_qfi}), with $G=\hat{a}+\hat{a}^{\dagger}$. Here, it is implicitly assumed that the phase of the displacement is known (i.e., the displacement could have been applied to $ae^{i\phi}+a^\dagger e^{-i\phi}$, but we have chosen $\phi=0$), hence, we refer to this case as phase-sensitive sensing.
The arguments of \cite{lang_optimal_2013} apply, and so this QFI is maximized by a squeezed vacuum state for a fixed photon number $\bar{n} = \bra{\psi}\hat{a}^{\dagger}\hat{a}\ket{\psi}$. 
Squeezed vacuum states are therefore optimal in the absence of noise. For a squeezing parameter $r$, $\bar n = \sinh^2{(r)}$, and
the QFI is found as $\mathcal{F}_Q = 4 e^{2r} = 8 (\bar n + 1/2 + \sqrt{\bar n (\bar n + 1)}) \approx 16 \bar{n}$ for $\bar{n} \gg 1$.

\subsubsection{Phase-sensitive sensing with decoherence (squeezed vacuum states are optimal)}
\label{subsec:noisy_phase_sensitive}

In the case of phase-sensitive sensing with loss only, the optimal state is known to be  squeezed vacuum \cite{latune_quantum_2013}. For completeness, we re-derive this fact below. The effect of loss can be simplified as follows. Loss modifies an annihilation operator $\hat{a}$ as
\begin{align}
    \hat{a} \rightarrow \sqrt{\eta} \, \hat{a} + \sqrt{1-\eta} \, \hat{b}
\end{align}
where $\hat{b}$ is the annihilation operator of the vacuum state. $\eta = e^{-\kappa t}$ is the transmission rate in terms of the loss rate $\kappa$, assuming that the channel is experienced for a time $t$.

We can have three scenarios: the displacement is encoded before the loss, simultaneously with the loss, or after the loss. As these are both Gaussian channels, it can be analytically shown that they commute up to a suppression factor in the displacement. More specifically, let us assume a channel where displacement and photon loss are interleaved for a total time $T$, where each loss channel lasts for $\tau$. From the effect of these operations on the annihilation and creation operators $\hat{a}$, $\hat a^\dagger$, it is possible to write
\begin{align}
    \mathcal{D}(\theta) \circ\Lambda_\text{loss}^\eta \circ \mathcal{D}(\theta) \circ\Lambda_\text{loss}^\eta \dots \approx \mathcal{D} \left(\theta \sum_{i=0}^{T/\tau-1} \eta^{i/2} \right) (\Lambda_\text{loss}^\eta)^{T/\tau},
\end{align}
where $\eta = e^{-\kappa \tau}$, and $\mathcal{D}(\theta) = e^{-i \theta (\hat{a} + \hat{a}^\dagger)}$ is the single-mode displacement channel. Defining the total displacement as $\epsilon = \theta T/\tau$ and taking the continuous-time limit $\tau \rightarrow 0$, 
\begin{align}
    \underset{\tau \rightarrow 0}{\text{lim}} \; \epsilon \frac{\tau}{T} \sum_{i=0}^{T/\tau-1} e^{-\kappa \tau i/2} = 2\epsilon \frac{(1-e^{-\kappa T/2} )}{\kappa T}.
\end{align}
Then, a displacement of $\epsilon$ during loss is equivalent to first experiencing the loss channel, then being displaced by $\beta\epsilon$, where $\tilde \eta \coloneqq e^{-\kappa T}$ and
\begin{equation}\label{eq:loss_conversion_factor}
\beta = 2(1-\sqrt{\tilde\eta})/|\log{\tilde\eta}|.
\end{equation}
Using this conversion factor $\beta$, we write the QFI expressions in the presence of loss assuming that the displacement occurs after the loss channel, and convert the expression to the ``loss during displacement" case easily.

In the presence of photon loss with a transmission rate $\eta$ (assuming displacement occurs after loss),
the QFI of the squeezed state is modified to
\begin{align}
\label{eq:qfi_squeezed_loss_only}
    \mathcal{F}_Q = \frac{4}{\eta \, e^{-2r} + 1-\eta}
\end{align}
where $\eta = e^{-\kappa t}$, and $t$ is the time for which the state evolves under the noise channel. 
In the high-loss limit ($\eta \,e^{-2r} \ll 1-\eta$), the QFI reduces to $\mathcal{F}_Q \approx 4/(1-\eta)$, and the scaling with the mean photon number is lost.
In order to find the optimal state in the presence of photon loss, we follow Refs. \cite{Escher2011, Demkowicz2012, Kolodynski_2013} and compute the channel QFI bound. For an evolution $\hat{\rho}(\epsilon) = \sum_l \hat{\Pi}_l(\epsilon) \hat{\rho}_0 \hat{\Pi}_l^\dagger(\epsilon)$, this bound states that
\begin{align}
    \mathcal{F}_Q(\hat{\rho}(\epsilon)) \leq \text{min}_{\{\hat{\Pi}_l(\epsilon)\}} C_Q[\hat{\rho}_0, \hat{\Pi}_l(\epsilon)]
\end{align}
where
\begin{align}
    &C_Q[\hat{\rho}_0, \hat{\Pi}_l(\epsilon)] = 4 \left[\langle \hat{H}_1 \rangle - \langle \hat{H}_2 \rangle^2\right], \label{eq:escher_variance} \\
    & \hat{H}_1 = \sum_l \frac{d \hat{\Pi}_l^\dagger(\epsilon)}{d\epsilon}  \frac{d \hat{\Pi}_l(\epsilon)}{d\epsilon}, \quad \hat{H}_2 = i\sum_l \frac{d \hat{\Pi}_l^\dagger(\epsilon)}{d\epsilon}   \hat{\Pi}_l(\epsilon)
\end{align}
In other words, the QFI is found from a minimization over the QFIs of possible purifications of the noisy state. Then, we can first write a purification of the state after the loss in the form of $\hat{U}_\text{BS}\ket{\psi}_S\ket{0}_E$, where $\ket{\psi}_S$ is an initial pure state, and $\ket{0}_E$ is the environment in its vacuum state. The Hamiltonian generating the displacement after the loss can be written as $\hat{H} = \hat{G}_\text{S} \otimes I_\text{E} + I_\text{S} \otimes \hat{h}_\text{E}$, with an arbitrary operator $\hat{h}_\text{E}$ acting on the environment. $\hat{h}_\text{E}$ does not modify the QFI in the system as it only acts on the environment, however, it can be used to optimize the total QFI of the purified state. In this way, a saturable bound on the QFI of the lossy system state can be obtained.

We choose $\hat{h}_\text{E} = \beta (\hat{b}^\dagger +  \hat{b})$, where $\hat{b}$ is the annihilation operator of the environment (and $\hat{a}$ is the annihilation operator of the system). This choice can be explained intuitively as displacing the environment such that the leaked information to the environment during the photon loss channel is recovered.
The QFI of the purified state is found as $4\text{var}_{\ket{\Psi}}(\hat H)$. In the Heisenberg picture, the evolutions of the annihilation operators of the system and the environment are given respectively by
\begin{align}
    \hat{a}^{H} &= \sqrt{\eta} \,\hat{a} + \sqrt{1-\eta} \, \hat{b} \nonumber \\
    \hat{b}^{H} &= \sqrt{\eta} \,\hat{b} - \sqrt{1-\eta} \, \hat{a}
\end{align}
where $\eta$ is the transmission rate of the photon loss channel. Then, we obtain
\begin{align}
    \langle \hat{H}\rangle &= \langle \hat{a}^{H} + (\hat{a}^{H})^\dagger + \beta(\hat{b}^{H} + (\hat{b}^{H})^\dagger) \rangle = \langle \hat{a}^\dagger + \hat{a} \rangle (\sqrt{\eta} - \sqrt{1-\eta} \, \beta) \nonumber \\
    \langle \hat{H}^2\rangle &= \langle (\hat{a}^\dagger + \hat{a})^2 \rangle (\sqrt{\eta} - \sqrt{1-\eta} \, \beta)^2 + (\sqrt{1-\eta} + \sqrt{\eta} \, \beta)^2
\end{align}
To maximize the variance (hence the QFI of the purified state), we set $\langle \hat{a}^\dagger + \hat{a} \rangle=0$. Then, we optimize over $\beta$ in order to get the tightest possible bound on the QFI. Taking the derivative of the variance with respect to $\beta$ and setting it to zero gives
\begin{align}
    \beta^* = \frac{(\langle (\hat{a}^\dagger + \hat{a})^2 \rangle-1)\sqrt{\eta}\,\sqrt{1-\eta}}{\langle (\hat{a}^\dagger + \hat{a})^2 \rangle(1-\eta) + \eta} = \frac{(V-1)\sqrt{\eta}\,\sqrt{1-\eta}}{V(1-\eta) + \eta}
\end{align}
where we defined $V = \langle (\hat{a}^\dagger + \hat{a})^2 \rangle$. Plugging this into the expression for the QFI of the purified state, we obtain
\begin{align}\label{eq:cqfi_V_quantity}
    \mathcal{F}_Q^{p} = \frac{4V}{V(1-\eta)+\eta}
\end{align}
This function increases monotonically with $V$. Then, maximizing the QFI is equivalent to maximizing $V$ with the photon number constraint $\langle\hat{a}^\dagger \hat{a} \rangle \leq \bar n$. This maximum is achieved for squeezed states, with $V = e^{2r}$, $\bar n = \sinh^2{r}$, and the QFI of the purified state is
\begin{align}
    \mathcal{F}_Q^{p} = \frac{4}{\eta \, e^{-2r} + 1-\eta}.
\end{align}
From Eq. (\ref{eq:qfi_squeezed_loss_only}), we observe that this is the exact expression obtained for the QFI of squeezed states in the presence of photon loss. Since the maximal QFI over all purifications for all input states is saturable by these states, we conclude that they are the optimal input states for this case.

\subsubsection{Lossless phase-insensitive sensing (Fock states are optimal)}
\label{subsec:phase_insensitive}


Let us now consider the high phase-noise limit.  We will first show that the optimal input state is a Fock state.  We will then compute the QFI of Fock states in the presence of loss.

Let us  write the two-stage sensing channel in terms of the channels $\Phi$ and $\Lambda_{\epsilon}$, which will denote the preparation phase noise channel and lossy displacement channel respectively.  In the limit of very high phase noise, $\Phi$ becomes the completely dephasing channel, which can be written 
\begin{equation}
    \Phi(\hat{\rho})=\int_{0}^{2\pi}\frac{d\phi}{2\pi}R(\phi)\hat{\rho} R^{\dagger}(\phi),
\end{equation}
where $R(\phi)=\exp\left(-i\phi\hat{a}^{\dagger}\hat{a}\right)$. The output state $\Phi(\hat{\rho})$ is invariant under rotations. To show that Fock states are the optimal state in certain regimes, we consider the following sequence of inequalities. Here, the maximum will be taken over all states $\rho$ such that $\textrm{Tr}(\hat{a}^{\dagger}\hat{a}\hat{\rho})\leq N$ for a fixed $N$. We will omit that constraint from the notation to reduce clutter. 

\begin{align}
    \max_{\hat{\rho}} \mathcal{F}_{Q}(\Lambda_{\epsilon}\circ\Phi(\hat{\rho}))\leq\max_{\hat{\rho}\, : \, R(\varphi)\hat{\rho} R^{\dagger}(\varphi)=\hat{\rho}\, \forall \, \varphi} \mathcal{F}_{Q}\left(\Lambda_{\epsilon}(\hat{\rho})\right).
\end{align}
We have used the fact that since $\Phi(\hat{\rho})$ is invariant under rotations, it's QFI is less than equal to the maximal QFI over all rotationally invariant states. Any state which is invariant under rotations can be written in the following form 
\begin{equation*}
    \hat{\rho}=\sum_{m=0}^{\infty}p_m\ket{m}\bra{m},
\end{equation*}
with $\sum_mp_m=1$.
We can then use the convexity of the QFI to further bound the previous inequalities. Note that this convexity bound is tight for Fock states.
\begin{align}
    \max_{\hat{\rho}\, : \, R(\varphi)\hat{\rho} R^{\dagger}(\varphi)=\hat{\rho}\, \forall \, \varphi} \mathcal{F}_{Q}\left(\Lambda_{\epsilon}(\hat{\rho})\right)&\leq \max_{p_m\, :\, \sum_mp_m=1}\sum_{m=0}^{\infty} p_m \mathcal{F}_Q\left(\Lambda_{\epsilon}(\ket{m}\bra{m})\right)
\end{align}
Recall that we maximize under an average photon number constraint, and therefore we only consider states such that $\sum_m mp_m\leq N$. In the case of no-loss, the QFI of a displaced Fock state is known.
\begin{align}
    \max_{\hat{\rho}\, : \, R(\varphi)\hat{\rho} R^{\dagger}(\varphi)=\hat{\rho}\, \forall \, \varphi} \mathcal{F}_{Q}\left( D(\epsilon)\hat{\rho} D^{\dagger}(\epsilon) \right) &\leq \max_{p_m\, :\, \sum_mp_m=1}\sum_{m=0}^{\infty} p_m \mathcal{F}_Q(D(\alpha) \ket{m}\bra{m} D^{\dagger}(\alpha)) \\
    &=\max_{p_m\, :\, \sum_mp_m=1}\sum_{m=0}^{\infty}p_m4(1+2m) 
    =4(1+2\langle\hat{n}\rangle).
\end{align}
Note that all these inequalities are saturated for Fock states in the absence of loss. We therefore conclude that Fock states are optimal in the presence of complete dephasing noise.

\subsubsection{QFI of Fock states in the presence of loss}

Let us now compute the QFI of a Fock state after experiencing the photon loss channel with transmission rate $\eta$. These results provide support to our numerical results that Fock states are optimal for low losses, but not for high losses. 

The photon loss channel will act only on the diagonal elements. For a Fock state $\ket{N}$, let us write the state at some time $t$ as
$\hat{\hat{\rho}} = \sum_{k=0}^N p_k(t) \ket{k} \bra{k}$. We find that $p_k(t) = \binom{N}{k} \eta^k \, (1-\eta)^{N-k}$.
Given this eigendecomposition of the state as a function of time, we use Eq. (\ref{eq:qfi_with_generator}) to compute the QFI. First, we find for $l < N$
\begin{align}
    p_l + p_{l+1} 
    = p_l \left(1  + \frac{N-l}{l+1} \frac{\eta}{1-\eta}  \right) ,\quad
(l+1)\frac{(p_{l+1} - p_l)^2}{p_{l+1} + p_l}  = p_l \frac{(l+1 - (N-l)r)^2}{l+1 + (N-l)r}
\end{align}
where we defined $r = \eta/(1-\eta)$. Then, we obtain from Eq. (\ref{eq:qfi_with_generator})
\begin{align}
\mathcal{F}_Q = 4 \sum_{l=0}^{N} p_l  \frac{(l+1 - (N-l)r)^2}{l+1 + (N-l)r} = \frac{4}{1-\eta} \sum_{l=0}^{N} \binom{N}{l} \eta^l (1-\eta)^{N-l}  \frac{(l+1 - (N+1)\eta)^2}{l+1+ \eta (N-2l-1)}
\end{align}
We can approximate this expression for some parameter regimes in order to observe the scaling of the QFI with respect to $N$ and $\eta$. First, let us assume that $N(1-\eta) \gg 1$, $N\eta \gg 1$, which is the moderate loss regime. In this regime, the binomial distribution can be approximated with a Gaussian distribution.
 Approximating also the QFI sum as an integral, we obtain:
\begin{align}
     \mathcal{F}_Q &\approx \frac{4}{(1-\eta)^{3/2} \sqrt{2\pi\eta}}  \int_{-\infty}^\infty dx \, e^{-\frac{x^2}{2\eta \, (1-\eta)}} \frac{x^2}{x(1-2\eta) + 2\eta(1-\eta)}  \approx \frac{2}{1-\eta}
     \label{FQlargeNfixeta}
\end{align}
Then, in this limit, we observe that the QFI is independent of $N$, and Heisenberg Scaling is lost.

In contrast, in the small loss regime where $\eta = 1-\epsilon$, $\epsilon \ll 1$, $N \epsilon \lesssim 1$, $N (1-\epsilon) \gg 1$, the binomial distribution can be approximated as a Poisson distribution with
\begin{align}
    \binom{N}{l} \eta^l (1-\eta)^{N-l} = \binom{N}{N-l} \epsilon^{N-l} (1-\epsilon)^{l} \sim \frac{\mu^{N-l} e^{-\mu}}{(N-l)!} 
\end{align}
for $\mu = N \epsilon$. Then,
\begin{align}
    \mathcal{F}_Q \approx \frac{4}{\epsilon} \sum_{l=0}^{N} \frac{\mu^{N-l} e^{-\mu}}{(N-l)!}  \frac{(\mu - (N-l))^2}{N-l - \mu (1-\frac{2l}{N})}
    \approx \frac{4 e^{-\mu}}{\epsilon} \sum_{k=0}^{N} \frac{\mu^k}{k!}  \left( \mu -3k + \frac{4\mu^2}{\mu + k} \right)
\end{align}
Evaluating the sum over the first two terms and converting the third into an integral, we obtain
\begin{align}
    \mathcal{F}_Q  \approx \frac{4}{\epsilon}\left(-2\mu + 4 \mu^2 e^{-\mu} \sum_{k=0}^N \frac{\mu^k}{k!} \int_0^1 dt \, t^{\mu+k-1} \right) 
     \approx 4N \left(-2 + 4 \mu \, e^{-\mu} \int_0^1 dt \, t^{\mu-1} e^{t\mu} \right)
\end{align}
Finally, by expanding the integral in powers of $\mu$ for $\mu \ll 1$, we can write
\begin{align}
    \mathcal{F}_Q  \approx 8N \frac{1-\mu}{1+\mu}. 
\end{align}
Then, in the small loss regime, Heisenberg Scaling is preserved. We do note that such a scaling might be indicated already from Eq.~\eqref{FQlargeNfixeta}, but that equation assumes $\eta$ is fixed and $N$ is large. 

\subsection{Convexity Bound}


In this section, we obtain two bounds on QFI using its convexity property.  Consider the displacement channel $\mathcal{D}_{\epsilon}$, the phase noise channel $\Phi$ and the pure loss channel $\Lambda$. We will consider the case where there is state preparation noise, a loss channel, and then a displacement channel. The computed QFI can be related to the QFI for simultaneous loss and displacement using the chain rule for QFI and the conversion factor of equation Equation~\eqref{eq:loss_conversion_factor}. In this setting, the QFI can be bounded from above as follows 
\begin{align}
    \mathcal{F}_Q(\mathcal{D}_{\epsilon}\circ \Lambda\circ\Phi(\hat{\rho}))&=\mathcal{F}_Q\left(\mathcal{D}_{\epsilon}\circ\Lambda\left(\int d\varphi P(\varphi)R(\varphi)\hat{\rho} R^{\dagger}(\phi)\right)\right) \\
    &=\mathcal{F}_Q\left(\int d\varphi P(\varphi) \mathcal{D}_{\epsilon}\circ\Lambda(R(\varphi)\hat{\rho} R^{\dagger}(\varphi))\right) \\
    &\leq \int d\varphi P(\varphi)\mathcal{F}_Q(\mathcal{D}_{\epsilon}\circ\Lambda(R(\varphi)\hat{\rho} R^{\dagger}(\varphi))).
\end{align}
The first equality is the operator-sum representation of the dephasing channel, the second is obtained by linearity of quantum channels, and the third is obtained using the convexity of the QFI. For simplicity, we will call such upper bounds convexity bounds.

\subsubsection{Convexity bound for Gaussian states}

Let us first consider the convexity bound for Gaussian states, which will indicate a suppression of QFI due to phase noise. We will consider the optimization over states $\hat{\rho}$ satisfying the constraint $\langle\hat{n}\rangle_{\hat{\rho}}\leq N_{target}$. We will drop explicit references to this constraint from the equations to help simplify the notation.

We first consider the optimization over all Gaussian states.
\begin{equation}
    \max_{\hat{\rho}\textrm{ Gaussian}}\mathcal{F}_Q(\mathcal{D}_{\epsilon}\circ \Lambda\circ\Phi(\hat{\rho}))\leq \max_{\hat{\rho}\textrm{ Gaussian}}\int d\varphi P(\varphi)\mathcal{F}_Q(\mathcal{D}_{\epsilon}\circ\Lambda(R(\varphi)\hat{\rho} R^{\dagger}(\varphi))).
\end{equation}
The state $\mathcal{D}_{\epsilon}\circ\Lambda\circ\Phi(\hat{\rho})$ may be non-Gaussian since $\Phi$ is a not a Gaussian channel. However, $\mathcal{D}_{\epsilon}\circ\Lambda(R(\varphi)\hat{\rho} R^{\dagger}(\varphi))$ is Gaussian, and so its QFI can be computed from its first and second moments \cite{fadel_quantum_2025,weedbrook_gaussian_2012}.
Consider the single mode Gaussian state $\ket{\psi}=\hat{D}(\alpha)\hat{S}(re^{i\gamma})$, whose covariance matrix we denote by $\Gamma$. Without loss of generality, we can take the maximal $\hat{\rho}=\ket{\psi}\bra{\psi}$ to be Gaussian due to the convexity of the QFI. One can show that after the rotation operators, the loss channel, and the displacement, the input covariance $\Gamma$ is transformed into the output covariance $\tilde{\Gamma}$ (the tilde denoting the covariance matrix after the channel), given by \cite{fadel_quantum_2025,gardner25_swe}
\begin{equation}
    \tilde{\Gamma}=\frac{\eta}{2}\begin{pmatrix}
        \cosh(2r)-\sinh(2r)\cos(\gamma+2\varphi) & -\sinh(2r)\sin(\gamma+2\varphi) \\
        -\sinh(2r)\sin(\gamma+2\varphi) & \cosh(2r)+\sinh(2r)\cos(\gamma+2\varphi)
    \end{pmatrix} + \frac{1-\eta}{2}\mathbb{I},
\end{equation}
and its mean vector is given by
\begin{equation}
    \mu=\left(\sqrt{2}\Re(\alpha e^{i\varphi}),\sqrt{2}\Im(\alpha e^{i\varphi})+\sqrt{2}\epsilon\right)^T.
\end{equation} 
For the first and second moments, we use the conventions of \cite{fadel_quantum_2025}. The QFI can then be computed using \cite{pinel_quantum_2013,fadel_quantum_2025}
\begin{equation}
    \mathcal{F}(\hat{\rho})=2(0,\ 1)\tilde{\Gamma}^{-1}\begin{pmatrix}
        0\\ 1
    \end{pmatrix}.
\end{equation}
Substituting $\tilde{\Gamma}$ into the equation yields
\begin{equation}
    \mathcal{F}_Q(\mathcal{D}_{\epsilon}\circ \Lambda(\hat{\rho}))=\frac{8\eta \bar{n}+8\eta\sqrt{\bar{n}(1+\bar{n})}\cos(\gamma+2\varphi)) +4}{1+4\eta(1-\eta)\bar{n}}.
\end{equation}
Taking the integral over the phase and optimizing over $\bar{n},\gamma$ leads to the Gaussian convexity bound. Assuming the phase is given by a zero-mean normal distribution with variance $\sigma_\phi^2$, then the bound becomes
\begin{equation}\label{eq:gaussian_convexity_bound}
    \max_{\hat{\rho}\textrm{ Gaussian}}\mathcal{F}_Q(\mathcal{D}_{\epsilon}\circ \Lambda\circ\Phi(\hat{\rho}))\leq \frac{8\eta \bar{n}+8\eta\sqrt{\bar{n}(1+\bar{n})}e^{-2\sigma_\phi^2} +4}{1+4\eta(1-\eta)\bar{n}}.
\end{equation}
However, it is important to note that Equation~\eqref{eq:gaussian_convexity_bound} is only an upper bound for the QFI of Gaussian states.

\subsubsection{Convexity bound for General States}

For general states, we get a weaker bound. More specifically, we relax the bound by ignoring the contribution of phase noise to once again obtain a computation using Gaussian states.
\begin{align}
    \max_{\hat{\rho}} \mathcal{F}_Q(\mathcal{D}_{\epsilon}\circ \Lambda\circ\Phi(\hat{\rho})) &\leq \max_{\hat{\rho}} \int d\varphi P(\varphi)\mathcal{F}_Q(\mathcal{D}_{\epsilon}\circ\Lambda(R(\varphi)\hat{\rho} R^{\dagger}(\varphi))) \\
    & \leq \int d\varphi P(\varphi)\max_{\hat{\rho}}\mathcal{F}_Q(\mathcal{D}_{\epsilon}\circ\Lambda(\hat{\rho})) \\
    & =\max_{\hat{\rho}}\mathcal{F}_Q(\mathcal{D}_{\epsilon}\circ\Lambda(\hat{\rho})).
\end{align}
We know from \cite{latune_quantum_2013} and Subsection~\ref{subsec:noisy_phase_sensitive} that the optimal state for displacement sensing in the presence of loss only is the squeezed vacuum state. Substituting these states in the bound yields
\begin{equation}\label{eq:general_convexity_bound}
    \max_{\hat{\rho}} \mathcal{F}_Q(\mathcal{D}_{\epsilon}\circ \Lambda\circ\Phi(\hat{\rho})) \leq \frac{8\eta \bar{n}+8\eta\sqrt{\bar{n}(1+\bar{n})} +4}{1+4\eta(1-\eta)\bar{n}}.
\end{equation}
Since this relaxation discards the phase-noise contribution, the resulting bound is loose---it reduces to the loss-only squeezed-vacuum QFI and therefore upper-bounds both the best Gaussian state and the optimized states. We use it only as a consistency check on the numerical optimization, and do not claim that the optimized states are globally optimal, only that they realize a quantifiable advantage over the best Gaussian state.

\subsection{The Variance Bound}\label{subsec:variance_bound}

The QFI can also be bounded by above by four times the variance of the generator \cite{toth_lower_2018,toth_extremal_2013}. For simplicity, we will refer to this bound as the variance bound. This inequality is tight when the state $\hat{\rho}$ is pure.
\begin{equation}
    \mathcal{F}_Q(\mathcal{D}_{\epsilon}(\hat{\rho}))\leq 4\left(\Delta \hat{G}\right)^2_{\hat{\rho}}= 8\left(\Delta\hat{X}\right)^2_{\hat{\rho}}.
\end{equation}
Here, $\hat{X}$ is the quadrature operator. We will consider the variance bound applied to the case of state preparation phase noise, but no photon loss.
\begin{equation}
    \frac{1}{8}\mathcal{F}_{Q}(\mathcal{D}_{\epsilon}\circ\Phi(\hat{\rho}))\leq \left(\Delta \hat{X}^2\right)_{\Phi(\hat{\rho})}.
\end{equation}
We compute the variance in the Heisenberg picture, considering the adjoint representation of the phase noise channel.
\begin{align}
    (\Delta\hat{X})^2_{\Phi(\hat{\rho})}&=\textrm{Tr}(\hat{X}^2\Phi(\hat{\rho}))-\textrm{Tr}(\hat{X}\Phi(\hat{\rho}))^2 \\
    &=\textrm{Tr}(\Phi_H(\hat{X}^2)\hat{\rho})-\textrm{Tr}(\Phi_H(\hat{X})\hat{\rho})^2,
\end{align}
where 
\begin{equation}
\Phi_H(\hat{X})=\int_0^{2\pi}d\varphi P(\varphi)R^{\dagger}(\varphi)\hat{X}R(\varphi).
\end{equation}
Further, we use the fact that the quadrature operators transform as follows under conjugation by rotation operators.
\begin{equation*}
    R^{\dagger}(\varphi)\hat{X}R(\theta)=\cos(\varphi)\hat{X}-\sin(\varphi)\hat{P},\quad R^{\dagger}(\varphi)\hat{P}R(\theta)=\sin(\varphi)\hat{X}-\cos(\varphi)\hat{P}.
\end{equation*}
This allows us to compute the Heisenberg picture operator evolutions. We'll assume that the phase noise is Gaussian distributed $\varphi\sim \mathcal{N}(0,\sigma_\phi^2)$.
\begin{align}
    \Phi_H(\hat{X})=\int_0^{2\pi}d\varphi P(\varphi)\left(\cos(\varphi)\hat{X}-\sin(\varphi)\hat{P}\right)=e^{-\sigma_\phi^2/2}\hat{X}.
\end{align}
\begin{align}
    \Phi_H(\hat{X}^2)&=\int_0^{2\pi}d\varphi P(\varphi) R^{\dagger}(\varphi)\hat{X}R(\varphi)R^{\dagger}(\varphi)\hat{X}R(\varphi) \\
    &=\int_0^{2\pi}d\varphi P(\varphi)\left(\cos^2(\varphi)\hat{X}^2+\sin(\varphi)\hat{P}^2 -\cos(\varphi)\sin(\varphi)\{\hat{X},\hat{P}\} \right) \\ 
    &=e^{-\sigma_\phi^2}\cosh(\sigma_\phi^2)\hat{X}^2+e^{-\sigma_\phi^2}\sinh(\sigma_\phi^2)\hat{P}^2.
\end{align}
And so,
\begin{align}
    (\Delta\hat{X})^2_{\Phi(\hat{\rho})}&=e^{-\sigma_\phi^2}\cosh(\sigma_\phi^2)\langle\hat{X}^2\rangle_{\hat{\rho}} + e^{-\sigma_\phi^2}\sinh(\sigma_\phi^2)\langle\hat{P}^2\rangle_{\hat{\rho}} - e^{-\sigma_\phi^2}\langle\hat{X}\rangle^2_{\hat{\rho}} 
\end{align}


Let $G=(\Delta\hat{X})^2_{\Phi(\hat{\rho})}$. We will want to find the state $\hat{\rho}$ that maximizes $G$ to obtain a general upper bound on the QFI in the presence of phase noise, subject to 
\begin{equation}
    1+2\langle \hat n\rangle_{\hat{\rho}}=\langle \hat{X}^2\rangle_{\hat{\rho}}+\langle \hat{P}^2\rangle_{\hat{\rho}}\,,\quad 0\leq \langle\hat{n}\rangle_{\hat{\rho}}\leq N\,,\quad (\Delta \hat{X})^2_{\hat{\rho}}(\Delta\hat{P})^2_{\hat{\rho}}\geq \frac{1}{4}\,,\quad (\Delta\hat{X})^2_{\hat{\rho}},(\Delta\hat{P})^2_{\hat{\rho}}>0.
\end{equation}
Note that the bound is saturated with $\langle \hat n \rangle_{\hat{\rho}} = N$ since we can simply scale $\hat X$ and $\hat P$ to maintain satisfaction of the constraint while increasing $G$.  We also note that the optimal configuration should have $\langle \hat X\rangle_{\hat\rho}^2=0$, since shrinking this and maintaining $\langle\hat X^2\rangle_{\hat\rho}$ constant will increase $G$ and satisfy the other constraints. We then obtain
\begin{equation}
    G= e^{-\sigma_\phi^2}\sinh(\sigma_\phi^2) (1+2N)+ e^{-2\sigma_\phi^2} (\Delta\hat{X}^2)_{\hat{\rho}}.
\end{equation}
This requires $(\Delta\hat{X}^2)_{\hat{\rho}}$ to be as large as possible. We then have, from Heisenberg Uncertainty,
\begin{equation}
    \langle \hat P^2\rangle_{\hat{\rho}}=1+2N-\langle \hat X^2\rangle_{\hat{\rho}}=1+2N-(\Delta\hat{X}^2)_{\hat{\rho}}\ge (\Delta\hat{P}^2)_{\hat{\rho}}\geq \frac{1}{4(\Delta\hat{X}^2)_{\hat{\rho}}}.
\end{equation}
Here equality sign is attained when $\langle \hat P\rangle_{\hat{\rho}}=0$ and the state is a minimum uncertainty state (a Gaussian state). This leads to the following bound on $G$,
\begin{equation}\label{eq:varbound_opt_G}
    G\leq \frac{1}{2}+N +e^{-2\sigma_\phi^2}\sqrt{N(N+1)} \,,
\end{equation}
and \begin{equation}
    \mathcal{F}_Q(\mathcal{D}_{\epsilon}(\hat{\rho}))\leq 4+8N+8e^{-2\sigma_\phi^2}\sqrt{N(N+1)}.
\end{equation}

Recall this variance bound was computed in the presence of phase noise only, and is valid for all states $\hat{\rho}$ such that $\langle\hat{n}\rangle_{\hat{\rho}}\leq N$. We will now add the effects of loss to the bound. First, it can be shown that for a pure loss channel $\Lambda$ with transmission rate $\eta$, we have that
\begin{equation}
    \langle\hat{n}\rangle_{\Lambda(\hat{\rho})}=\eta\langle\hat{n}\rangle_{\hat{\rho}}.
\end{equation}
From this, we obtain the inequality 
\begin{equation}
    \mathcal{F}_Q(\mathcal{D}_{\epsilon}\circ\Lambda\circ\Phi(\hat{\rho}_0))\leq\max_{\hat{\rho}\ : \langle{\hat{n}}\rangle_{\hat{\rho}}\leq \eta N} \mathcal{F}_Q(\mathcal{D}_{\epsilon}\circ\Phi(\hat{\rho})),
\end{equation}
Since $\Lambda(\hat{\rho}_0)$ is a state with average photon number $\eta\bar{n}$, it's QFI is necessarily smaller than or equal to the maximum QFI over all states with the same average photon number. Here we have also implicitly used the commutation of the photon loss and phase noise channels. Further, we have assumed that the displacement occurs after the loss channel, but as seen in Section~\ref{subsec:noisy_phase_sensitive}, the QFI for simultaneous loss and displacement is the same as that of loss followed by a displacement, where the displacement's amplitude is multiplied a constant factor $\beta$. For a fixed photon number, we can then bound the quantity $\mathcal{F}(\Phi(\hat{\rho}))$ of a state subjected to phase noise using the above variance bound.

Putting it all together, we obtain the general bound for a state $\hat{\rho}$ subjected to Gaussian phase noise with variance $\sigma_\phi^2$ followed by a loss channel with transmission rate $\eta$,
\begin{equation}\label{eq:var_bound_final}
    \mathcal{F}_Q(\mathcal{D}_{\epsilon}\circ\Lambda\circ\Phi(\hat{\rho}))\leq 4+8\eta N+8e^{-2\sigma_\phi^2}\sqrt{\eta N(\eta N+1)}.
\end{equation}
Once again, although the channels are applied sequentially here they can be related to the simultaneous displacement and loss setting of the main text using the conversion factor of Equation~\eqref{eq:loss_conversion_factor}.

\subsection{Combined Channel Bound}

The channel QFI bound derived in Subsection~\ref{subsec:noisy_phase_sensitive} can be combined with the results of Subsection~\ref{subsec:variance_bound} yielding a QFI upper bound in the presence of loss and phase noise. This combined bound is obtained once again by following the prescription of  \cite{Escher2011, Demkowicz2012, Kolodynski_2013}. We consider the bound given by Equation~\eqref{eq:cqfi_V_quantity}, but rather than optimize over all pure states, we optimize over states of the form $\Phi(\rho_0)$, where $\Phi$ is the phase noise channel. 
The derivation of Equation~\eqref{eq:cqfi_V_quantity} in \cite{Escher2011} and Subsection~\ref{subsec:noisy_phase_sensitive} required the state be pure. This is because the bound is obtained by optimizing the variance, which is proportional to the QFI for a pure state. The optimization is over operators $H_1$ and $H_2$ (See Equation~\eqref{eq:escher_variance}). For a mixed state, the variance is an upper bound of the quantum Fisher information, as seen in the previous subsection. Optimizing over mixed states rather than pure states therefore yields a looser upper bound. Otherwise, the calculations of Subsection~\ref{subsec:noisy_phase_sensitive} are the same for mixed states.
We compute
\begin{align}
    \max_{\Phi(\rho_0)}\mathcal{F}_Q^p(\rho_0)&=\max_{V_*}\frac{4V_*}{V_*(1-\eta)+\eta} \\
    &= \frac{4\max_{\Phi(\rho_0)}V_*}{\max_{\Phi(\rho_0)}V_*(1-\eta)+\eta},
\end{align}
where $V_*=\textrm{Tr}\left[(\hat{a}+\hat{a}^{\dagger})^2\Phi(\rho_0)\right]$. The second equality is due to the monotonicity of Equation~\eqref{eq:cqfi_V_quantity} in $V$ for $\eta\in [0,1]$. For states with $\langle\hat{a}+\hat{a}^{\dagger} \rangle_{\Phi(\rho_0)}=0$, which is the case here, we have that $V_*\propto \langle\hat{X}^2\rangle_{\Phi(\rho_0)}=(\Delta\hat{X})^2_{\Phi(\rho_0)}$. Therefore, finding the state that maximizes $V_*$ is equivalent to finding the state that maximizes the variance of the $\hat{X}$ quadrature in the presence of phase noise. Such states were found in Section~\ref{subsec:variance_bound}. The expectation value $V_*$ can be obtained by rescaling Equation~\eqref{eq:varbound_opt_G} by a factor of 2, yielding
\begin{equation}
    V_*=1+2\bar{n}+2e^{-2\sigma_\phi^2}\sqrt{\bar{n}(\bar{n}+1)}.
\end{equation}
Substituting this quantity back in, our QFI upper bound becomes
\begin{equation}
    \label{eq:combined_bound}\max_{\Phi(\rho_0)}\mathcal{F}_Q^p(\rho_0)=\frac{4+8\bar{n}+8e^{-2\sigma_\phi^2}\sqrt{\bar{n}(\bar{n}+1)}}{1+2(1-\eta)\bar{n}+2(1-\eta)e^{-2\sigma_\phi^2}\sqrt{\bar{n}(\bar{n}+1)}}.
\end{equation}
For $\eta=1$, this bound is equivalent to the variance bound of Equation~\eqref{eq:var_bound_final} and the Gaussian convexity bound of Equation~\eqref{eq:gaussian_convexity_bound}. 

\begin{figure}
    \centering
    \includegraphics[width=\linewidth]{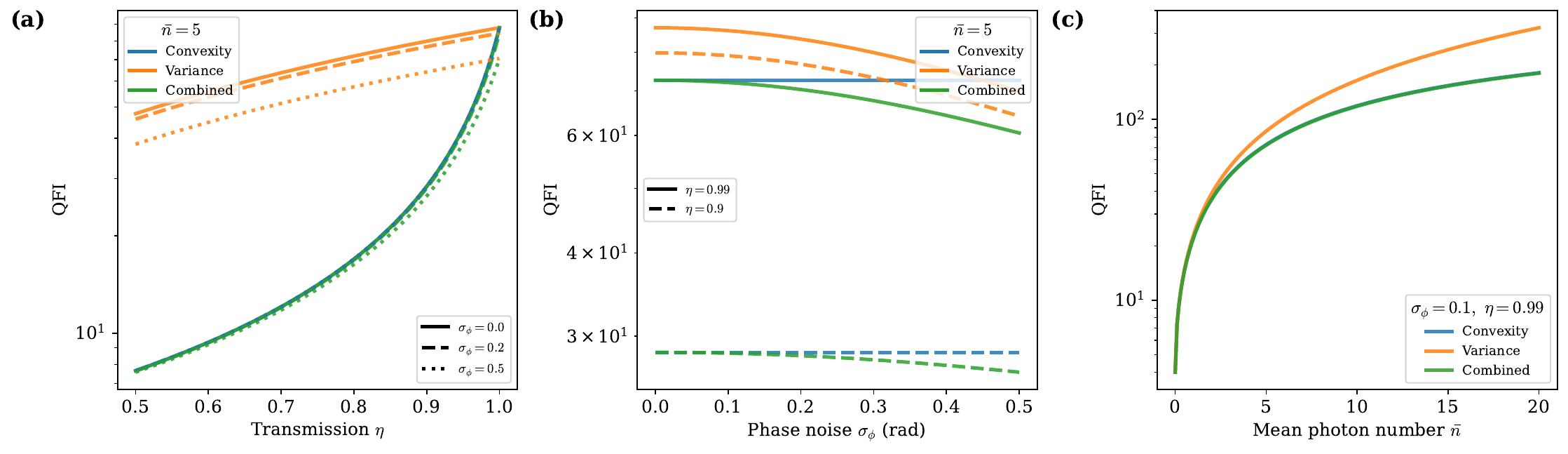}
    \caption{Comparison of the analytical upper bounds for the QFI. We plot the convexity bound (Eq.~\eqref{eq:general_convexity_bound}), variance bound (Eq. ~\eqref{eq:var_bound_final}), and the combined channel bound (Eq. ~\eqref{eq:combined_bound}), as functions of the transmission coefficient $\eta$, phase noise standard deviation $\sigma_\phi$, and mean photon number $\bar n$. For all cases, the combined channel bound outperforms the convexity and variance bounds. a) Performance as a function of $\eta$ for $\sigma_\phi \in [0, 0.2, 0.5]$, $\bar n = 5$. b) Performance as a function of $\sigma_\phi$ for $\eta \in [0.9, 0.99]$, $\bar n = 5$. c) Performance as a function of $\bar n$ for $\eta =0.99$, $\sigma_\phi = 0.1$.}
    \label{fig:bounds_comparison}
\end{figure}

\subsection{Comparison of the Analytical Bounds}

Here, we compare the different analytical upper bounds to the QFI that we computed in the previous sections. We refer to these bounds as the convexity bound (Eq.~\eqref{eq:general_convexity_bound}), variance bound (Eq. ~\eqref{eq:var_bound_final}), and the combined channel bound (Eq. ~\eqref{eq:combined_bound}). We plot the comparison in Fig. \ref{fig:bounds_comparison}. First, in Fig. \ref{fig:bounds_comparison}a, we fix the mean photon number as $\bar n = 5$, and plot the bounds as a function of the transmission coefficient $\eta$ for a phase noise standard deviation $\sigma_\phi \in [0, 0.2, 0.5]$. For all of these cases, the convexity and combined channel bounds seem to be tighter than the variance bound, which scales approximately linearly with $\eta$. Next, in Fig. \ref{fig:bounds_comparison}b, we again fix the mean photon number as $\bar n=5$, and plot the bounds as a function of $\sigma_\phi$ for $\eta \in [0.9, 0.99]$. We observe that the combined channel bound performs the best (in terms of tightness) in this regime. Finally, we fix $\sigma_\phi=0.1$, $\eta = 0.99$, and plot the QFI as a function of the mean photon number $\bar n$. The convexity and combined channel bounds perform very similarly (as they converge to the same limit for $\eta \rightarrow 1$), and outperform the variance bound. 

\subsection{Iterative Algorithm for Global QFI Optimization}


Here, we outline the iterative algorithm for the global optimization of the QFI of Refs. \cite{macieszczak2013_qfi, Macieszczak2014}. This algorithm can be used to refine the QFI optimization, after an initial coarse global numerical search. For this purpose, let us assume that the parameter $\epsilon$ is encoded to an initial state $\hat{\rho}$ in the following way: $\hat{\rho}_\epsilon = e^{-i\epsilon \hat{G}}\Lambda(\hat{\rho}) e^{i\epsilon \hat{G}}$. Here, $\hat{G}$ is the generator belonging to the $L^2(\mathcal{H})$ space of Hermitian Hilbert–Schmidt operators in $\mathcal{H}$. $\Lambda$ represents any decoherence process after the initial state preparation, and before the parameter encoding. 
The QFI of a state $\hat{\rho}_\epsilon$ can be found from
\begin{align}
    \mathcal{F}_Q = \text{Tr}[\hat{\rho}_\epsilon \hat{L}^2], \quad \frac{1}{2}\{\hat{L}, \hat{\rho}_\epsilon\} = \frac{\partial \hat{\rho}_\epsilon}{\partial \epsilon}
\end{align}
where $\hat{L}$ is the symmetric logarithmic derivative, also defined in Eq. (\ref{eq:symm_log_derivative}). For the parameter encoding given above, $\hat{L}$ satisfies the equation $\frac{1}{2}\{\hat{L}, \hat{\rho}\} = -i[\hat{G}, \hat{\rho}]$.
The basis of the algorithm is the following variational expression of the QFI:
\begin{equation}
\mathcal{F}_Q^\text{max} = \sup_{X\in L^2(\mathcal{H})}\sup_{|\psi\rangle}\,\langle\psi|\, \Lambda^\dagger\left( -\hat{X}^2+2i[\hat{G},\hat{X}] \right)   |\psi\rangle,\label{eq:supQF}
\end{equation}
where $\ket{\psi}$ is a normalized vector in $\mathcal{H}$.
$\mathcal{F}_Q^\text{max}$ is thus the supremum of the QFI over all input states. To see this, consider that the supremum with respect to the operator $\hat{X}$ results in $\frac{1}{2}\{\hat{X}, \Lambda(\ket{\psi}\bra{\psi})\} = -i[\hat{G}, \Lambda(\ket{\psi}\bra{\psi})]$, which is the equation for the symmetric logarithmic derivative $\hat{L}$. Then, for a given $\ket{\psi}$, the supremum over $\hat{X}$ gives the QFI for $\ket{\psi}\bra{\psi}$. The convexity of the QFI guarantees that the maximal QFI will be achieved for a pure input state, thus, the second supremum over $\ket{\psi}$ globally maximizes the QFI. 

This result motivates the following iterative algorithm: let us start with an input state $\ket{\psi_0}$, $\hat{\rho}_0 = \Lambda(\ket{\psi_0}\bra{\psi_0})$. We can compute the symmetric logarithmic derivative for this state as $\hat{L}_0$. Then, we choose the next input state $\ket{\psi_1}$ for the algorithm as the eigenvector corresponding to the maximum eigenvalue of $\Lambda^\dagger(\hat{O}(\hat{L}_0))$, $\hat{O}(\hat{X}) \coloneqq -\hat{X}^2 + 2i [\hat{G}, \hat{X}]$. We repeat this step until the algorithm converges to a QFI. 
Ref. \cite{macieszczak2013_qfi} proves that the algorithm achieves a global maximum for a finite dimensional $\mathcal{H}$ and when at each step, the eigenvector with the maximal eigenvalue, as well as its symmetric logarithmic derivative, is unique. The uniqueness of the symmetric logarithmic derivative is achieved when $\hat{\rho} = \Lambda(\ket{\psi}\bra{\psi})$ is full-rank, whereas the uniqueness of the maximum eigenvalue of $\hat{L}$ depends on the symmetries of the problem that might introduce degeneracies.

For displacement sensing with bosonic states, the Hilbert space $\mathcal{H}$ is infinite-dimensional. However, one can truncate the space in e.g. the Fock basis, by defining the basis as $\{\ket{n}\}$, $0 \leq n\leq N_\text{max}$. We can choose $N_\text{max}$ such that $N_\text{max} \gg \bar n$, where $\bar n$ is the mean photon number $\bar n = \bra{\psi} \hat{a}^\dagger \hat{a} \ket{\psi}$. Furthermore, we want to introduce a photon number constraint in the QFI maximization in the form of $\bra{\psi} \hat{a}^\dagger \hat{a} \ket{\psi} \leq \bar n$. We achieve this by modifying Eq. (\ref{eq:supQF}) as
\begin{equation}
\mathcal{F}_Q^\text{max} = \inf_{\mu \geq 0}\sup_{X\in L^2(\mathcal{H})}\sup_{|\psi\rangle}\,\langle\psi|\, \Lambda^\dagger\left( -\hat{X}^2+2i[\hat{G},\hat{X}] \right) - \mu\hat{a}^\dagger \hat{a}  |\psi\rangle +\mu \bar n.
\label{eq:supQF_constrained}
\end{equation}
Note that this modification does not change the supremum with respect to the operator $\hat{X}$ for a given $\ket{\psi}$: the result of the optimization $\hat{X}$ is still the symmetric logarithmic derivative. For the optimization over $\ket{\psi}$, one needs to find the eigenvector of $\Lambda^\dagger\left( -\hat{X}^2+2i[\hat{G},\hat{X}] \right) - \mu\hat{a}^\dagger \hat{a}$ that corresponds to the maximum eigenvalue. Lastly, for the optimization over $\mu$, we need to run the algorithm for increasing values of $\mu$, starting from zero.

\clearpage
\section{Optimization details \label{sup:opt_details}}

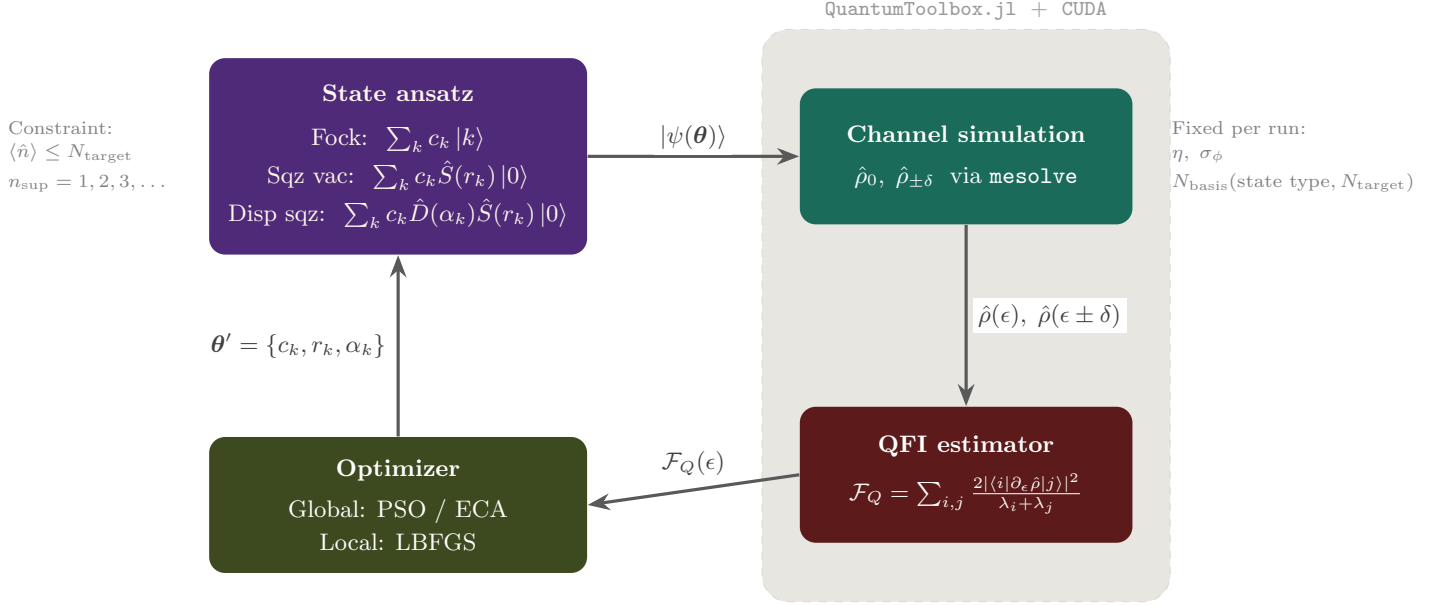
\begin{figure*}[htbp]
    \centering
 
\begin{tikzpicture}[
    >=Stealth,
    box/.style={
        draw=none, rounded corners=6pt, minimum height=1.8cm,
        minimum width=4.4cm, align=center, text=white
    },
    ansatz/.style={box, fill=cPurple, minimum width=5.0cm, minimum height=2.6cm, text=cCream},
    channel/.style={box, fill=cTeal},
    qfi/.style={box, fill=cMaroon},
    optim/.style={box, fill=cOlive, minimum width=5.0cm},
    arr/.style={->, line width=1.2pt, >=Stealth, color=black!65},
    lbl/.style={font=\small, fill=white, inner sep=2pt, text=black!75},
    note/.style={font=\scriptsize, text=black!50, align=left},
]
 
\node[ansatz] (ansatz) {
    {\textbf{State ansatz}}\\[6pt]
    {\small Fock: $\;\sum_k c_k \ket{k}$}\\[2pt]
    {\small Sqz vac: $\;\sum_k c_k \hat{S}(r_k)\ket{0}$}\\[2pt]
    {\small Disp sqz: $\;\sum_k c_k \hat{D}(\alpha_k)\hat{S}(r_k)\ket{0}$}
};
 
\node[channel, right=2.8cm of ansatz] (channel) {
    \textbf{Channel simulation}\\[5pt]
    {\small $\hat{\rho}_0,\; \hat{\rho}_{\pm\delta}$\; via \texttt{mesolve}}
};
 
\node[qfi, below=2.4cm of channel] (qfi) {
    \textbf{QFI estimator}\\[5pt]
    {\small $\mathcal{F}_Q = \sum_{i,j}\frac{2|\langle i|\partial_\epsilon\hat{\rho}|j\rangle|^2}{\lambda_i + \lambda_j}$}
};
 
\node[optim, below=2.4cm of ansatz] (optim) {
    \textbf{Optimizer}\\[5pt]
    {\small Global: PSO / ECA}\\[2pt]
    {\small Local: LBFGS}
};
 
\draw[arr] (ansatz.east) -- (channel.west)
    node[lbl, midway, above] {$\ket{\psi(\boldsymbol{\theta})}$};
 
\draw[arr] (channel.south) -- (qfi.north)
    node[lbl, midway, right, xshift=2pt] {$\hat{\rho}(\epsilon),\;\hat{\rho}(\epsilon\pm\delta)$};
 
\draw[arr] (qfi.west) -- (optim.east)
    node[lbl, midway, above, yshift=3pt] {$\mathcal{F}_Q(\epsilon)$};
 
\draw[arr] (optim.north) -- (ansatz.south)
    node[lbl, midway, left, xshift=-2pt] {$\boldsymbol{\theta}' = \{c_k, r_k, \alpha_k\}$};
 
\node[note, anchor=east] at ([xshift=-0.4cm]ansatz.west) {
    Constraint:\\[1pt]
    $\langle \hat{n} \rangle \leq N_\mathrm{target}$\\[3pt]
    $n_\mathrm{sup} = 1,2,3,\dots$
};
 
\node[note, anchor=west] at ([xshift=0.4cm]channel.east) {
    Fixed per run:\\[1pt]
    $\eta,\;\sigma_\phi$\\[3pt]
    $N_\mathrm{basis}(\text{state type}, N_\mathrm{target})$
};
 
\begin{scope}[on background layer]
    \node[draw=black!15, dashed, rounded corners=10pt,
          fill=cSimBg,
          fit=(channel)(qfi),
          inner xsep=14pt, inner ysep=22pt] (simbox) {};
\end{scope}
\node[font=\footnotesize\sffamily, text=black!40, anchor=south]
    at (simbox.north) {\texttt{QuantumToolbox.jl}\; +\; \texttt{CUDA}};
 
\end{tikzpicture}
    \caption{ Optimization schematic. The states obtained from the channel evolution are used to calculate the QFI, which is then fed into the optimizer to update the state parameters.}
    \label{fig:optimization}
\end{figure*}

We maximize the Quantum Fisher Information (QFI) for single-mode displacement sensing under loss and phase noise. The optimization uses a two-phase hybrid strategy: a global evolutionary search followed by gradient-based local refinement, with warm-starting from previously computed solutions at nearby parameter values.

\subsection{State parametrization}

The three state ansatze are defined in the main text. Here we specify the parameter bounds used during optimization. Each state is expressed as a normalized superposition of $n_\mathrm{sup}$ components with complex coefficients $c_k = c_k^\mathrm{re} + i\,c_k^\mathrm{im}$, where $c_k^\mathrm{re}, c_k^\mathrm{im} \in [-1, 1]$. The coefficient vector is normalized after unpacking.

\begin{itemize}
    \item \textbf{Fock superpositions:} 3 real parameters per component, which are the complex amplitude $c_k$ and the Fock index $k$. The Fock index $k \in [0, N_\mathrm{basis}-1]$ is treated as continuous during the global search and rounded to the nearest integer for local refinement.
    \item \textbf{Squeezed-vacuum superpositions:} 4 real parameters per component, which are the complex amplitude $c_k$ and squeezing parameter $r_k$. The complex squeezing parameter $r_k = r_k^\mathrm{re} + i\,r_k^\mathrm{im}$ with $r_k^\mathrm{re}, r_k^\mathrm{im} \in [-r_\mathrm{max}, r_\mathrm{max}]$, where $r_\mathrm{max} = \max(\mathrm{asinh}(2N_\mathrm{target}),\; 2)$.
    \item \textbf{Displaced squeezed-vacuum superpositions:} 6 real parameters per component, which are the complex amplitude $c_k$, squeezing parameter $r_k$, and displacement $\alpha_k$. Displacement $\alpha_k = \alpha_k^\mathrm{re} + i\,\alpha_k^\mathrm{im}$ with $\alpha_k^\mathrm{re}, \alpha_k^\mathrm{im} \in [-\sqrt{N_\mathrm{target}}, \sqrt{N_\mathrm{target}}]$, and squeezing $r_k$ bounded as above.
\end{itemize}

\subsection{Quantum channel \label{sup:quantum_channel}}

The probe state $\rho_0$ undergoes single-mode Lindblad evolution in three sequential stages:

\begin{enumerate}
    \item \textbf{Input noise} (duration $t=1$): Photon loss at rate $\kappa_\mathrm{in} = -\ln(\eta_\mathrm{in})$ via collapse operator $\sqrt{\kappa_\mathrm{in}}\,\hat{a}$, and phase diffusion at rate $\chi_\mathrm{in} = \sigma_{\phi,\mathrm{in}}^2$ via $\sqrt{\chi_\mathrm{in}}\,\hat{N}$. If both $\eta_\mathrm{in}=1$ and $\sigma_{\phi,\mathrm{in}}=0$, this stage is skipped.
    \item \textbf{Sensing channel} (duration $t_\mathrm{final}=1$): Hamiltonian
    \begin{align}
        H = \Delta\,\hat{N} + \epsilon_p(\hat{a}^\dagger + \hat{a}) + i\epsilon_a(\hat{a}^\dagger - \hat{a}),
    \end{align}
    with channel loss $\kappa_\mathrm{ch} = -\ln( \eta_\mathrm{ch})$ and channel phase noise $\chi_\mathrm{ch} = \sigma_{\phi,\mathrm{ch}}^2$.
    \item \textbf{Output noise} (duration $t=1$): Same form as input noise with parameters $\eta_\mathrm{out}$, $\sigma_{\phi,\mathrm{out}}$.
\end{enumerate}

Each stage is solved via \texttt{mesolve()} in \texttt{QuantumToolbox.jl}, with the output density matrix of one stage serving as the input to the next. The framework supports independent loss and phase noise parameters at each of the three stages (input, channel, and output), allowing flexible modeling of where decoherence occurs. All results presented in this work use channel-stage loss and input-stage phase noise only ($\eta_\mathrm{ch} < 1$, $\sigma_{\phi,\mathrm{in}} > 0$), with all other noise parameters set to their ideal values. Other configurations (e.g., output detection loss, channel phase noise) are available for future studies.

\subsection{QFI computation}

The QFI is computed via the symmetric logarithmic derivative (SLD). The sensed parameter $\theta$ can in principle be any channel parameter ($\Delta$, $\epsilon_a$, $\epsilon_p$, $\eta$, $\sigma_\phi$, etc.); in this work we use $\theta = \epsilon_p$ (phase-quadrature displacement). Given the output density matrix $\rho(\theta)$:
\begin{enumerate}
    \item The derivative $\partial\rho/\partial\theta$ is computed by central finite differences with step size $h = 10^{-5}$.
    \item The density matrix without displacement is eigendecomposed: $\rho = \sum_i \lambda_i |u_i\rangle\langle u_i|$.
    \item The derivative is rotated into the eigenbasis: $(\partial\rho/\partial\theta)_\mathrm{rot} = U^\dagger (\partial\rho/\partial\theta)\, U$.
    \item The QFI is evaluated as
    \begin{align}
        \mathcal{F}_Q = \sum_{\substack{i,j \\ \lambda_i+\lambda_j > 10^{-8}}} \frac{2\,\left|(\partial\rho/\partial\theta)^\mathrm{rot}_{ij}\right|^2}{\lambda_i + \lambda_j}\,.
    \end{align}
\end{enumerate}

\subsection{Objective function}

The optimizer minimizes
\begin{align}
    f(\mathbf{x}) = -\mathcal{F}_Q[\psi(\mathbf{x})] + w_n \max\!\big(0,\; \langle\hat{N}\rangle - N_\mathrm{target}\big)^2 + w_b \sum_i \left[\max(0,\; x_i - x_i^\mathrm{ub})^2 + \max(0,\; x_i^\mathrm{lb} - x_i)^2\right],
\end{align}
where $w_n = 10^6$ penalizes mean photon number exceeding the target $N_\mathrm{target}$, and $w_b = 10^4$ penalizes parameter bound violations for the local optimization stage when the bounds are not explicitly set. $w_b = 0$ for the global optimization stage. Evaluations returning NaN, $\pm\infty$, or arising from zero-norm states are assigned a cost of $10^{10}$.

Although the energy constraint is primarily used to eliminate truncation errors because of the finite basis size used in the numerical simulation, we can also view this as an energy constraint in state preparation. For a large enough basis size where a numerically accurate squeezed state can be constructed that saturates the loss limit, we can remove the $N_{target}$ constraint and find optimal states in the unconstrained case (not presented in this work).

\subsection{Two-phase hybrid optimization}

\textbf{Phase~1 --- Global search (ECA).}
We use the Evolutionary Centers Algorithm (ECA) from \texttt{Metaheuristics.jl} with population size $20\,n_\mathrm{sup}$ and $500\,n_\mathrm{sup}$ iterations (where $n_\mathrm{sup}$ is the number of superposition components), capped at $2\times10^5$ function evaluations. When a warm-start guess is available, it seeds the initial population.

\textbf{Phase~2 --- Local refinement (L-BFGS).}
The best solution from Phase~1 is refined using L-BFGS (\texttt{Optim.jl}). For Fock superpositions, the Fock indices are rounded to integers and held fixed; only the complex coefficients are optimized (200 iterations). For squeezed-state families, all continuous parameters are refined ($100\,n_\mathrm{sup}$ maximum iterations). Convergence tolerances are $f_\mathrm{reltol} = 5\times10^{-3}$ and $g_\mathrm{tol} = 5\times10^{-3}$. 

\subsection{Warm-starting}

To efficiently sweep over the physical parameters ($N_\mathrm{target}$, $\eta$, $\sigma_\phi$, $n_\mathrm{sup}$), previously optimized results are reused as initial guesses with the following priority:
\begin{enumerate}
    \item \textbf{Exact match:} same $(\eta,\, \sigma_\phi,\, n_\mathrm{sup})$, closest $N_\mathrm{target}$.
    \item \textbf{Padded:} result at $n_\mathrm{sup}{-}1$ components, zero-padded with an additional component.
    \item \textbf{Cross-type:} parameters from a simpler ansatz (e.g., squeezed-vacuum) embedded into a richer one (e.g., displaced squeezed-vacuum) by inserting zeros for the missing degrees of freedom.
    \item \textbf{Nearby parameters:} $|\Delta\eta| < 0.05$, $|\Delta\sigma_\phi| < 0.05$, same $n_\mathrm{sup}$.
    \item \textbf{None:} full global search from scratch.
\end{enumerate}
Results are saved only when the new QFI exceeds the previously cached value.

\begin{figure}[ht]
    \centering
    \includegraphics[width=0.99\linewidth]{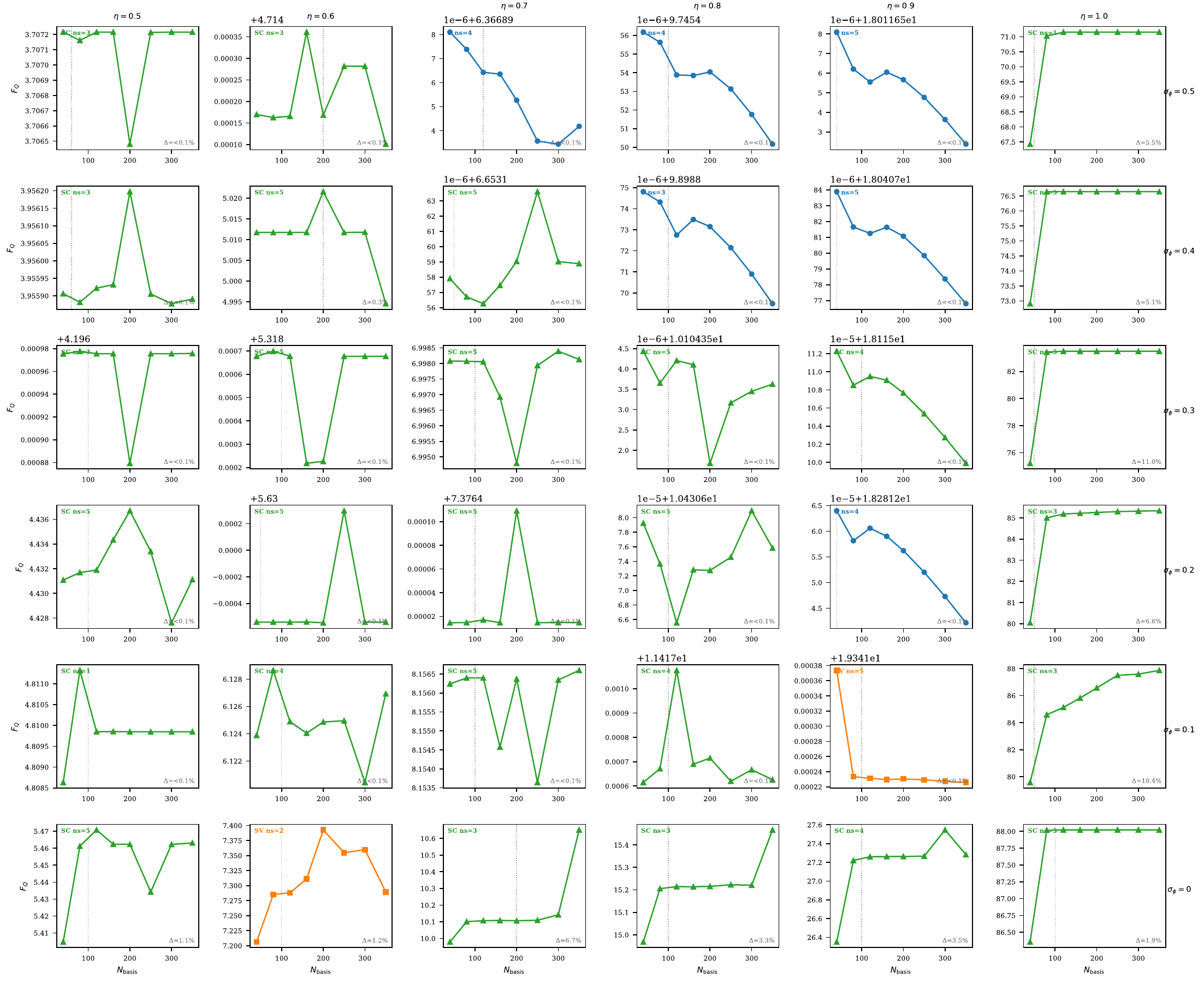}
    \caption{QFI convergence with Fock basis size for the optimized states at $\bar{N}_\mathrm{target}=5$. Each panel shows a different $(\eta,\,\sigma_\phi)$ point from the non-Gaussian advantage grid of Fig. 2 of the main text.}
    \label{fig:basis_convergence_nga5}
\end{figure}

\subsection{Ballpark Estimate of Truncation Error}

In this subsection, we'll compute a back-of-the-envelope estimate for the truncation errors in the numerical results. The truncation error highly depends on the state we are trying to represent. For this reason, it is hard to characterize. We can, however, get a ballpark estimate for the truncation error by studying how it affects the fidelity of squeezed vacuum states. The numerical results show that this class of states is optimal for the sensing task in the low loss and low-phase noise regime. In the absence of noise, they are optimal. We choose to study this state because these regimes are particularly conducive to Fock space truncation errors, as the loss channel reduces the support of a quantum state on higher photon number basis states, which are eliminated by the truncation.
We'll consider a Fock space truncation of $N_{\textrm{basis}}$ and an average photon number of $\bar{n}=5$ to be consistent with the upper-limits of the numerics. The squeezed-vacuum state is given by
\begin{equation}
    \ket{\psi}=\frac{1}{\sqrt{\cosh(r)}}\sum_{n=0}^{\infty}\tanh^2(r)\frac{\sqrt{(2n)!}}{2^nn!}\ket{2n},
\end{equation}
and the truncated-squeezed vacuum state is given by
\begin{equation}
    \ket{\psi_{\textrm{trunc.}}}=\frac{1}{\sqrt{Z}}\frac{1}{\sqrt{\cosh(r)}}\sum_{n=0}^{\lfloor\frac{N}{2}\rfloor}\tanh^n(r)\frac{\sqrt{(2n)!}}{2^nn!}\ket{2n},\label{eq:trunc_vac_state}
\end{equation}
where $Z$ is the norm of the truncated vector
\begin{equation}
    Z=\frac{1}{\cosh(r)}\sum_{n=0}^{\lfloor \frac{N_{\textrm{basis}}}{2}\rfloor }\tanh^{2n}(r)\frac{(2n)!}{4^n(n!)^2}.
\end{equation}
The state vector is re-normalized after truncation so that we are always working with a quantum state.
The fidelity between these two states is given by 
\begin{align}
    F\left(\ket{\psi},\ket{\psi_{\textrm{trunc.}}}\right)&=\frac{1}{Z}\left|\braket{\psi\, |\,\psi_{\textrm{trunc.}}}\right|^2 \\
    &=\frac{1}{Z}\left[\frac{1}{\cosh(r)}\sum_{n=0}^{\lfloor \frac{N_{\textrm{basis}}}{2}\rfloor }\tanh^{2n}(r)\frac{(2n)!}{4^n(n!)^2}\right]^2 \\
    &=\frac{1}{\sqrt{\cosh(r)}}\sum_{n=0}^{\lfloor\frac{N}{2}\rfloor}\tanh^n(r)\frac{\sqrt{(2n)!}}{2^nn!}\ket{2n} \\
    &=\frac{1}{\sqrt{1+\bar{n}}}\sum_{n=0}^{\lfloor\frac{N_{\textrm{basis}}}{2}\rfloor} \frac{\bar{n}^n}{(1+\bar{n})^n}\frac{(2n)!}{4^n(n!)^2}. \label{eq:trunc_fidelity}
\end{align}

\begin{figure*}[ht]
    \centering
    \begin{subfigure}{0.5\linewidth}
        \includegraphics[width=\linewidth]{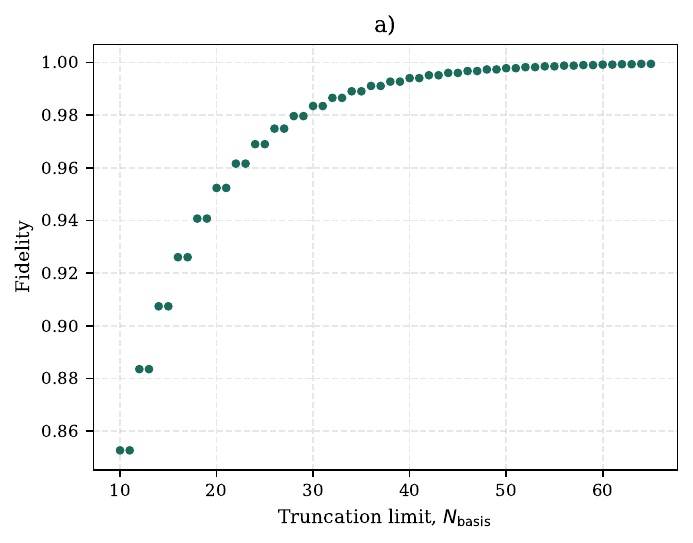}
        \phantomcaption
    \end{subfigure}\hfill
    \begin{subfigure}{0.5\linewidth}
        \includegraphics[width=\linewidth]{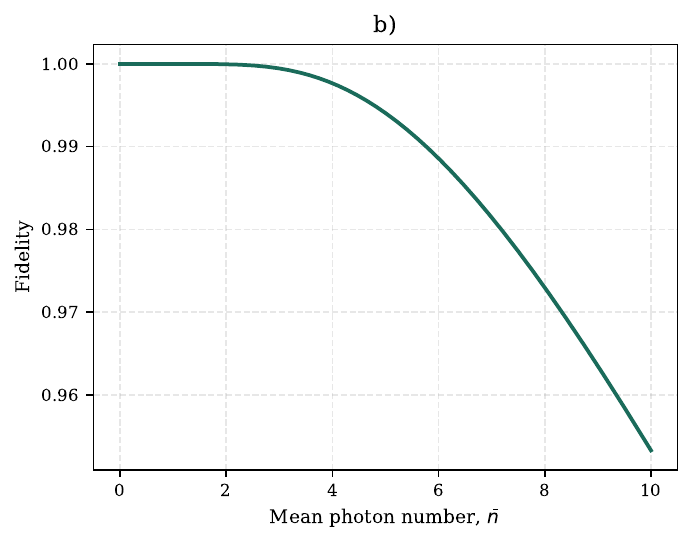}
        \phantomcaption
    \end{subfigure}\hfill
    \caption{Fidelity between a squeezed vacuum state and the same state with Fock basis truncation, as given by Equation~\eqref{eq:trunc_fidelity}. a) The mean photon number is fixed to $\bar{n}=5$, and the truncation value $N_{\textrm{basis}}$ is varied. b) The truncation value is fixed $N_{\textrm{basis}}$ is fixed, and the mean photon number $\bar{n}$ is varied.}
    \label{fig:trunc_fidelity}
\end{figure*}

Figure~\ref{fig:trunc_fidelity} shows the fidelity as a function of both $\bar{n}$ and $N_\textrm{basis}$ for the squeezed vacuum state. As expected, increasing the truncation limit increases the fidelity, and increasing the mean photon number decreases the fidelity. At $N_\textrm{basis}=40$ and $\bar{n}=5$, the fidelity of the state is $\approx 99.41$. Numerical truncation errors behave quite differently to simply having a different state that is off by some infidelity. Indeed, the unitary operators and jump operators are also truncated, which can lead to further errors. However, this formula let's us get an estimate of the order of magnitude of the errors, and is consistent with the relative error around 1\% seen in the results of the main text. 
Further, superpositions of squeezed vacuum can also be sensitive to truncation error. This is because states with discrete rotational symmetries can have quite sparse Fock basis support \cite{gutman_squeezed-vacuum_2026}. So errors of this order of magnitude may persist among the other families, even if the loss rate is higher. 

\subsection{Basis convergence}

 All quantum states are represented in a truncated Fock basis of dimension $N_\mathrm{basis}$ during the optimization process. 

 For optimization,
  $N_\mathrm{basis}$ is chosen to be large enough to faithfully represent the state at the target mean photon number $\bar{N}$ while keeping the computational cost manageable. The required basis size depends on both $\bar{N}$ and the state family: Fock-sparse superpositions  require smaller bases (e.g., $N_\mathrm{basis} = 40$ for $\bar{N} \leq 5$, scaling to 250 for $\bar{N} = 80$), while squeezed and
  squeezed-coherent states have broader Fock-space tails and require larger bases at the same $\bar{N}$ (e.g., $N_\mathrm{basis} = 100$ for $\bar{N}
   \leq 5$, scaling to 250 for $\bar{N} \leq 80$). These values are set conservatively for each sweep block, and correctness is verified post-hoc
  by the basis convergence procedure described below. Since the state is parameterized independently of the basis, $N_\mathrm{basis}$ can be increased after optimization without re-running the optimization.

QFI values are verified for Hilbert-space truncation convergence. Starting from $N_\mathrm{basis} = \max(N_\mathrm{basis}^\mathrm{opt},\, 40)$, the QFI is recomputed at progressively larger bases following the sequence $N_\mathrm{next} = \lceil 1.5 \times N_\mathrm{current} \rceil$ up to $N_\mathrm{max} = 500$. Convergence is declared when the relative change satisfies $|F_Q^{(i)} - F_Q^{(i-1)}|/|F_Q^{(i-1)}| < 1\%$. If the QFI becomes negative at any basis size (which happens in rare cases when the basis is very large and, the matrix being near singular, the eigendecomposition becomes inaccurate), the last valid positive value is retained. All QFI values reported in the paper are basis-converged according to this procedure. The Wigner functions are plotted at $N_\mathrm{basis} = 200$.

\begin{figure}[ht]
    \centering
    \includegraphics[width=0.99\linewidth]{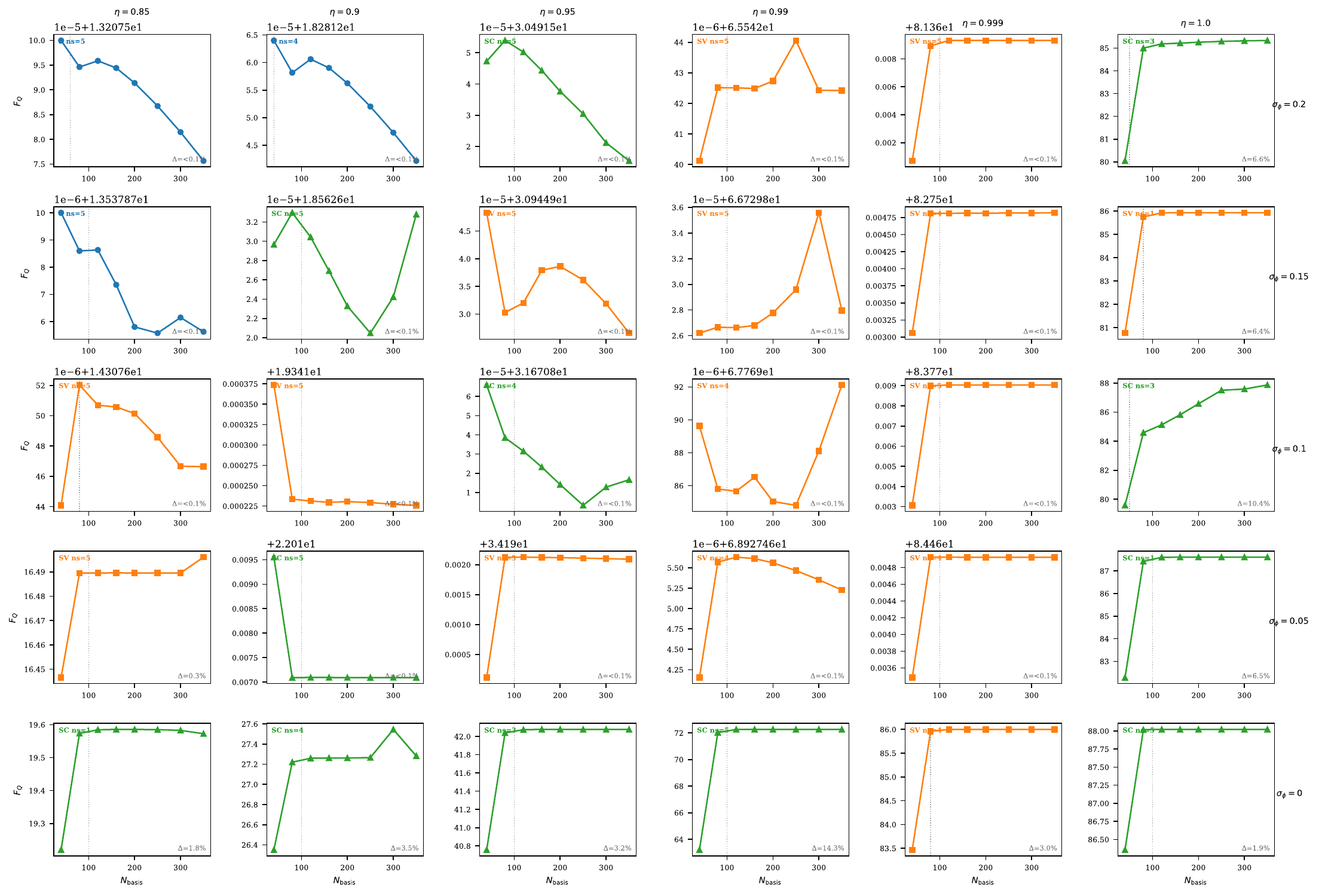}
    \caption{QFI convergence with Fock basis size for the optimized states at $\bar{N}_\mathrm{target}=5$ in the low-loss, low-phase-noise regime. Each panel shows a different $(\eta,\,\sigma_\phi)$ point from the non-Gaussian advantage grid of Fig. 2 in the main text.}
    \label{fig:basis_convergence_nga5_ll}
\end{figure}

Figures~\ref{fig:basis_convergence_nga5} and~\ref{fig:basis_convergence_nga5_ll} show this convergence for
representative $(\eta,\sigma_\phi)$ points from the non-Gaussian advantage grid of the
main text (Fig. 2), for the coarse grid and the low-loss, low-phase-noise regime respectively.
In every case the QFI settles well before $N_{\max}$, confirming that the reported
advantages are basis-converged and not numerical artifacts.

\subsection{Software}

The optimization pipeline is implemented in Julia using \texttt{QuantumToolbox.jl} for quantum state construction and Lindblad evolution, \texttt{Metaheuristics.jl} for ECA, and \texttt{Optim.jl} for L-BFGS. GPU-accelerated variants using \texttt{CUDA.jl} are employed for the master equation dynamics and QFI computation when the basis size exceeds $\sim100$ for a single mode or $\sim20$ per mode for the two-mode case. The optimization algorithm implemented on the CPU. The code is modular and can be adapted to other optimization algorithms, quantum channels, and sensing tasks.

\section{Non-Gaussian Advantage Calculation \label{sup:ng_adv_calc}}
The advantage grid in Fig.~\ref{fig:ng_adv} of the main paper is calculated with reference to the \textit{best} Gaussian state under $N_{target}$ photons, which includes displaced squeezed states with $\bar{N} \leq N_{target}$. In the presence of both large loss and phase noise, a displacement is advantageous. Therefore, the baseline chosen to compute non-gaussian advantage is the result of the following optimization procedure:
\begin{enumerate}
    \item A single displaced squeezed-vacuum state $|\alpha, r\rangle = \hat{D}(\alpha)\hat{S}(r)|0\rangle$ is parameterized by the displacement $\alpha \in \mathbb{C}$ and squeezing $r \in \mathbb{C}$ (i.e., $n_\mathrm{sup}=1$ in the displaced squeezed-vacuum ansatz of Sec.~\ref{sup:opt_details}).
    \item For each grid point $(\eta, \sigma_\phi)$, the two-phase hybrid optimization (ECA global search followed by L-BFGS local refinement) is run to maximize $\mathcal{F}_Q$ subject to the photon number constraint $\langle \hat{N} \rangle \leq N_\mathrm{target}$, using the penalized objective function of Sec.~\ref{sup:opt_details}. We use the same optimization procedure for consistency despite knowing that a global search is not necessary for this simple ansatz.
    \item Warm-starting from the best squeezed vacuum state is used as described in Sec.~\ref{sup:opt_details}, so that the optimizer can efficiently explore the trade-off between displacement and squeezing at each noise configuration.
\end{enumerate}

This procedure yields a potentially different Gaussian state at each $(\eta, \sigma_\phi)$ point, reflecting how the optimal balance between displacement and squeezing depends on the specific noise environment.

\subsection{$\bar{N}=5$ advantage grid details}
\label{subsec:grid_details}

\begin{figure}[ht]
    \centering
    \includegraphics[width=0.85\linewidth]{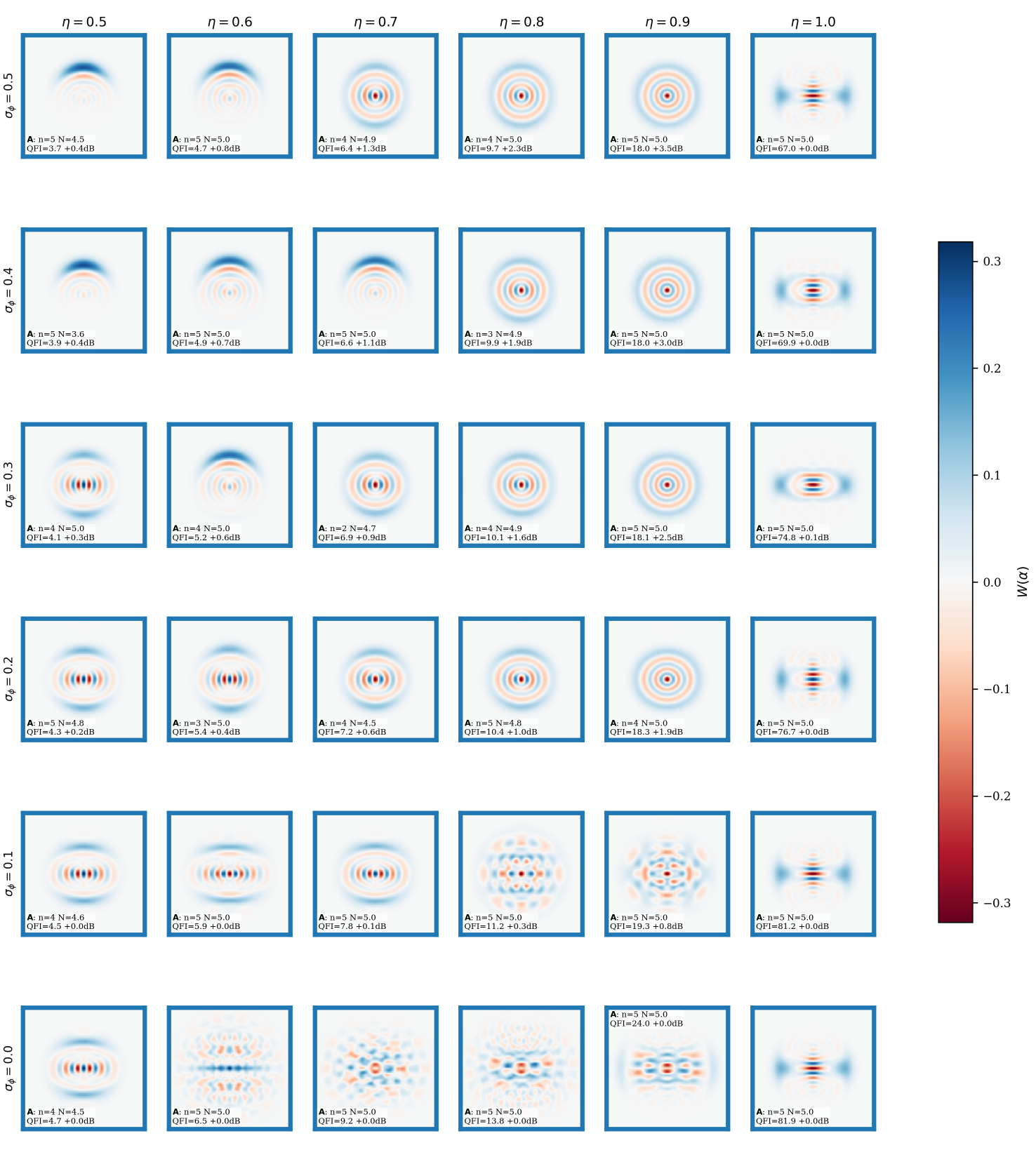}
    \caption{Wigner functions of the optimized Fock-superposition states across the $(\eta,\,\sigma_\phi)$ grid at $\bar{N}=5$. Each panel shows the best state (over all $n_\mathrm{sup}$) at that grid point.}
    \label{fig:best_fock_N5}
\end{figure}
\begin{figure}[ht]
    \centering
    \includegraphics[width=0.85\linewidth]{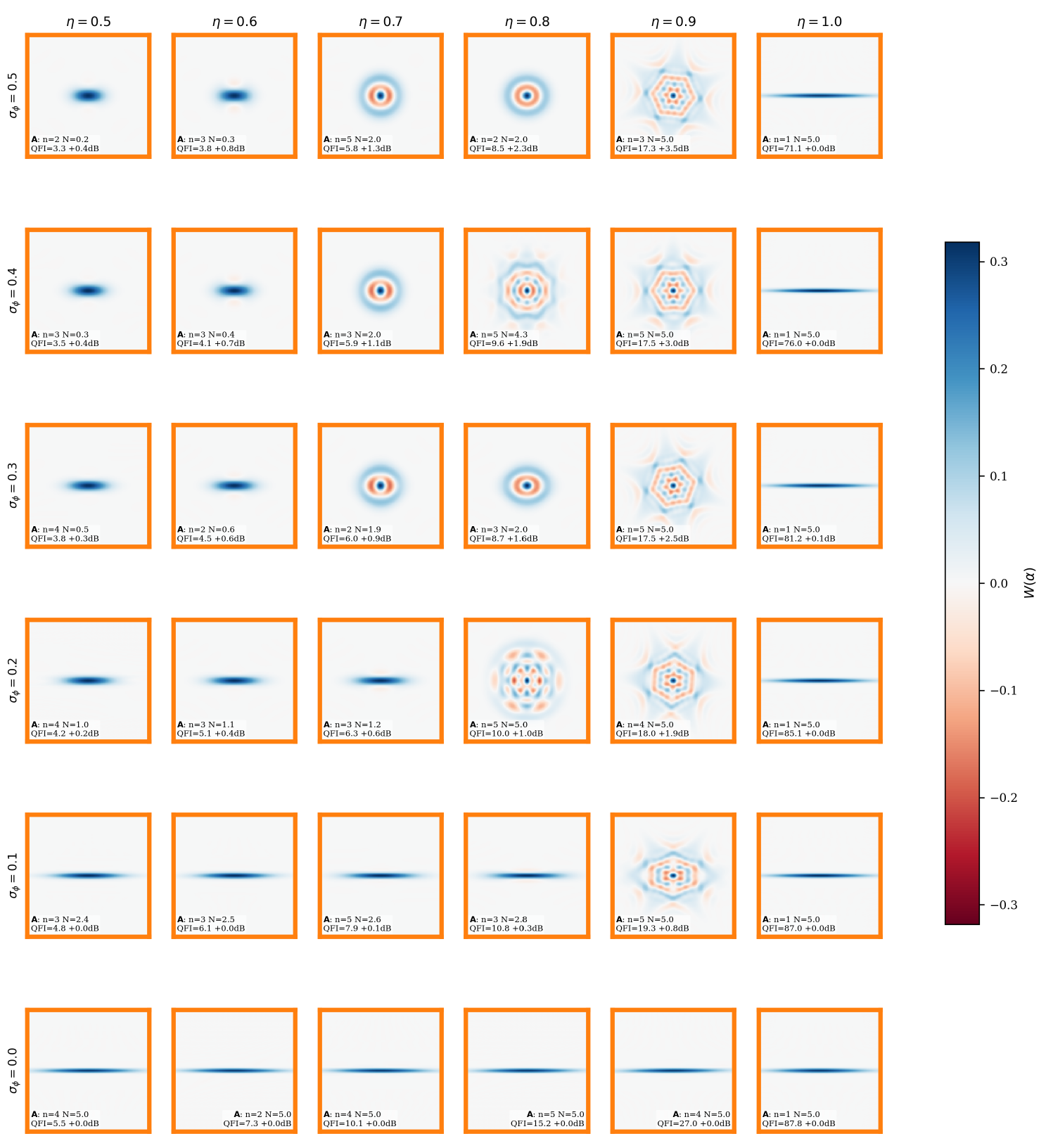}
    \caption{Wigner functions of the optimized squeezed-vacuum superposition states across the $(\eta,\,\sigma_\phi)$ grid at $\bar{N}=5$. Layout as in Fig.~\ref{fig:best_fock_N5}.}
    \label{fig:best_sqzvac_N5}
\end{figure}
\begin{figure}[ht]
    \centering
    \includegraphics[width=0.95\linewidth]{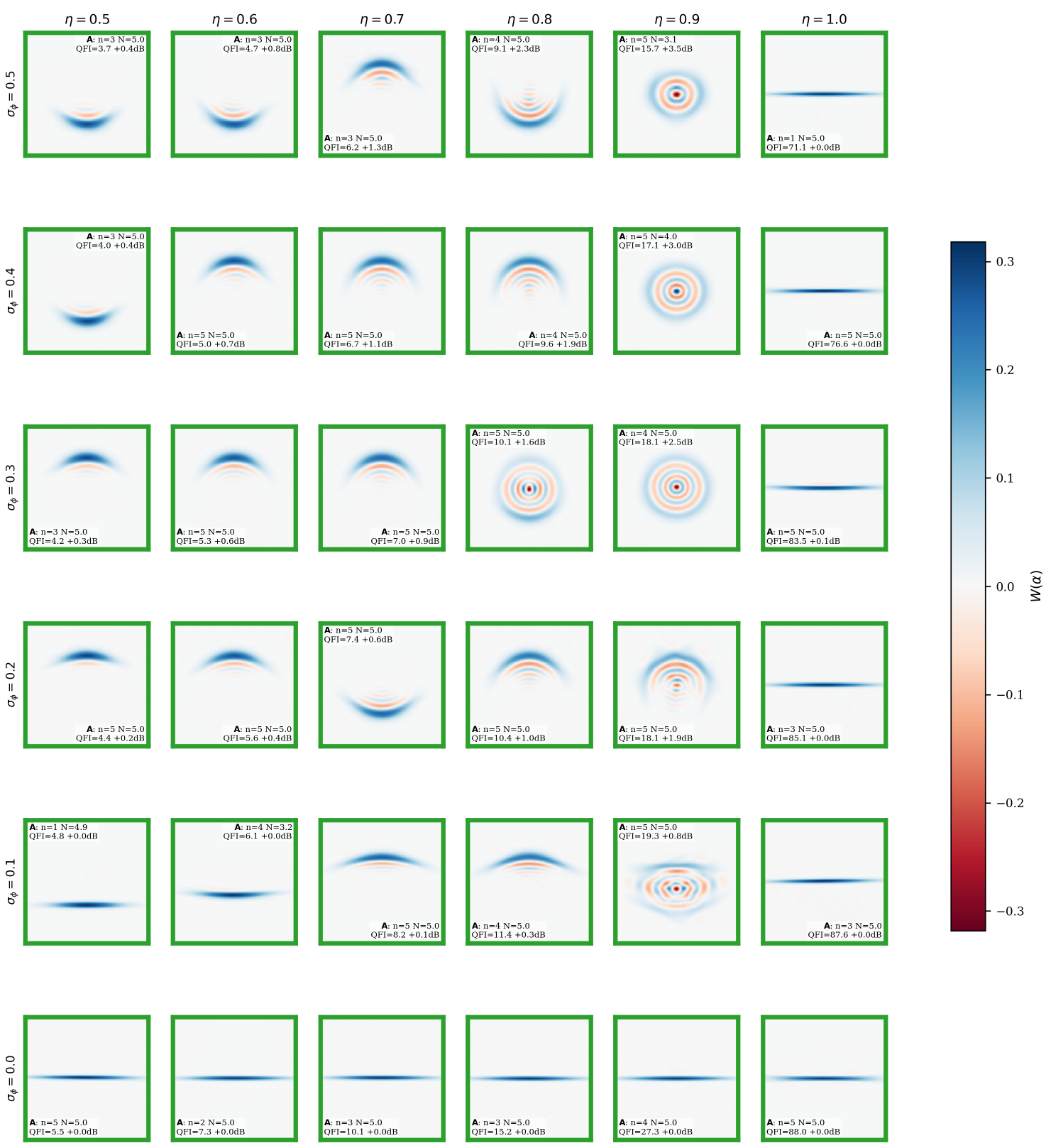}
    \caption{Wigner functions of the optimized displaced squeezed-vacuum superposition states across the $(\eta,\,\sigma_\phi)$ grid at $\bar{N}=5$. Layout as in Fig.~\ref{fig:best_fock_N5}.}
    \label{fig:best_sqzcoh_N5}
\end{figure}

For each state ansatz, the corresponding optimized Wigner functions across the $(\eta,\,\sigma_\phi)$ grid are shown in Figures~\ref{fig:best_fock_N5},~\ref{fig:best_sqzvac_N5}, and~\ref{fig:best_sqzcoh_N5}.




\subsection{Advantage of displacement for Gaussian inputs}

\begin{figure}
    \centering
    \includegraphics[width=0.5\linewidth]{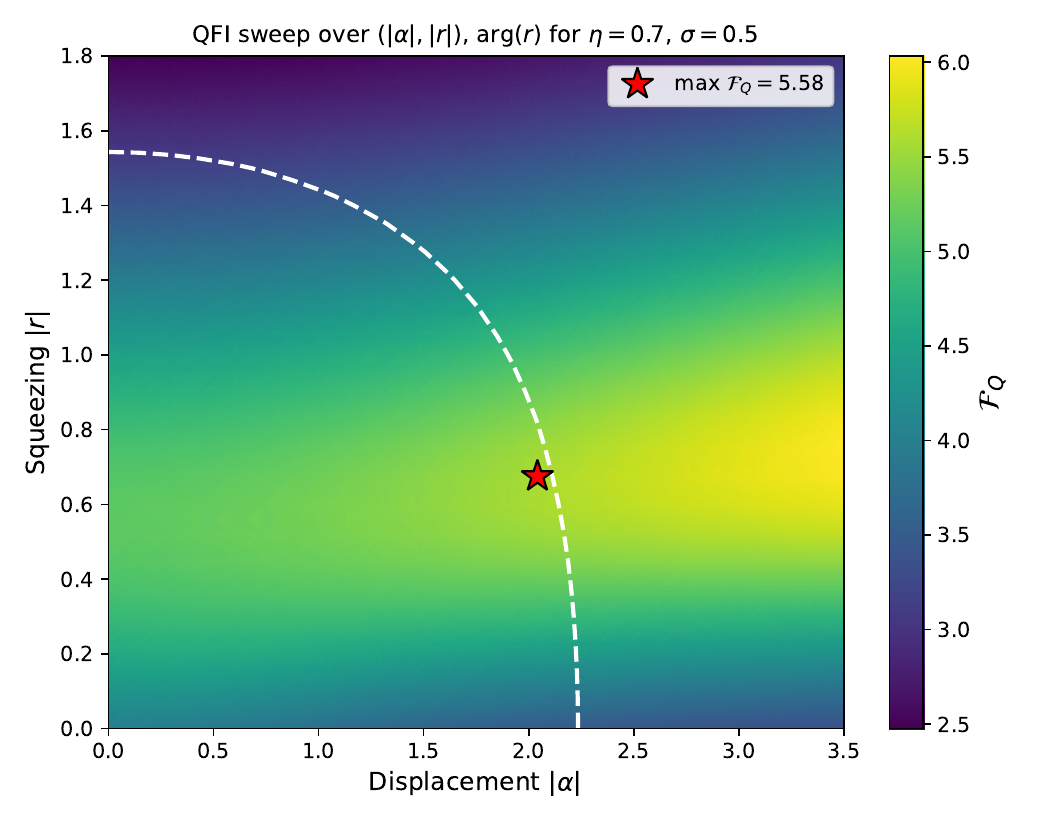}
    \caption{QFI of a displaced, squeezed state $|\alpha, r\rangle = \hat{D}(\alpha)\hat{S}(r)|0\rangle$ as a function of the squeezing $|r|$, and displacement $|\alpha|$. We assume a photon number budget of $\bar n \leq 5$, which is indicated with the white dashed lines ($\bar n \leq 5$ corresponds to the southwest of the dashed line). The angles of squeezing and displacement are chosen such that the $\hat p$ quadrature is squeezed, and the displacement is along this quadrature as well. The optimal parameters that maximize the QFI are marked with a red star, for which $|r^*| = 0.675$, and $|\alpha^*|= 2.04$.}
    \label{fig:displaced_gaussian_grid_theory}
\end{figure}

\begin{figure}
    \centering
    \includegraphics[width=\linewidth]{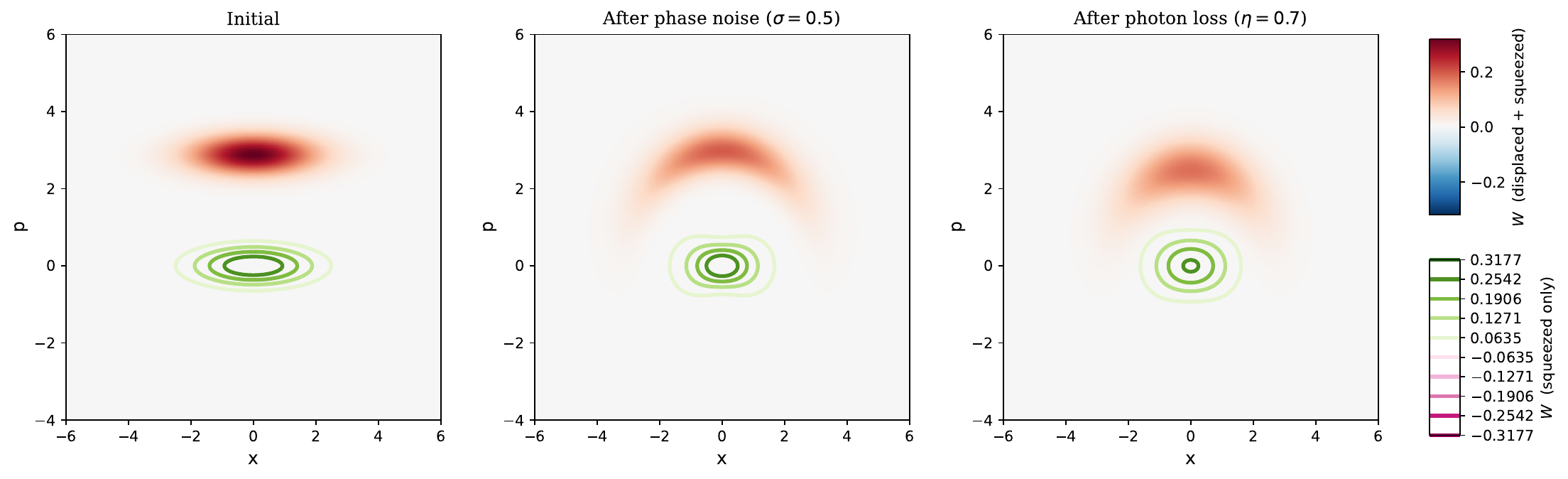}
    \caption{Wigner functions of two Gaussian states with the same squeezing level $|r| = 0.675$. One of the states is displaced by $|\alpha|= 2.04$ from the origin. We plot this state with the red shadows. In contrast, the undisplaced state is plotted with the green contour lines. The Wigner functions are computed for three stages: the initial state preparation, after experiencing the phase noise channel, and finally, after experiencing the photon loss. The photon loss channel has a transmission rate of $\eta = 0.7$, whereas the phase noise channel has a standard deviation $\sigma_\phi = 0.5$. }
    \label{fig:wigner_plots_disp_gaussian_theory}
\end{figure}


In the numerical optimization of the QFI with squeezed and displaced vacuum states, we have observed that the optimal states are not displaced from the origin in the absence of phase noise. This is in accordance with the optimality of squeezed states in the presence of photon loss only, as shown in Sec. \ref{subsec:noisy_phase_sensitive}. For this case, displacing the state increases the number of photons without providing any increase in the sensitivity with respect to the displacement that we aim to sense.
In contrast, for relatively high phase noise as well as photon loss, the optimal states are displaced from the origin, where the displacement depends on the relative strengths of the environmental noises, as well as the photon number budget. Here, we examine this effect. 

First, we show the optimality of displaced squeezed states for an example case of a photon loss with $\eta=0.7$, phase noise with $\sigma_\phi=0.5$ in Fig. \ref{fig:displaced_gaussian_grid_theory}. We plot the QFI with respect to the magnitude of the squeezing $|r|$, and displacement $|\alpha|$. We consider a photon number budget of $\bar n \leq 5$, which is indicated with the dashed white lines ($\bar n \le 5$ corresponds to $(|r|, |\alpha|)$ to the southwest of the boundary). The optimal QFI for this budget is 5.58, which occurs at $|r^*| = 0.675$, $|\alpha^*|= 2.04$, indicated with the red star on the Figure.  We compare this state to the undisplaced squeezed state with the same squeezing level $|r^*|$ in order to understand the advantage in the QFI. For this purpose, we draw the Wigner functions of these states in
Fig. \ref{fig:wigner_plots_disp_gaussian_theory} for three stages: (i) after the initial, noiseless preparation, (ii) after experiencing the phase noise channel, and (iii) after experiencing the photon loss channel. The undisplaced squeezed state is plotted with the green contour lines, whereas the displaced state is plotted with the red shadows.

From the Figure, we notice that the displaced state shows significant non-Gaussian features after the photon loss: the Wigner function has deviated from the elliptical shape associated with Gaussian states. In contrast, the undisplaced state is only slightly distorted. This is because the phase noise creates a mixture of rotated states, which results in a smearing of the Wigner function along the angular direction. For a displaced state, the same phase noise produces a larger spatial distortion, because the rotations act along an arc whose length scales with the displacement $|\alpha|$. This distortion result in increased sensitivity to displacements, as the radial width of the highly smeared Wigner function of the displaced Gaussian state is not altered significantly, compared to the undisplaced state (c.f. the rightmost plot in Figure \ref{fig:wigner_plots_disp_gaussian_theory}).

This small radial width can be exploited with, e.g. photon counting. To demonstrate this, we calculate the CFI obtained from homodyne detection and photon counting with both of these states. We observed that the CFIs for homodyne detection for the displaced and undisplaced states are 4.04 and 4.36, respectively, whereas the CFIs for photon counting are 5.25 and 4.29, respectively. Comparing these to their QFIs, 5.58 and 5.09, we observe that homodyne detection allows to reach $\approx 86\%$ of the QFI for the undisplaced state, and surpasses the CFI of the displaced state. Note that average number of photons in the undisplaced state is 0.529, compared to 4.70 in the displaced state. Then, we can obtain a higher CFI with homodyne detection using much less photons with the undisplaced state. However, the quantum advantage of displacement is apparent when we compare the CFIs for photon counting: we can reach $\approx 94\%$ of the QFI of the displaced state, and outperform the homodyne+squeezed state strategy. Thus, the optimality of the displaced Gaussian state results from the increased non-Gaussianity due to the phase noise channel, which can be exploited by non-Gaussian measurements, such as photon counting.

\clearpage
\section{Analyzing solutions that converge \label{sup:convergence_examples}}

As the number of superposition components $n_\mathrm{sup}$ increases, in some cases, the optimized states from different ansatze converge toward a common state, suggesting the existence of an underlying optimal state. 
This section shows four such examples of convergence: Sec. \ref{sup:subsec:cubic_convergence} shows convergence to a cubic phase state at low loss and moderate phase noise,  Sec. \ref{sup:subsec:binomial_convergence} shows convergence to a binomial-amplitude state in the same region, Sec. \ref{sup:subsec:fock_convergence} shows convergence to a Fock state at loss $\eta = 0.9$ and phase noise $\sigma_\phi = 0.3$ rad, , and Sec. \ref{sup:subsec:hexagon_convergence} shows convergence to a hexagonal grid state at low loss and moderate phase noise.


\subsection{Convergence to a cubic-phase state \label{sup:subsec:cubic_convergence}}



\begin{figure}[H]
    \centering
    \includegraphics[width=0.9\linewidth]{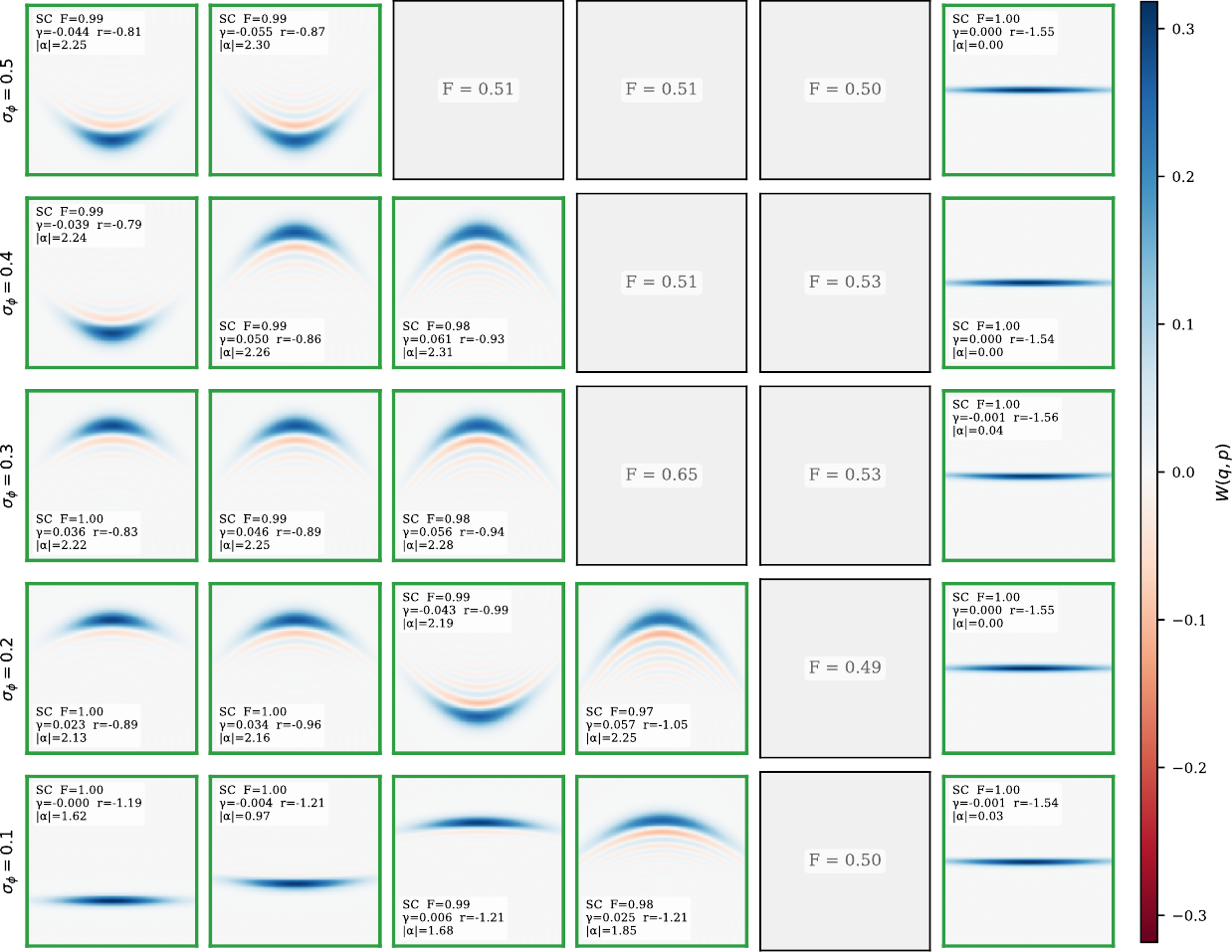}
    \caption{Wigner functions of the cubic phase with best fidelity match to states in 
    at each $(\eta, \sigma_\phi)$ point of the coarse scan at $\bar{N}=5$. Only Wigner functions at cells passing the $F > 0.95$ cubic-phase fidelity with optimized states threshold are shown. 
    These
cells define the cubic-phase-like class. Because the family reduces to squeezed vacuum
as $\gamma\to0$, near-zero-cubicity fits appear as squeezed Wigner functions, while
finite-$\gamma$ fits show the characteristic sheared fringes.
    The cubic phase states are parameterized as in Eq.~\eqref{eq:cubic-phase-state} with the optimized parameters annotated in the figure.}
    \label{sup_fig:opt-overlap-cubic}
\end{figure}

The cubic-phase reference family follows \cite{Yanagimoto2020} and carries four parameters,
\begin{equation}\label{eq:cubic-phase-state}
    \ket{\gamma, r, \theta, \alpha} \;=\; \hat D(\alpha)\,\hat R(\theta)\,e^{i\gamma\hat q^3}\,\hat S(r)\,\ket{0},
\end{equation}
with cubicity $\gamma$, squeeze $r$, rotation $\hat R(\theta) = e^{i\theta\hat n}$, and complex displacement $\alpha$; all four are jointly optimized at each $(\eta, \sigma_\phi)$ grid point to maximize the fidelity with the optimized state $\langle\hat n\rangle \leq \bar{N}$. Figure~\ref{sup_fig:opt-overlap-cubic} shows the best-matching Wigner function from this family at each grid point that clears a fidelity threshold of $F > 0.95$. 
The family contains pure squeezed vacuum as the $\gamma \to 0$ limit, so cells where the best fit returns a near-zero cubicity appear as squeezed Wigners and cells where the optimum has finite $\gamma$ show the sheared fringes.

\subsection{Convergence to a state with binomial-amplitude distribution} \label{sup:subsec:binomial_convergence}
\begin{figure}
    \centering
    \includegraphics[width=\linewidth]{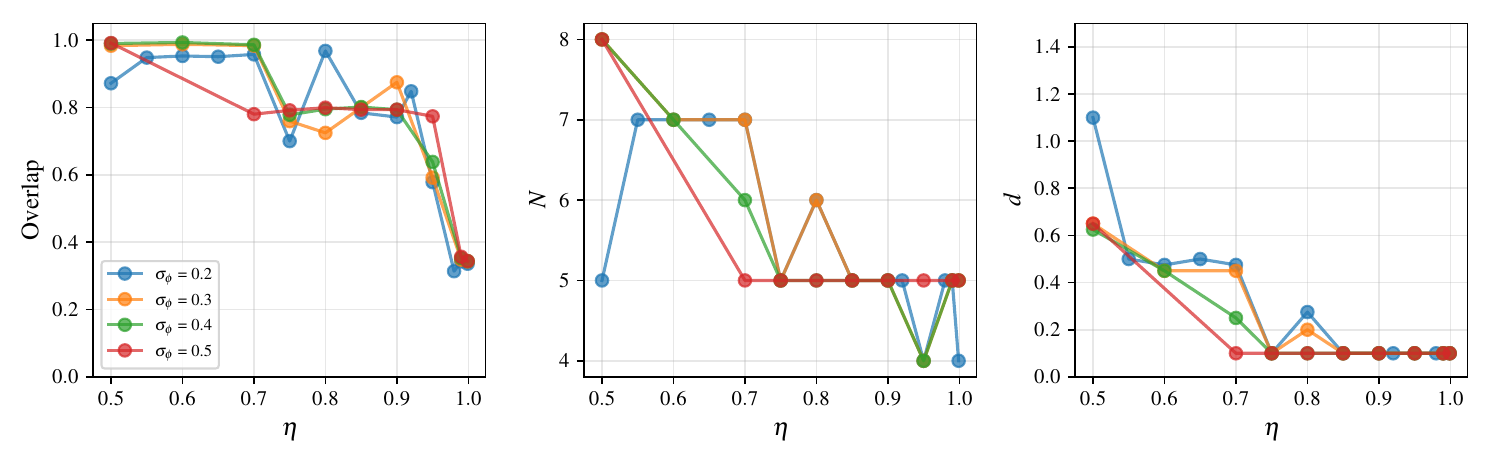}
    \caption{Optimization to maximize the overlap $|\braket{\psi_\text{opt}|\psi_{N,d}}|^2$ between the maximal QFI state $\ket{\psi_\text{opt}}$, and a binomial-amplitude state $\ket{\psi_{N,d}}$. We plot (i) the maximal overlap on the leftmost plot, (ii) corresponding $N$ parameter in the center plot, and (iii) corresponding $d$ parameter in the rightmost plot. For the definition of the binomial-amplitude state, see Ref. \cite{Brunelli2019}. We observe that these states show almost perfect overlap with the maximal QFI state for the moderate phase noise and photon loss regime of $\sigma_\phi \sim 0.3$, $\eta \sim 0.6$. }
    \label{fig:binomial_amplitude_overlap_theory}
\end{figure}

In the moderate phase noise and photon loss regime of $\sigma_\phi \sim 0.3$, $\eta \sim 0.6$, we observe a trend where the Wigner function of the state that maximizes the QFI has a main lobe with a crescent or a banana-like shape. This lobe is aligned with the direction of the displacement, along which there are also interference fringes. In order to classify this state, we chose to look into the following family of states that contains superposition of Fock states:
\begin{align}
    \ket{\psi_{N, d}} \propto \sum_{k=0}^N \binom{N}{k} d^{-k} \ket{k} 
\end{align}
which we refer to as the binomial-amplitude state. These states arise in the context of linear and quadratic reservoir engineering for dissipative preparation of pure, non-Gaussian states \cite{Brunelli2019}. We optimize over the parameters $N$, $d$ to maximize the overlap between a binomial-amplitude state and a given maximal QFI state. The results can be found in Fig. \ref{fig:binomial_amplitude_overlap_theory}, where we see excellent overlap for $0.3\leq \sigma_\phi \leq 0.4$, and $0.5 \leq \eta \leq 0.7$.

\begin{figure}
    \centering
    \includegraphics[width=0.6\linewidth]{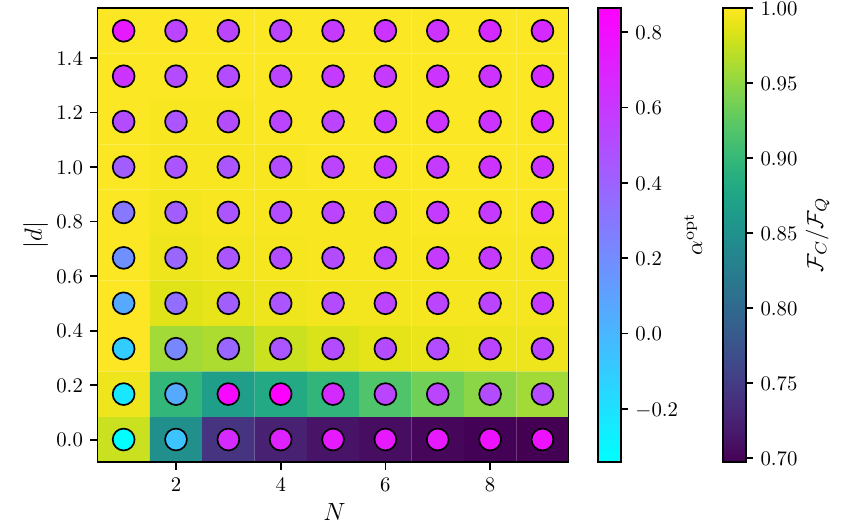}
    \caption{Sensitivity achieved with the binomial-amplitude state input for optimized photon counting $\mathcal{F}_C$, normalized by the QFI $\mathcal{F}_Q$, as a function of the binomial-amplitude state parameters $N$ and $|d|$. We displace the state right before the measurement by $\alpha^\text{opt}$ along the direction of the sensed displacement. This measurement can recover most of the QFI for a large part of the parameter space. }
    \label{fig:binomial_photon_counting_grid_theory}
\end{figure}

Intuitively, the reason why these states are optimal in this region of the parameter space arises from the following: the initial arc-like structure in the Wigner function protects against the phase noise channel, as the radial structure is mostly preserved. The photon loss results in the smearing of the main lobe in the radial direction, reducing the sensitivity. However, we can still benefit from this lobe structure by e.g. performing photon counting on the resulting state. Specifically, one can optimize over the photon counting measurement by first displacing the state by $\alpha$ in the $\hat{p}$ direction, then performing photon counting (i.e. by performing displaced PNR). In this way, we can recover most of the QFI by using optimal $\alpha$. 
This is demonstrated in Fig. \ref{fig:binomial_photon_counting_grid_theory}: we plot the CFI (obtained from the optimized photon counting) normalized by the QFI, $\mathcal{F}_C/\mathcal{F}_Q$, as well as the optimal displacement $\alpha^\text{opt}$, as a function of the binomial-amplitude state parameters $N$ and $|d|$. The phase of $d$ is selected to align the thinnest part of the main lobe in the $\hat{x}$ direction. 

\subsection{Convergence to a sparse Fock state \label{sup:subsec:fock_convergence}}

\begin{figure}[H]
    \centering
    \includegraphics[width=\linewidth]{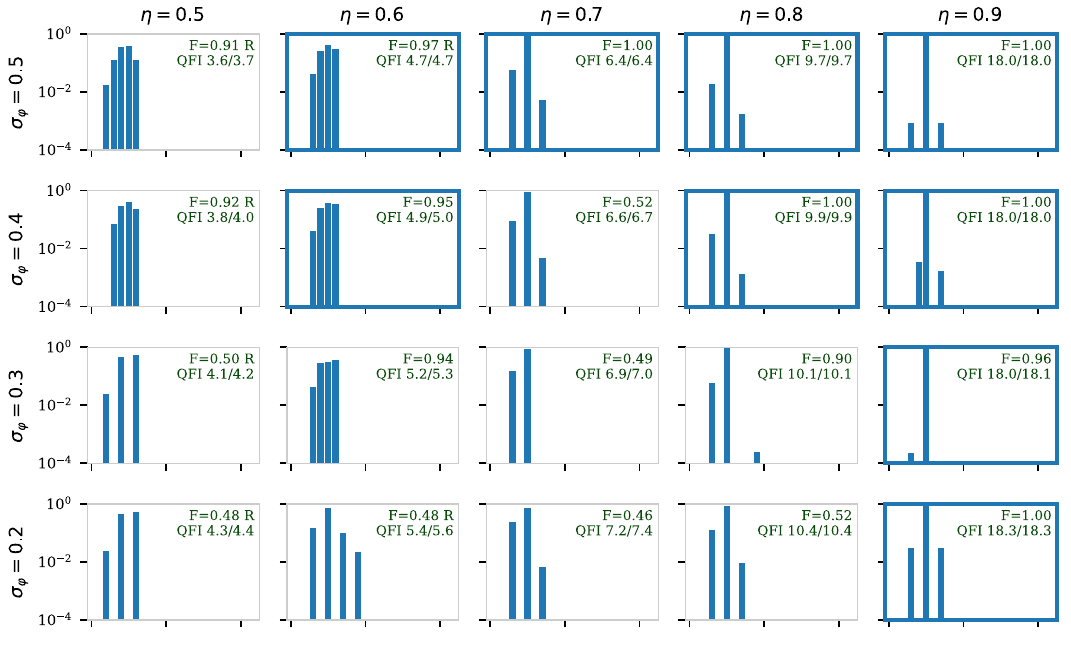}
    \caption{The Fock distribution of the optimized state belonging to the Fock superposition ansatz class
    at each $(\eta, \sigma_\phi)$ point of the coarse scan at $\bar{N}=5$. The probability of each Fock component $|c_n|^2$ is indicated on a log-scale. Only cells passing the $F > 0.95$ Fock-state fidelity with best overall optimized states threshold are outlined in blue. 
    }
    \label{sup_fig:opt-overlap-fock}
\end{figure}

While the lower left triangular region of the $(\eta, \sigma_\phi)$ grid has $>95\%$ overlap with a cubic phase state, the upper right region of the grid has high overlap with a state that is a sparse superposition of Fock states, with the dominant component being a single Fock state $|n\rangle$ with $n \approx \bar{N}$. The amplitudes of the coefficients of these superpositions optimized for each pair $(\eta, \sigma_\phi)$ is shown in Fig. \ref{sup_fig:opt-overlap-fock}, where the superpositions that achieve a fidelity higher than 0.95 are outlined in blue.

\begin{figure}[H]
    \centering
    \includegraphics[width=\linewidth]{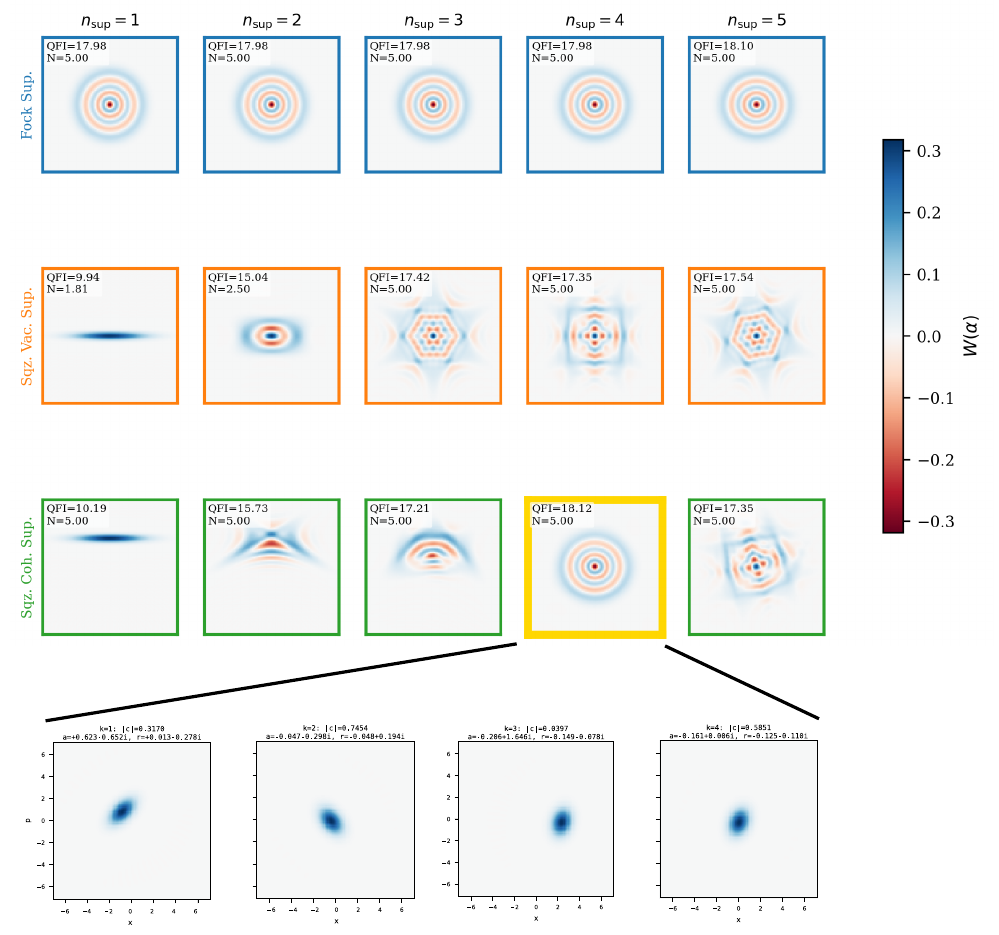}
    \caption{Evolution of the optimized state with increasing $n_\mathrm{sup}$ at $\eta=0.9$, $\sigma_\phi=0.3$~rad, $\bar{N}=5$. Even though the optimal state has a Wigner function that resembles a Fock state, we found this state via our squeezed-coherent superposition ansatz. The four components of the squeezed-coherent superposition are shown in the lower panel of the figure. Even though each component does not bear any resemblance to a Fock state, their superposition results in a Fock-like Wigner function via interference.}
    \label{fig:fock_nsups_evo}
\end{figure}

Figure~\ref{fig:fock_nsups_evo} shows the evolution of the optimized state as $n_\mathrm{sup}$ increases for each ansatz. The following figures show the component decomposition of these states at $\eta=0.9$, $\sigma_\phi=0.3$~rad, and $\bar{N}=5$.

\clearpage
\subsection{Convergence to a state with discrete rotational symmetry \label{sup:subsec:hexagon_convergence}}

\begin{figure}[H]
    \centering
    \includegraphics[width=\linewidth]{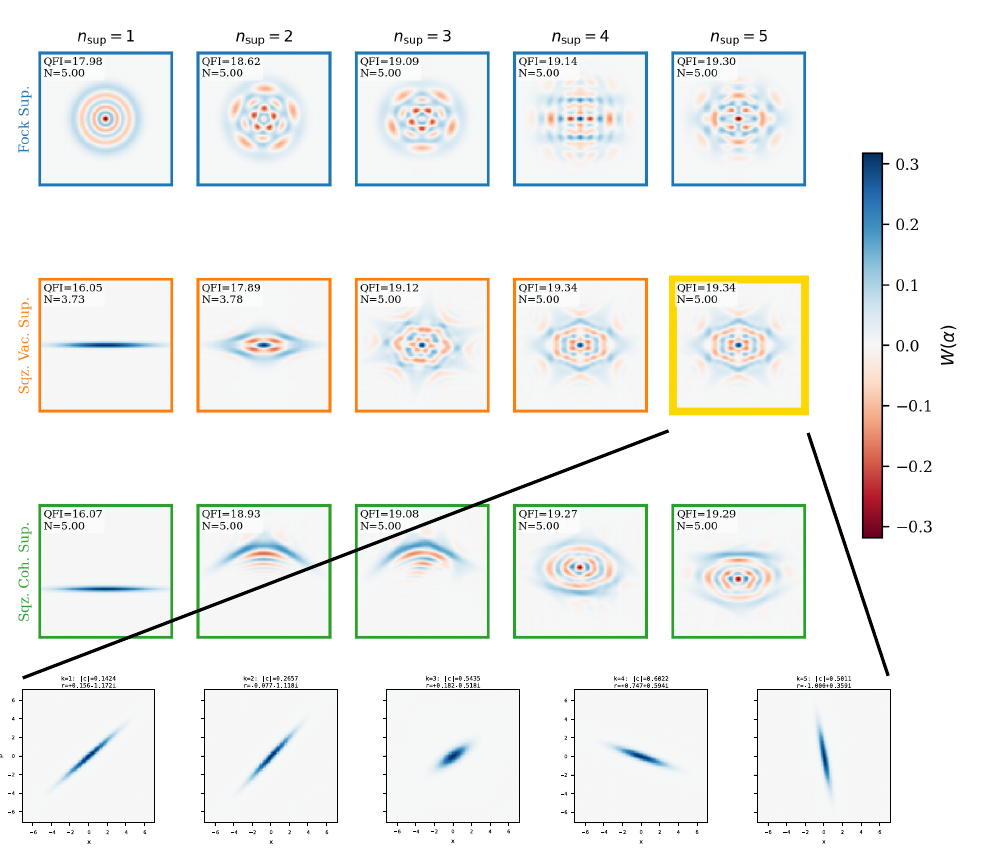}
    \caption{Evolution of the optimized state with increasing $n_\mathrm{sup}$ at $\eta=0.9$, $\sigma_\phi=0.1$~rad, $\bar{N}=5$. In the lower panels, we show the five components of the squeezed-vacuum superposition ansatz which achieved the maximum QFI.  We can see that five different squeezed angles were superposed to achieve a nearly hexagonal pattern in the Wigner function. }
    \label{fig:hex_nsups_evo}
\end{figure}

At low loss ($\eta=0.9$) and moderate phase noise ($\sigma_\phi=0.1$~rad) with $\bar{N}=5$, the optimized states exhibit hexagonal symmetry in their Wigner functions. Figure~\ref{fig:hex_nsups_evo} shows the evolution of the optimized state as $n_\mathrm{sup}$ increases for each ansatz.  Even though the individual components within each ansatz do not resemble the final optimal state, their superposition managed to achieve such a nearly hexagonal pattern. We also note that the optimal QFI achieved within each ansatz are comparable to each other.  In the lower panel, we show the five components of the squeezed-vacuum superposition ansatz which achieved the maximum QFI.  We can see that five different squeezed angles were superposed to achieve a nearly hexagonal pattern in the Wigner function. 

%

%

%

\clearpage

\section{Obtainable advantage with known measurements and states}\label{sup:cfi_section}

\subsection{Advantage with cat states}

\begin{figure}[H]
    \centering
    \includegraphics[width=0.5\linewidth]{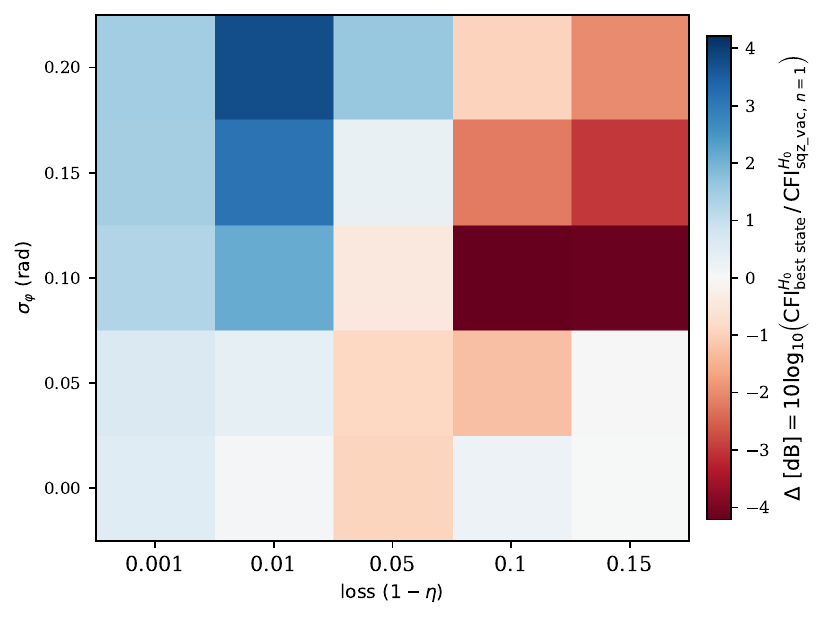}
    \caption{Classical Fisher information (CFI) advantage, under balanced homodyne at fixed local-oscillator phase $\theta_{\textrm{LO}}=0$, of the best optimized state (the one maximizing the QFI) over the best squeezed vacuum, across the low-loss $(\eta,\,\sigma_\phi)$ grid at $\bar{N}=5$. The color encodes $\Delta\,[\mathrm{dB}] = 10\log_{10}\!\left(\mathrm{CFI}^{H_0}_{\textrm{best}}/\mathrm{CFI}^{H_0}_{\textrm{sqz.\,vac.}}\right)$. Blue (positive) regions mark a non-Gaussian advantage realizable with homodyne, while red (negative) regions indicate that the QFI-optimal state, read out with fixed homodyne, underperforms the squeezed vacuum, so that its advantage requires a non-Gaussian measurement.}
    \label{fig:cat_homodyne_grid_cfi}
\end{figure}

\begin{figure}[H]
    \centering
    \includegraphics[width=0.8\linewidth]{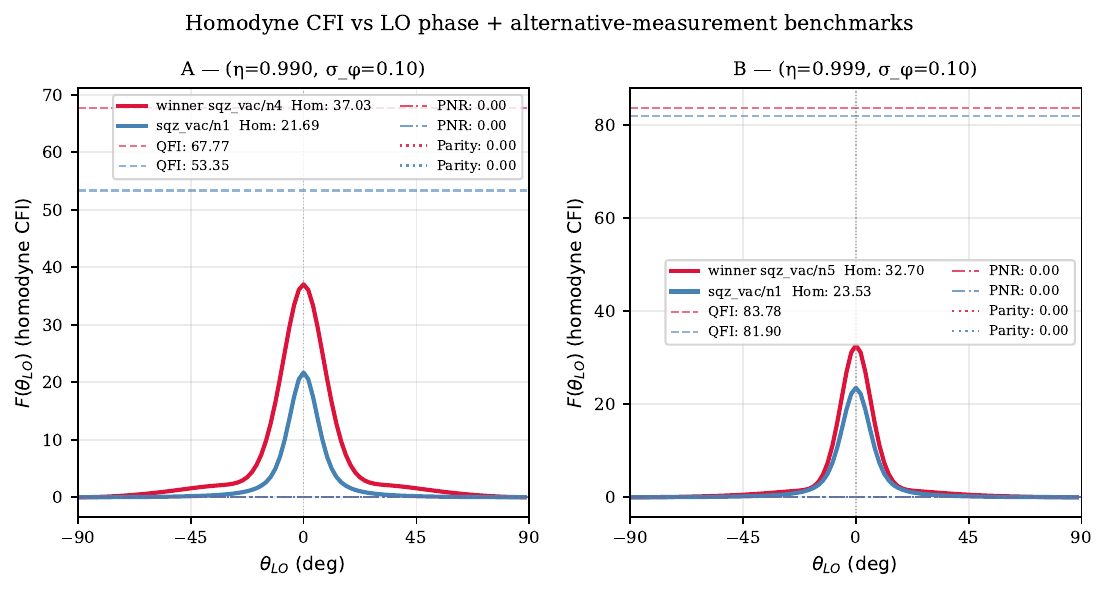}
    \caption{CFI vs homodyne angle for the cat state and the best squeezed vacuum state at $\eta=0.99, 0.999$, $\sigma_\phi=0.1$~rad, $\bar{N}=5$. The cat state shows a clear advantage over the squeezed vacuum state across all homodyne angles. Since the QFI is higher than the CFI for both states, there is potential for further improvement with more general measurements.}
    \label{fig:cat_vs_sqzvac_benchmarks}
\end{figure}

We now examine how much of the non-Gaussian advantage is accessible with balanced homodyne detection---the readout used in current detectors. Figure~\ref{fig:cat_homodyne_grid_cfi} maps the homodyne CFI of the QFI-optimal state relative to the best squeezed vacuum across the low-loss grid, while Fig.~\ref{fig:cat_vs_sqzvac_benchmarks} shows that experimentally accessible cat states already realize an advantage over the squeezed vacuum under homodyne readout.

\subsection{Advantage for various measurement strategies}

\begin{figure*}
    \centering
    \includegraphics[width=\linewidth]{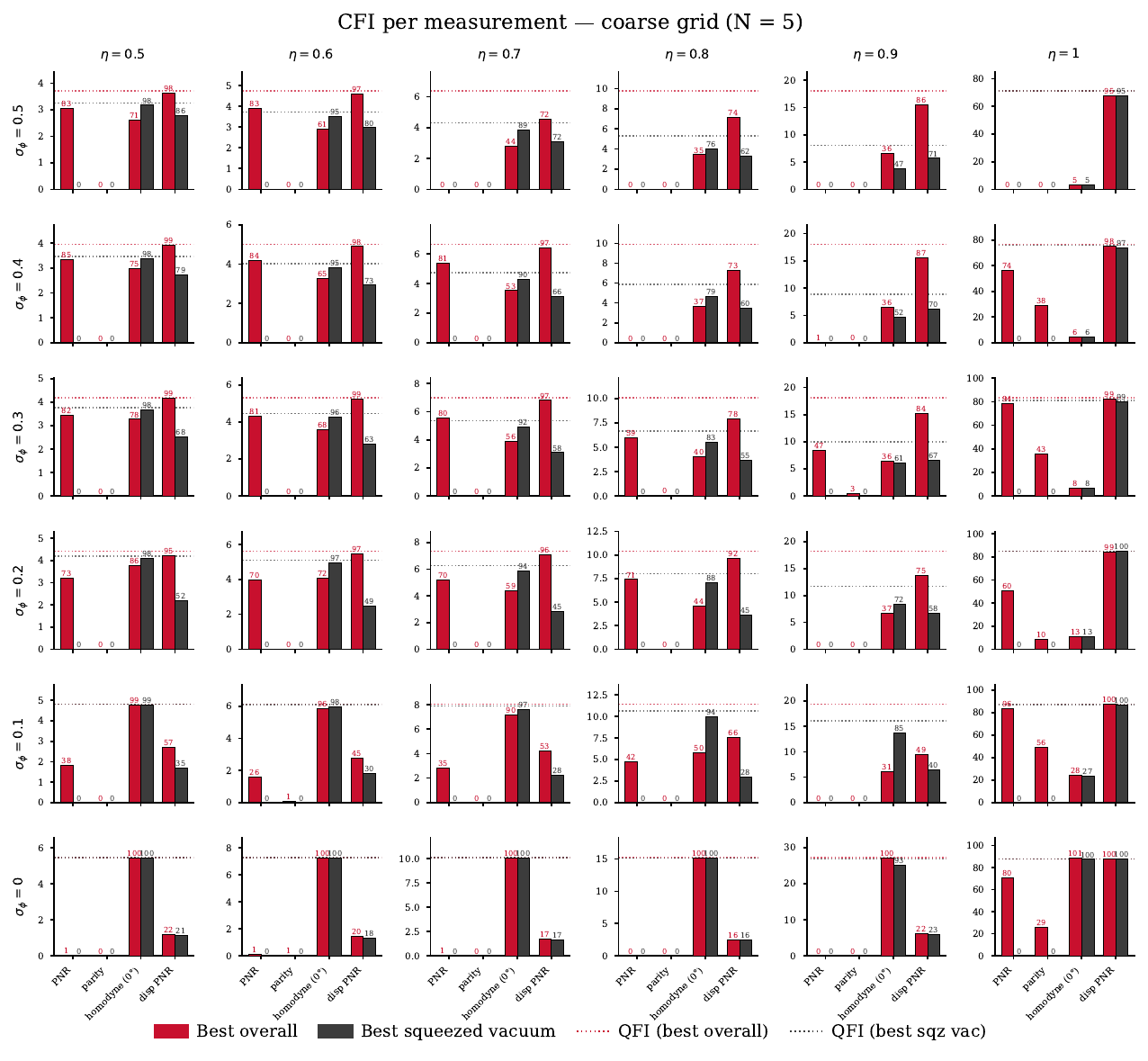}
    \caption{Classical Fisher information (CFI) for several measurement strategies, as a fraction of the QFI, across the coarse $(\eta, \sigma_\phi)$ grid.}
    \label{em_fig:cfi_histograms_coarse}
\end{figure*}

\begin{figure*}[!t]
    \centering
        \includegraphics[width=0.8\linewidth]{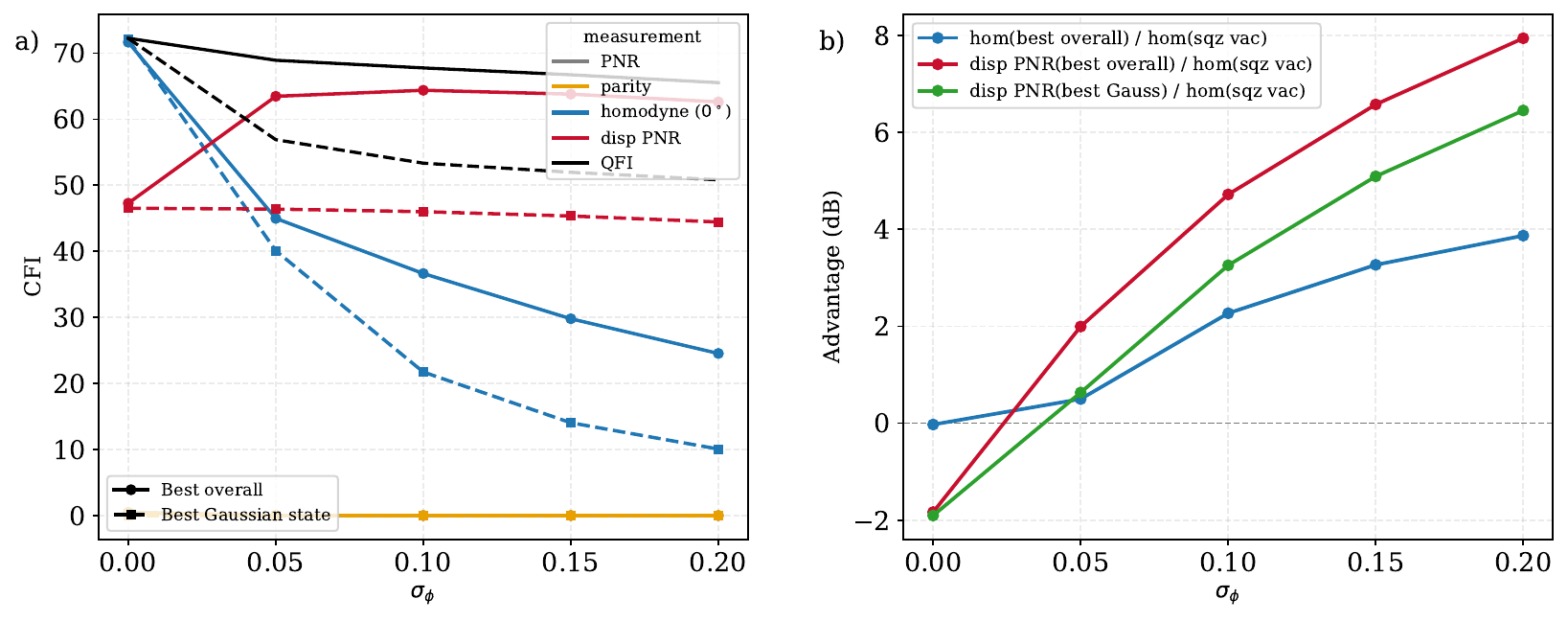}
    \begin{subfigure}{0pt}\phantomcaption\label{em_fig:cfi_lines_eta0p99}
    \end{subfigure}

    \vspace{0.5em}

    \begin{subfigure}{0pt}        \phantomcaption\label{em_fig:cfi_advantage_eta0p99}
    \end{subfigure}
    \caption{\textbf{Classical Fisher information of feasible measurements at $\eta=0.99$.} (a) CFI of each measurement (PNR, parity, homodyne at optimized angle $\theta=0^\circ$, displaced PNR, displaced parity) and the QFI (dotted) versus phase noise $\sigma_\phi$ at $\eta=0.99$, $N=5$, for the best overall state and the best Gaussian state. (b) Metrological advantage (dB) of the displaced-PNR readout relative to homodyne of the squeezed vacuum, versus $\sigma_\phi$.}
    \label{em_fig:cfi_finite_loss}
\end{figure*}

\bibliography{apssamp}